\documentclass[11pt]{article}

  \usepackage{amsmath, amssymb}
  \usepackage{graphicx}
  \usepackage{enumerate}
  \usepackage{natbib}
  \usepackage{url}
  \usepackage{bm}   
  \IfFileExists{bbm.sty}{\usepackage{bbm}}{\providecommand{\mathbbm}{\mathbb}} 
 \usepackage{booktabs}

  \renewcommand{\harvardurl}[1]{\textbf{URL:} \url{#1}}
 \usepackage[margin=1in]{geometry}

\newcommand{\argmin}{\mathop{\mathrm{arg\,min}}}

\begin{document}
  
  \def\spacingset#1{\renewcommand{\baselinestretch}%
  {#1}\small\normalsize} \spacingset{1}

    \title{\bf Heterogeneity and Subgrouping in Modern Statistics and AI}
    \author{%
  \begin{tabular}{cc}
    Xuan Bi\textsuperscript{*} & Spencer Hilligoss\textsuperscript{*} \\
    \small University of Minnesota & \small University of California, Irvine \\[0.6em]
    Cadence Pinkerton\textsuperscript{*} & Xiwei Tang\textsuperscript{*} \\
    \small University of California, Santa Barbara & \small University of Texas at Dallas \\[0.6em]
    Fei Xue\textsuperscript{*} & Annie Qu\textsuperscript{*} \\
    \small Purdue University & \small University of California, Santa Barbara
  \end{tabular}}
    \date{}
    \maketitle

  \begingroup
\renewcommand{\thefootnote}{*}
\footnotetext{These authors contributed equally to this work.}
\endgroup

  
  \bigskip
  \begin{abstract}
  Heterogeneity is a defining feature of modern data: individuals, subpopulations, sites, modalities, tasks, and time points may differ in distributions, associations, treatment effects, or decision objectives. This review develops a unified perspective in which heterogeneous analysis is a problem of structured information borrowing—deciding what should be shared, what should remain distinct, and how uncertainty about that structure should enter inference and decisions. We connect classical stratification, interaction models, finite mixtures, mixed-effects models, and empirical Bayes with subgroup discovery, individualized learning, synthetic data generation, large language models, federated learning, privacy, security, and mobile health. Precision medicine provides a central example: subgroup rules offer interpretable approximations to individualized response surfaces, whereas partially pooled models stabilize patient-specific estimation. Across applications, the same tensions recur: population averages can conceal consequential variation, unrestricted personalization is unstable, adaptive subgroup discovery creates selection bias, and borrowing across incompatible sources causes negative transfer. These difficulties are amplified in contemporary AI, where sparse user histories, entangled representations, heterogeneous preferences, feedback loops, distribution shift, and privacy constraints make reliable personalization and evaluation especially challenging. We conclude with an agenda for calibrated personalization that integrates multilevel representations, selective borrowing, subgroup-aware evaluation, and valid uncertainty quantification.

  \end{abstract}

  \noindent%
  \vfill

  \newpage
  \spacingset{1.45} 


\section{Introduction}\label{sec:introduction}

Modern data rarely come from units that are interchangeable for every scientific purpose. Patients with the same diagnosis may follow different disease courses or respond differently to treatment; hospitals may measure similar variables under different protocols; and users may expect different answers from the same language model. A population average can summarize these data, but it can conceal variation that matters for explanation, prediction, and decisions.

The statistical problem is to distinguish systematic heterogeneity from sampling noise and then decide how it should affect analysis. Three choices are linked: \textbf{representation} specifies what varies and at what level; \textbf{borrowing} determines which information can be shared across units; and \textbf{inference} quantifies uncertainty in both estimated effects and any learned structure. Classical stratification, interactions, mixtures, and random effects, as well as modern transfer, multi-task, and federated learning, can all be viewed through these choices.

Heterogeneity can reveal structure hidden by an average, but searching many candidate groups can produce unstable labels and exaggerated effects. The goal is therefore calibrated personalization: retain differences relevant to the target while borrowing enough information for reliable estimation. Neither complete pooling nor a separate model for every unit is generally adequate.

\subsection{Heterogeneity in conventional statistical analysis}\label{heterogeneity-in-conventional-statistical-analysis}

A common model is often a useful starting point. It defines a shared estimand and supports efficient estimation when the modeled relationship is sufficiently stable. Classical methods also provide established ways to represent departures: stratification describes known groups, interactions allow effects to vary with observed covariates, mixed-effects models allow unit-specific coefficients, and finite mixtures represent latent populations \citep{laird1982random,mclachlan2000finite}. These formulations differ in whether variation is observed or latent, discrete or continuous, and fixed or random, but each balances a common structure against local differences.

The target determines whether a departure matters. A group with high outcome risk need not have a different treatment effect, and a cluster that is distinct in covariate space need not be useful for a decision. Longitudinal data add another level: subjects may differ in baseline, trajectory, or response to an exposure. A continuous random-effects distribution and a latent-class model impose different structures on these differences; neither is a default description of nature. Section~\ref{sec:foundational-concepts-and-classic-methods} develops these distinctions.

Data-adaptive subgroup discovery also changes the inferential problem. When the same data are used to find and evaluate groups, conventional uncertainty calculations can ignore selection and overstate evidence. Small groups make this especially difficult because they are both hard to estimate and easy to discover by chance. Section~\ref{sec:highdim-clustering-inference} reviews inference after discovery, and Section~\ref{sec:uq} treats the broader uncertainty questions.

\subsection{Heterogeneity in modern data analysis and AI}\label{heterogeneity-in-modern-data-analysis-and-ai}

High-dimensional data make richer forms of heterogeneity observable through profiles, images, molecular measurements, text, wearable trajectories, and multiple modalities. Flexible learners can capture complex interactions, yet a highly individualized prediction does not establish that the differences it uses are stable, causal, or transportable.

Modern learning methods differ in how they share information. Multi-task and transfer methods share representations across related tasks while adapting to a target; federated learning combines information across sites that may differ in population or measurement; multimodal methods seek common signals while retaining modality-specific information. In each case, the central statistical choice is which sources or components are informative for the target and which may cause negative transfer. Section~\ref{sec:federated} examines this choice across sites.

Foundation models and large language models make the same issue visible at scale. Pretraining pools information across domains, tasks, languages, and users, often optimizing average performance. Preference alignment can likewise compress diverse judgments into a common objective. These shared representations are useful, but aggregate success can coexist with poor performance for rare users, specialized tasks, or less represented languages. Personalization is difficult because histories are sparse and selective, preferences may change or conflict, and feedback depends on the system's earlier responses. Evaluation must therefore consider the target group and interaction context, not only an overall benchmark score. Sections~\ref{sec:ai} and~\ref{sec:heterogeneous-rl} discuss adaptation and sequential decisions.

\subsection{Applications across disciplines}\label{applications-across-disciplines}

Precision medicine illustrates why the distinctions matter. Patients may differ in prognosis, treatment response, toxicity, and preferences. Molecular and clinical data can support subgroup descriptions, while an individualized treatment rule uses patient characteristics to select an action. A conditional average treatment effect describes mean effect variation across covariate profiles; it is not an individual's realized contrast between two potential outcomes. Estimating either a treatment effect or a treatment rule requires causal identification and uncertainty assessment, discussed in Section~\ref{sec:causal}.

Data integration presents a related borrowing problem. Multi-study consortia, health systems, and biobanks combine sources with different populations, protocols, devices, and missingness patterns. Pooling can increase precision, but a source useful for one target or model component may be irrelevant or harmful for another. The analysis must state what is shared and for which target; when records cannot be centralized, communication and privacy constraints also matter.

Personalized AI combines these concerns. A model must reuse population-level knowledge while adapting to a user, task, or context with limited observations. It must also evaluate performance for groups and use cases that contribute little to average metrics. Across medicine, data integration, and AI, the recurring question is how to preserve relevant departures without treating noisy or selected differences as established structure.

\subsection{Scope and organization of this review}\label{scope-and-organization-of-this-review}

Throughout, we ask when sharing information improves estimation and when it risks obscuring consequential differences. Section~\ref{sec:foundational-concepts-and-classic-methods} introduces the core representations and partial pooling. Sections~\ref{sec:highdim-clustering-inference}--\ref{sec:heterogeneous-rl} cover subgroup discovery, causal effects, synthetic data, AI personalization, and sequential learning. Sections~\ref{sec:uq}--\ref{sec:mhealth} examine uncertainty, federated learning, security, and mobile-health applications; Sections~\ref{sec:open_problems} and~\ref{sec:conclusion} close with open problems and conclusions. Throughout, subgrouping is treated as one useful representation of variation rather than proof that a population consists of discrete natural types.

\section{Foundational Concepts and Classical Methods}\label{sec:foundational-concepts-and-classic-methods}

A heterogeneous analysis should specify four objects: the \textbf{unit} that may differ (person, subgroup, site, modality, task, or time); the \textbf{component} that varies (distribution, association, treatment effect, measurement, or decision objective); the \textbf{structure} of variation; and the \textbf{target} for estimation or prediction. Without these choices, a method may describe variation at the wrong level or answer a different question from the one intended.

Let $r$ index the unit and let $\theta_r$ denote its parameter or predictive function. A useful representation is
\[
  \theta_r=\theta_0+\delta_r,
\]
where $\theta_0$ is shared structure and $\delta_r$ is a unit-specific departure. Homogeneity sets $\delta_r=0$; separate models leave departures unrestricted. Most heterogeneous methods lie between these limits by grouping, shrinking, or smoothing departures. Excessive pooling can bias the target, while too little pooling makes estimates unstable.

\subsection{Representations of heterogeneity}\label{homogeneous-model-versus-heterogeneous-model}

For example, a common regression model assumes $\mathbb{E}(Y_i\mid X_i=x)=m(x;\theta_0)$. This does not require identical covariate distributions: it states that the conditional outcome relationship is shared. A heterogeneous model allows the relationship to depend on a modifier, group, or unit. The appropriate form depends on how units are identified and on the target of the analysis.

\begin{table}[ht]
\centering
\small
\setlength{\tabcolsep}{4pt}
\begin{tabular}{p{2.6cm}p{4.0cm}p{4.1cm}p{4.0cm}}
\toprule
Representation & How variation is represented & Typical borrowing & Main interpretive caution\\
\midrule
Observed groups & Membership is defined by measured characteristics or a prespecified rule. & Estimate group-specific effects, often with shrinkage toward a common mean. & The rule defines the target; a prognostic group need not be a treatment-benefit group.\\
Latent classes & Units belong to one of several unobserved components. & Share information within estimated classes. & A fitted class is not automatically a distinct population or a stable label for future units.\\
Random effects & Unit-specific coefficients vary around a population distribution. & Partial pooling is stronger for units with less information. & Distributional assumptions shape shrinkage; uncertainty for a particular unit differs from population-average uncertainty.\\
Mixed membership & A unit combines multiple shared components, for example $\theta_r=\sum_g\pi_{rg}\beta_g$. & Borrow across components in proportions that may vary by unit or model component. & Membership can depend on the outcome, context, and target; one global label may be inadequate.\\
\bottomrule
\end{tabular}
\caption{Common representations of heterogeneity and their borrowing assumptions. These structures are alternatives to consider, not interchangeable guarantees.}
\label{tab:heterogeneity-representations}
\end{table}

Pooling can be expressed as regularization. If $L_r(\theta_r)$ is the loss for unit $r$, an estimator may minimize
\[
  \sum_{r=1}^R L_r(\theta_r)+\lambda\,\mathcal{P}(\theta_1,\ldots,\theta_R),
\]
where the penalty $\mathcal{P}$ encodes which departures should be small, equal, smooth, or shared through lower-dimensional structure. The tuning parameter controls the granularity of the fit. Its choice should reflect the target: optimizing average prediction alone can conceal errors for a small subgroup or a designated site.

\subsection{Subgroup analysis}\label{subgroup-analysis}

A subgroup rule $G(X)$ defines a set of units for a particular target, such as a mean, risk, association, treatment effect, or decision value. Three distinctions guide interpretation. First, prespecified groups and data-adaptive groups are different inferential objects. In the latter, selecting a partition and estimating its contrasts on the same data can create selection bias. Stability is useful evidence of reproducibility, but does not by itself establish valid post-selection coverage or show that the selected groups correspond to latent populations. Section~\ref{sec:highdim-clustering-inference} discusses the distinction between selected clusters and latent-component targets; Section~\ref{sec:uq} reviews inferential adjustments.

Second, prognostic and predictive subgroups answer different questions. Prognosis concerns outcome level under a reference condition; treatment-effect heterogeneity concerns differences in the contrast between treatments. The conditional average treatment effect is a covariate-conditional mean and does not identify an individual's realized treatment contrast. Causal interpretation also requires assumptions appropriate to the design and target, developed in Section~\ref{sec:causal}.

Third, a partition is a representation, not proof that nature is divided into types. Covariate clusters need not be relevant to an outcome or decision, and a treatment-benefit rule may overlap substantially in covariate space with other patients. Hard groups can make a complex effect surface interpretable, but cut points may be unstable and individuals near a boundary may be similar. The analysis should therefore state what the group means, how a future unit would be assigned, and whether evaluation concerns the selected rule or an underlying population structure.

\subsection{Mixed-effects models and partial pooling}\label{mixed-effects-models-and-partial-pooling}

For repeated measurements on subject $i$ at occasion $j$, a linear mixed model is
\[
  Y_{ij}=X_{ij}^{\mathsf T}\beta+Z_{ij}^{\mathsf T}b_i+\varepsilon_{ij},
  \qquad b_i\sim G,\quad \mathbb{E}(b_i)=0.
\]
The population coefficient $\beta$ describes shared effects, while $b_i$ represents subject-specific departures. Extensions to generalized outcomes, time-varying coefficients, and nested site and patient effects retain the same basic idea \citep{laird1982random}.

Partial pooling combines each unit's data with the estimated population distribution. A unit with a short or noisy history is shrunk more toward the shared component; a unit with more informative observations can receive a more individualized estimate. This is the classical form of calibrated personalization and is closely related to empirical Bayes estimation \citep{morris1983parametric}. For a new unit, prediction relies on the population distribution and whatever measured information is available to explain its departure.

Random-effects and subgroup models differ chiefly in the structure imposed on the distribution of departures. A continuous distribution represents graded variation; a discrete distribution yields latent classes; mixtures of continuous distributions allow both class-level and within-class variation. These assumptions affect interpretation and uncertainty. A stable population distribution does not eliminate uncertainty about an individual effect when that person's history is sparse, and partial pooling alone does not address confounding or model misspecification. Section~\ref{sec:uq} distinguishes population-average guarantees from uncertainty statements for particular groups or individuals.

The foundations yield three principles for the rest of the review: define heterogeneity relative to a component and target; treat subgrouping and personalization as structured borrowing; and account for uncertainty in the learned sharing structure when making individual, subgroup, or downstream decisions.

\section{High-Dimensional Clustering and Inference}
\label{sec:highdim-clustering-inference}

Clustering is rarely the final scientific object in a high-dimensional analysis. Investigators also ask which variables define the estimated groups, which cluster pairs differ on a feature, how stable the assignments are, and whether these differences support calibrated inferential claims. These questions are especially visible in single-cell genomics, where one expression matrix may be reused for variable-gene screening, representation learning, graph construction, clustering, marker discovery, and annotation. Interpretation and inference therefore depend on how the subgroup structure was learned.

Let $X=(X_1,\ldots,X_n)^\top\in\mathbb R^{n\times p}$. A generic adaptive clustering rule can be written as
\[
  \widehat z=\mathcal A\!\left(X;\widehat K,\widehat\lambda,
  \widehat{\mathcal R}\right).
\]
where $\widehat K$ is an estimated number of clusters or graph resolution, $\widehat\lambda$ is a tuning parameter, and $\widehat{\mathcal R}$ is an estimated feature set, subspace, embedding, or graph. The resulting target depends on the formulation: a finite-mixture model represents latent distributional components, $K$-means targets a population quantization, and a graph pipeline identifies communities in an estimated graph \citep{FraleyRaftery2002,KiselevEtAl2019}. Two broad workflows organize the literature. The first estimates clusters and then performs inference on features or contrasts using the estimated labels. The second estimates subgroup structure together with informative features or a discriminative subspace. An accompanying schematic of these workflows is provided in Online Supplement~A.

\subsection{Post-Clustering Inference}
\label{subsec:two-stage-post-clustering}

A two-stage analysis first obtains $\widehat z=\mathcal A(X)$ and then tests a contrast such as
$
  H_{0,j}^{kk'}(\mathcal T):\quad
  \mu_{kj}^{\mathcal T}=\mu_{k'j}^{\mathcal T},
$
where $\mathcal T$ must identify the inferential target. Three common targets are
\[
  \mu_{kj}^{\mathrm{sel}}
  =|\widehat C_k|^{-1}\!\sum_{i\in\widehat C_k}\mathbb E(X_{ij}),
  \qquad
  \mu_{kj}^{g}=\mathbb E\{X_j\mid g(X)=k\},
  \qquad
  \mu_{kj}^{\mathrm{mix}}=\mathbb E(X_j\mid Z=k).
\]
They respectively describe the average mean parameter attached to the selected observation indices, a population region defined by a fixed rule $g$, and a latent mixture component $Z$. The first is conditional on the data-dependent collection $\widehat C_k$; the second becomes a population quantity only after a rule is fixed; and the third depends on an identifiable latent-variable model. Inference calibrated for one target does not automatically apply to another. In particular, a valid contrast between selected clusters is not automatically inference for latent biological component means.

For example, in a standard single-cell pipeline, normalization and variable-gene screening are followed by dimension reduction or integration, nearest-neighbor graph construction, community detection, and marker analysis \citep{AndrewsEtAl2021,ButlerEtAl2018,HaoEtAl2021,WolfEtAl2018}. A conventional two-sample, rank-based, or regression test then treats the estimated labels as fixed. Yet the labels were constructed to produce separation in the same variables, or in a representation learned from them. Even under a homogeneous population, a clustering algorithm returns a partition and can create apparently large between-group contrasts. Reusing the same data for both clustering and testing can therefore inflate Type~I error.

Two forms of adaptivity are useful to distinguish. \emph{Label adaptivity} arises because $\widehat z$ is estimated from $X$. \emph{Hypothesis adaptivity} arises when the feature, cluster pair, number of groups, graph resolution, or scientific contrast is selected after inspecting the fitted structure. The second issue can remain even when an independent sample is used for a final test, because the scientific claim itself may have been chosen from many candidate contrasts. Correcting only the feature-wise test therefore does not necessarily validate a conclusion that also reflects prior variable screening, repeated resolutions, batch integration, cluster merging, or annotation.

Conditional selective inference addresses the first workflow by conditioning on an event that describes the clustering output or algorithmic path. Under their respective models, \citet{ZhangKamathTse2019} derive a truncated-normal test for post-clustering differential analysis, \citet{ChenWitten2023} construct finite-sample selective $p$-values after $K$-means, and \citet{GaoBienWitten2024} develop selective tests after several forms of agglomerative hierarchical clustering. \citet{ChenGao2025} further study single-feature testing after $K$-means or hierarchical clustering. These procedures directly alter the reference distribution to reflect that the groups were selected rather than externally observed. Their guarantees, however, are tied to a specified algorithm, feature model, target, and conditioning event; conditioning can also remove information and reduce power.

A second family of approaches separates the information used for subgroup discovery from that used for inference. Ordinary splitting of observations is awkward in unsupervised analysis because a partition learned in one sample must be transferred to another, rare groups may disappear from one split, and the contrast itself may remain adaptive. For count data, count splitting instead divides each observed count into approximately independent pieces under a Poisson model, using one piece to estimate latent structure and the other for downstream inference \citep{NeufeldEtAl2024Biostatistics}. Data thinning and generalized thinning extend this principle to broader distributional families \citep{DharamshiEtAl2025}. These methods are comparatively modular with respect to the clustering algorithm, but divide information and rely on a valid decomposition model; overdispersion, feature dependence, normalization, and preprocessing can affect the intended independence.

A third strategy calibrates the complete cluster-then-test pipeline under a fitted homogeneous null. ClusterDE, for example, generates synthetic one-cluster data, reruns clustering and marker analysis, and uses the induced discoveries to calibrate the observed workflow \citep{SongEtAl2023ClusterDE,LeeEtAl2026ClusterDE}. Such calibration can accommodate a complicated black-box pipeline, but its validity depends on whether the synthetic null reproduces the distributional features that generate spurious separation. Conditional inference, splitting or thinning, and synthetic-null calibration therefore solve the double-use problem by constructing different reference experiments. Exact results for $K$-means or hierarchical clustering do not automatically validate a graph-based pipeline in which variable screening, integration, graph construction, modularity optimization, resolution selection, and merging are all adaptive.

\subsection{Joint Feature Selection, Subgrouping, and Subspace Search}
\label{subsec:joint-clustering-feature-selection}

The second direction treats interpretation as part of the clustering problem rather than as a test applied after labels have been fixed. When $p$ is large, many coordinates may contribute noise while only a small subset carries subgroup information. Sparse $K$-means and related procedures estimate feature weights or penalized centers jointly with the partition, while sparse convex clustering combines feature selection with a fusion-based clustering objective \citep{WittenTibshirani2010,SunWangFang2012,WangEtAl2018SparseConvex, tang2021individualized, zhu2021longitudinal}. These methods seek a globally informative set of coordinates, but global relevance does not reveal which populations a feature separates.

A related class of methods searches for a lower-dimensional signal representation. IF-PCA screens influential features before principal-component estimation and clustering, whereas CHIME estimates sparse discriminant directions within an EM framework for high-dimensional Gaussian mixtures \citep{JinWang2016,CaiMaZhang2019}. A discriminative subspace can substantially improve subgroup recovery, especially when signal is distributed across correlated variables, but its coordinates need not map directly to an interpretable list of marker effects in the original feature space. Thus, improved clustering and scientific interpretation are related but distinct targets.

For a $K$-component location-mixture model, let
$
  U=(\mu_1,\ldots,\mu_K)^\top
    =(\mu_{kj})_{1\leq k\leq K,\,1\leq j\leq p},\; \mu_{kj}=\mathbb E(X_j\mid Z=k).
$
The global informative-feature set $\mathcal J(U)$ contains indices $j$ for which the component means are not all equal. A higher-resolution pairwise map records which component pairs differ on each feature:
\[
  \Xi(U)
  =\{(k,k',j):\mu_{kj}\neq\mu_{k'j},\ 1\leq k<k'\leq K\}.
\]
The projection of $\Xi(U)$ onto the feature index is $\mathcal J(U)$, but the triplets identify which feature distinguishes which pair of latent components. This distinction is important in single-cell analysis, where a gene may be globally heterogeneous, specific to one cell type, shared by a lineage, or informative only for one selected pair.

Pairwise variable selection for Gaussian model-based clustering provides an early formulation of this goal \citep{GuoEtAl2010}. Emerging work such as PARSE estimates the partition and the pair-specific mean map jointly under an elliptical-mixture model \citep{ZhangEtAlPARSE2026}. Its relevance here is conceptual: it treats marker structure as part of the unsupervised estimation problem and distinguishes global feature relevance from cluster-pair-specific interpretation. Because this work is still emerging, we do not rely on its method-specific rates or application findings as established conclusions; further details are provided in Online Supplement~A.

Joint estimation does not remove the inferential problem. Selection consistency, such as
$
  \Pr\{\widehat\Xi=\Xi(U^\star)\}\rightarrow1,
$
is an asymptotic support-recovery statement under a specified model, sparsity level, signal condition, and initialization regime. It does not automatically provide finite-sample $p$-values, confidence intervals, false-discovery control, or calibrated uncertainty for the selected partition. Conversely, a valid selective $p$-value for one contrast need not imply stable clustering or a scientifically useful explanation. A fuller interpretation should distinguish the selected features from estimated effect sizes,
$
  \widehat\Delta_{kk',j}
  =\widehat\mu_{kj}-\widehat\mu_{k'j},
$
and should assess sensitivity to tuning parameters, initializations, resolutions, and samples. Joint feature selection and valid post-selection inference are therefore complementary rather than competing goals.

\begin{table}[h]
\centering
\caption{Representative strategies for interpreting and testing high-dimensional clusters.}
\label{tab:clustering-workflows}
\footnotesize
\setlength{\tabcolsep}{3pt}
\renewcommand{\arraystretch}{0.95}
\begin{tabular}{@{}p{0.22\textwidth}p{0.36\textwidth}p{0.34\textwidth}@{}}
\toprule
\textbf{Strategy} & \textbf{Statistical insights} & \textbf{Representative references} \\
\midrule
Conditional selective inference & Condition on the clustering event when testing a selected contrast. & \citet{ZhangKamathTse2019}; \citet{ChenWitten2023}; \citet{GaoBienWitten2024}. \\
\midrule
Splitting or thinning & Use separate observations or stochastic data pieces for discovery and inference. & \citet{NeufeldEtAl2024Biostatistics};  \citet{DharamshiEtAl2025}. \\
\midrule
Synthetic-null calibration & Reapply the analysis pipeline to data generated under a no-cluster null. & \citet{SongEtAl2023ClusterDE}; \citet{LeeEtAl2026ClusterDE}. \\
\midrule
Sparse global feature selection & Estimate the partition together with a globally informative coordinate set. & \citet{WittenTibshirani2010}; \citet{SunWangFang2012}; \citet{WangEtAl2018SparseConvex}. \\
\midrule
Discriminative subspace search & Screen features or estimate sparse directions that preserve subgroup signal. & \citet{JinWang2016}; \citet{CaiMaZhang2019}. \\
\midrule
Pairwise high-resolution modeling & Estimate the partition and feature-specific contrasts across cluster pairs. & \citet{GuoEtAl2010}; \citet{ZhangEtAlPARSE2026}. \\
\bottomrule
\end{tabular}
\end{table}


\section{Causal Inference and Treatment Effect Heterogeneity}
\label{sec:causal}

In the potential-outcomes framework \citep{rubin1974estimating, imbens2015causal}, each unit $i$ has potential outcomes $Y_i(1)$ and $Y_i(0)$ under treatment and control, and hence its own treatment effect $Y_i(1)-Y_i(0)$. Nothing in the framework requires these unit-level effects to agree and their distribution across the population is the treatment effect heterogeneity. Because each unit reveals only one of its two potential outcomes, however, no unit-level effect is ever observed, and inference must target summaries of the effect distribution instead. Such summaries are expectations of the counterfactual contrast $Y(1)-Y(0)$, not averages of observed outcomes. For every unit, one term of the contrast is missing, and it is the identification conditions stated below that express these summaries as functionals of observable data. Which summary to target is itself a decision about heterogeneity. Each choice preserves some of the effect distribution and averages the rest away.

For summaries of the mean effect, this choice can be written in a single form. The average, group average, and conditional average treatment effects are all instances of
\begin{equation}
  \tau(V) = \mathbb{E}\{Y(1)-Y(0) \mid V\},
  \label{eq:causal-estimand-family}
\end{equation}
where $V$ is conditioning information selected by the analyst. Taking $V$ degenerate gives the average treatment effect (ATE) $\mathbb{E}\{Y(1)-Y(0)\}$, which reduces the effect distribution to its mean and preserves no heterogeneity. Taking $V$ to be membership in a pre-specified or discovered subgroup gives group average treatment effects (GATEs), which preserve between-group differences while suppressing variation inside each group. Taking $V = X$, the covariate vector, gives the conditional average treatment effect (CATE) $\tau(x) = \mathbb{E}\{Y(1)-Y(0) \mid X = x\}$: the CATE preserves every difference in effects that measured covariates can explain and averages away only the residual heterogeneity remaining within covariate strata. The individual treatment effect (ITE) $Y_i(1)-Y_i(0)$ lies outside this family. In a superpopulation formulation, conditioning on a unit's identity does not reveal the realized contrast, thus making the ITE the unit-level contrast itself rather than a conditional expectation. It is a fixed unit-level quantity only under a finite-population view that treats the potential outcomes as fixed, and it is not point-identified without strong additional assumptions. Across these summaries, richer conditioning preserves more heterogeneity but is harder to learn. The ATE is a single parameter estimable at parametric rates, while the CATE is an unknown function whose estimation difficulty grows with the dimension of $X$.

The CATE is therefore the finest conditional mean-effect summary that observed covariates can support. It does not capture all distributional heterogeneity, as units sharing the same $x$ may still differ in their effects. A treatment rule converts this heterogeneity into a decision, for example $d(x) = \mathbb{I}\{\tau(x) > 0\}$ for two treatments when larger outcomes are preferable and treatment costs and harms are equal or already included in the outcome. Identifying any of the causal effects discussed from data requires three conditions. Consistency requires that the observed outcome equal the potential outcome under the received treatment. Exchangeability requires that treatment assignment be independent of the potential outcomes given the conditioning covariates. Positivity requires that every covariate profile have a positive probability of each treatment. Randomization can establish exchangeability, whereas observational adjustment requires substantive justification. Causal formulations of algorithmic fairness, in which the heterogeneity of interest lies in a model's behavior across protected subgroups, are reviewed in Online Supplement~B.

\subsection{Heterogeneous treatment effects}
\label{sec:cate}

Modern CATE estimation has developed through three overlapping approaches, the first of which repurposes flexible outcome models. \citet{hill2011bayesian} fit Bayesian additive regression trees (BART) to the response surface $\mathbb{E}[Y \mid X, A]$ and estimated effects as differences of posterior predictions under the two treatment assignments. The proposal showed that a strong generic learner with no causal customization can outperform purpose-built estimators in nonlinear settings, and it raised the point that drives much of the later literature: predictive accuracy for outcomes is not the same as accuracy for effect differences, which may be far smoother or sparser than the outcome surfaces themselves.

The second targets the effect directly with partitioning schemes. Causal trees \citep{athey2016recursive} modify the CART splitting criterion to seek splits with heterogeneous effects and introduce honesty. Disjoint subsamples select the partition and estimate effects within leaves, so leaf-level estimates admit valid confidence intervals despite data-driven subgroup selection. This addresses the selection bias behind the multiplicity concern of the classical subgroup literature (Section~\ref{sec:rct}), though it does not by itself provide simultaneous error control across many leaves, outcomes, or subsequently selected claims. Causal forests \citep{wager2018estimation} aggregate honest trees and establish pointwise asymptotic normality of the resulting CATE estimates.

The third decouples the estimand from the learner through meta-learners and orthogonalization. \citet{kunzel2019metalearners} systematized the basic meta-learners: the S-learner (one model with treatment as a feature), the T-learner (separate models per arm), and the X-learner, which performs well under unbalanced designs by imputing individual effects and regressing them on covariates. A more general development treats CATE estimation as a two-stage semiparametric problem. Nuisance functions (outcome regressions and propensity scores) are estimated flexibly, then combined in an orthogonalized second stage so that first-stage errors enter only in products. The R-learner \citep{nie2021quasi} implements this through a residualized loss and achieves quasi-oracle rates, estimating the CATE nearly as well as if the nuisances were known. The DR-learner \citep{kennedy2023towards} regresses doubly robust pseudo-outcomes on covariates and characterizes when this is minimax optimal. Both build on the sample-splitting and Neyman-orthogonality machinery of double/debiased machine learning \citep{chernozhukov2018double}.

Neural network estimators implement one of the three approaches above with a neural network as the learner, and the distinctive contribution of this literature is how the network is regularized. \citet{johansson2016learning} observed that counterfactual inference is a covariate-shift problem, in which a model trained on treated units is implicitly evaluated on the control distribution, so representation learning should trade predictive fit against distributional distance between treated and control representations \citep{shalit2017estimating}. The architectures that develop this idea are reviewed in Online Supplement~B. Critical assessments of this literature have proved as valuable as the architectures. \citet{curth2021inductive} showed on careful semi-synthetic benchmarks that inductive biases matter more than architectural novelty. When true heterogeneity is limited, unconstrained models produce spurious heterogeneity, and estimators that shrink toward homogeneity dominate. In heterogeneity estimation the greater danger is usually too much flexibility, not too little.

\subsection{Randomized controlled trials with subgroup analyses}
\label{sec:rct}

The trials literature combines classical skepticism about subgroup claims with newer methods that make data-driven subgroup discovery valid. \citet{yusuf1991analysis} articulated the standard cautions. Subgroup analyses in trials powered for overall effects are underpowered, subject to multiplicity, and prone to producing qualitative results that fail to replicate. \citet{gail1985testing} distinguished quantitative interactions (effects vary in magnitude but not sign) from qualitative interactions (effects reverse sign across subgroups) and derived a likelihood ratio test for effect crossing. When treatment choice depends only on the sign of the effect, only qualitative interactions threaten a uniform recommendation. Once accessibility, toxicity, burden, or a minimum meaningful benefit enter the decision, quantitative heterogeneity can also change the optimal treatment. The ICH E9(R1) estimands framework \citep{ich2019estimands} made this discipline a regulatory requirement. Any claim about a subpopulation must specify in advance the population, variable, intercurrent-event strategy, and summary measure, that is, the precise estimand targeted.

More recent methods retain validity under data-driven discovery. \citet{ding2016randomization} showed that randomization alone supports exact tests of whether treatment effect variation exceeds what constant effects could produce, with no modeling assumptions. For high-dimensional discovery, \citet{chernozhukov2025generic} provide generic machine-learning inference on features of heterogeneity that remain estimable with valid inference even when the underlying machine-learning proxy is poor. These features include the best linear predictor of the CATE, sorted group average effects, and classification analysis of who benefits most. Adaptive enrichment designs \citep{simon2013adaptive} carry the logic into the design phase, allowing a trial to restrict enrollment to a benefiting subpopulation identified at interim analyses while preserving error control. Data-driven subgroup claims require inference that accounts for how the subgroups were selected. General-purpose selective-inference tools for data-discovered subgroups are reviewed in Section~\ref{sec:uq}.

\subsection{Heterogeneity in observational studies}
\label{sec:observational}

In observational data, heterogeneity analysis must also contend with confounding. Under the identification conditions stated above, adjustment for the propensity score $e(x) = \Pr(A=1 \mid X=x)$ \citep{rosenbaum1983central} removes bias, and effect modification can be examined within propensity strata or matched sets, separating variation in treatment effects from variation in treatment propensity, two quantities that raw comparisons conflate. Positivity is the condition most exposed by heterogeneity analysis, since overlap that is adequate in the full sample can fail inside the covariate regions that define a subgroup. The covariate balancing propensity score \citep{imai2014covariate} estimates $e(x)$ by directly targeting the balance it is meant to deliver, improving robustness to propensity model misspecification.

Instrumental variables (IV) show that heterogeneity changes what is estimable. With constant effects, the IV estimand is the treatment effect. Once effects vary across units, \citet{imbens1994identification} and \citet{angrist1996identification} showed that the same estimand becomes the local average treatment effect (LATE), the average effect among compliers, units whose treatment status is moved by the instrument. Heterogeneity thus forces a reinterpretation of the target as subgroup-specific, where the subgroup is defined by unobservable compliance behavior. Mendelian randomization, the framework's most active modern application, and moderated mediation, in which heterogeneity resides in the mechanism rather than the total effect, are reviewed in Online Supplement~B.

\subsection{Treatment decisions, reliability, and transport}
\label{sec:causal-recent}

Recent work on treatment effect heterogeneity has moved in three directions. The first is decision-making, where heterogeneous effects matter chiefly because they justify treating different units differently. \citet{athey2021policy} formulated policy learning over a constrained class of treatment rules using doubly robust scores, establishing regret bounds at the semiparametric-efficient rate. The framework is the natural consumer of the CATE estimates of Section~\ref{sec:cate}. Its sequential extensions, including dynamic treatment regimes, are discussed in Section~\ref{sec:heterogeneous-rl}, and their mobile-health applications, including micro-randomized trials, in Section~\ref{sec:mhealth}.

The second direction asks when CATE estimates can be trusted. Conformal methods construct prediction intervals for a potential outcome or an individual effect, which are distinct from confidence intervals for the conditional mean $\tau(x)$. \citet{lei2021conformal} obtain finite-sample average coverage in randomized settings, and their weighted variants extend to observational data with guarantees that depend on how well the propensity score or the relevant outcome quantiles are estimated. Coverage of this kind is an average property and does not point-identify any individual effect. When confounding cannot be bounded away, \citet{yadlowsky2022bounds} derive sharp bounds on CATEs under explicit sensitivity models, replacing point identification with partial identification. Calibration offers a complementary notion of trustworthiness. Causal isotonic calibration \citep{vanderlaan2023causal} post-processes any CATE estimator into a calibrated one, with sample splitting or cross-fitting recommended when the pseudo-outcomes rely on estimated nuisances. The general conformal and calibration machinery is developed in Section~\ref{sec:uq}.

The third direction concerns new data regimes. Because effects are heterogeneous and populations differ in their distributions of effect modifiers, effects estimated in one population rarely transfer unchanged to another. The generalizability and transportability literature \citep{degtiar2023review} formalizes when and how trial-estimated heterogeneous effects extend to target populations. Federated and multi-site variants of this program, in which heterogeneous effects are combined across sites under communication and privacy constraints, are discussed in Section~\ref{sec:federated}. Foundation models are also being applied to treatment effect estimation. Prior-data fitted networks, which are pre-trained large collections of simulated causal datasets \citep{ma2025foundation}, deliver zero-shot CATE estimation in a single forward pass. Whether such models encode appropriate inductive biases for effect heterogeneity, or instead import the spurious-heterogeneity pathologies of Section~\ref{sec:cate} at scale, is an open question for the AI methods reviewed in Section~\ref{sec:ai}.


\section{Heterogeneous Synthetic Data Generation}
\label{sec:synthetic}

Synthetic data generation aims to reproduce selected features of a target population. When synthetic data will be used for subgroup analyses or decisions, a generator should preserve the scientifically relevant subgroup structure, minority modes, and mixed variable types; matching aggregate marginals and low-order moments alone may not suffice. A model can match these summaries while distorting small subpopulations. The term \emph{subgroup} can refer to three different objects: a latent component in a fitted model, a rare mode in the data distribution, or a scientifically defined group based on external criteria (Section~\ref{subgroup-analysis}). These objects may coincide, but preserving one does not guarantee preservation of the others; a generator may cover every distributional mode while misrepresenting a clinically defined subgroup.

\subsection{Classical synthesis and the combining rules}
\label{sec:synthetic-classical}

Statistical disclosure control has developed a principled framework for synthetic data. This began with \citet{rubin1993statistical}, who proposed releasing fully synthetic microdata generated from the posterior predictive distribution of a model fit to confidential data, which in effect is multiple imputation with everything treated as missing. Crucially, \citet{raghunathan2003multiple} placed the proposal on inferential footing by deriving combining rules under which analyses of multiple synthetic datasets yield valid point estimates and variance estimates for population quantities. \citet{reiter2003inference} developed the partially synthetic variant, in which only sensitive values are replaced, with its own combining rules. Sequential, tree-based, and copula synthesizers in this tradition are reviewed in Online Supplement~C.

Classical synthesis makes heterogeneity explicit through mixture components, hierarchy levels, or stratifying variables. Its combining rules support valid downstream inference when the synthesis and analysis models satisfy their stated assumptions, although these methods can be difficult to scale to high-dimensional data with complex dependence. Deep generative models can represent richer dependence, but subgroup fidelity and inferential validity require evaluation or correction; neither follows from the architecture alone.

\subsection{Deep generative families}
\label{sec:synthetic-deep}

Table~\ref{tab:synthetic-families} summarizes three common deep generative families and potential subgroup failures. A variational autoencoder (VAE) with a simple unimodal latent prior may blur distinctions among subpopulations; mixture priors are one remedy. In mixture VAEs, the encoder estimates posterior component probabilities, an operation reminiscent of the E-step, although the analogy is not exact and familiar identifiability concerns remain. A well-known failure mode of generative adversarial networks (GANs) is mode collapse: minority modes may be underrepresented or absent even when aggregate metrics appear satisfactory. Diffusion models offer an alternative based on iterative denoising, but their minority-mode and subgroup coverage depends on the model, training distribution, guidance, and sampling procedure. Architecture alone does not guarantee subgroup fidelity, which requires the per-subgroup evaluation discussed in Section~\ref{sec:synthetic-fidelity}. For conditional diffusion models, the guidance scale controls a trade-off between fidelity to the specified condition and sample diversity \citep{ho2022classifier}. Stronger guidance can produce samples that more closely reflect the condition while reducing within-subgroup variation. Tabular, longitudinal, and electronic-health-record variants of these families are reviewed in Online Supplement~C.

\begin{table}[t]
\caption{Deep generative families and their characteristic subgroup failures.}
\label{tab:synthetic-families}
\centering
\renewcommand{\baselinestretch}{1}\small
\setlength{\tabcolsep}{4pt}
\begin{tabular}{@{}p{3.6cm}p{3.3cm}p{4.4cm}p{4.0cm}@{}}
\toprule
\raggedright\textbf{Family} & \raggedright\textbf{Where heterogeneity lives} & \raggedright\textbf{Failure for subgroups} & \raggedright\textbf{Remedy} \tabularnewline
\midrule
\raggedright VAE \citep{kingma2014auto} & \raggedright latent space under a unimodal prior & \raggedright blends distinct subpopulations & \raggedright mixture priors \citep{jiang2017variational} \tabularnewline[5pt]
\raggedright GAN \citep{goodfellow2014generative} & \raggedright implicit generator trained against a discriminator & \raggedright mode collapse: minority modes dropped while aggregate metrics stay healthy & \raggedright oversample rare categories in training \citep{xu2019modeling} \tabularnewline[5pt]
\raggedright Diffusion \citep{ho2020denoising} & \raggedright iterative denoising & \raggedright minority-mode coverage is empirical, not guaranteed & \raggedright per-subgroup coverage evaluation (Section~\ref{sec:synthetic-fidelity}) \tabularnewline
\bottomrule
\end{tabular}
\end{table}

\subsection{Subgroup fidelity, fairness, and valid inference}
\label{sec:synthetic-fidelity}
All three generative families leave the same statistical questions open. These concern the composition of the synthetic population, the evaluation of subgroup fidelity, the privacy risk borne by small subgroups, and the validity of inference on synthetic data. When the training data themselves are biased, faithful generation reproduces the bias, and the composition of the synthetic population inherits whatever underrepresentation was already present. \citet{choi2020fair} use a small unbiased reference set to importance-reweight training of a generator on a large biased dataset, correcting representation without subgroup labels and making the composition of the synthetic population an explicit design target rather than an artifact of the training distribution.

A single distributional distance cannot distinguish a generator that produces beautiful majority-mode samples from one that covers the full population, and this ambiguity has consequences for how subgroup performance is assessed. Precision-recall curves for distributions \citep{sajjadi2018assessing} separate sample quality (precision) from mode coverage (recall), and \citet{alaa2022faithful} refine these into $\alpha$-precision, $\beta$-recall, and authenticity, the last detecting memorization. The important recommendation is to audit these metrics per subgroup, since aggregate scores can conceal exactly the minority-mode failures that matter. Subgroup fidelity is also tied to privacy risk: minority groups may be harder to evaluate, and underrepresented regions may be more vulnerable to memorization. \citet{vanbreugel2023membership} show that membership-inference attacks against synthetic data succeed preferentially where the generator has locally overfit, with such overfitting occurring disproportionately for underrepresented subgroups. Privacy and fidelity can therefore be in tension: a generator should retain useful group-level structure without reproducing rare individuals too closely. Section~\ref{sec:federated} discusses this trade-off under differential privacy.

A natural question is what happens when an analyst runs standard inference on synthetic data as if it were real. \citet{decruyenaere2024real} explored this question and found bias and severely miscalibrated confidence intervals, since synthetic data are draws from an estimated distribution and naive inference ignores that extra layer of uncertainty. In follow-up work, the same authors proposed debiasing corrections that address these issues and restore near-nominal coverage \citep{decruyenaere2024debiasing}, closing the circle back to the multiple-imputation combining rules of Section~\ref{sec:synthetic-classical}.

\section{Heterogeneity in AI}
\label{sec:ai}

Large language models can serve heterogeneous subjects through personalized prompts and context, adapted parameters, and specialization of AI agents. Each offers a particular way to borrow strength across users while retaining heterogeneity that is critical for prediction, generation, or decision making.

\subsection{Prompt personalization}
Prompt personalization adapts a shared LLM to an individual user by supplying user-specific information in the context or prompt. Profile-based methods construct textual context \citep{zhang2024guided}, retrieval-based methods select records to include \citep{salemi2024lamp}, and embedding-based methods construct continuous prompt vectors \citep{doddapaneni2024userembedding}. The same LLM can therefore serve multiple users without maintaining a separate model for each user.

Text-based methods summarize or select evidence from a user's history. Profiles condense a user's history into a reusable textual description \citep{zhang2024guided}. Retrieval-based methods instead rank the user's historical records for the current request, for example by lexical or semantic similarity, and insert selected records into the LLM's context \citep{salemi2024lamp}. The selected records can differ across requests, and the retrieval model can be optimized using the quality of the resulting personalized responses \citep{salemi2024optimization}. When the target user's history is sparse, records from other users may provide additional information for personalization. Persona-DB matches users through higher-level personal information and incorporates matched users' records into the target user's context \citep{sun2025personadb}.

Embedding-based methods learn compact representations of user histories. The same encoder maps each user's history to user-specific soft prompts \citep{doddapaneni2024userembedding,hebert2024persoma}. Difference-aware Embedding-based Personalization (DEP) supplements the target user's history representation with embeddings of differences between that user's responses and those of other users facing similar content \citep{qiu2025latent}.

The effectiveness of these methods depends on how much user information is available and whether it remains relevant to the current request. The Language Model Personalization (LaMP) benchmark reports gains from retrieving user histories over a no-retrieval baseline, with separate evaluations for new users and later interactions of existing users \citep{salemi2024lamp}. When preferences change, however, historical records may no longer reflect what the user currently wants. Because LaMP's time-based evaluation does not isolate preference changes, its results do not establish whether these gains persist as preferences evolve.


Personalization can also reveal information unrelated to the user's current request. Historical text can expose sensitive attributes through indirect clues \citep{staab2024beyond}. The information used for adaptation must therefore be assessed for both relevance and unintended disclosure.

\subsection{Fine-tuning for domain and task specialization}

Fine-tuning adapts an LLM by updating all or part of its parameters. The training data and objective specify the intended specialization. Domain-adaptive continued pretraining uses domain text \citep{gururangan2020dont,cheng2024adaptllm}. Task-specific supervised fine-tuning uses examples of a particular input-output problem, while multitask instruction tuning trains a common model across several problems \citep{chung2024scaling}. These strategies adapt to domains or tasks and need not distinguish users performing the same task. Parameter sharing can then take two forms: retaining a common backbone with specialized modules, or combining several modules for the current target \citep{houlsby2019adapters,hu2022lora,pfeiffer2021adapterfusion}.

A shared backbone allows specialization without fitting a complete model for every target. Adapters introduce trainable modules, while LoRA learns low-rank weight updates with the pretrained weights fixed \citep{houlsby2019adapters,hu2022lora}. A module trained for one task is shared
	by its users. Personalized low-rank adaptation (PLoRA) instead incorporates user representations \citep{zhang2024plora}. 

Combining modules allows a target to borrow from several learned specializations. AdapterFusion trains a fusion mechanism over fixed task adapters \citep{pfeiffer2021adapterfusion}. HydraLoRA learns a shared low-rank factor, specialized factors, and a router that combines them according to the input \citep{tian2024hydralora}. Here, selection occurs at the input level rather than assigning each user permanently to one expert.

The empirical evidence for these methods comes from evaluations at different levels of specialization. For domain adaptation, \citet{gururangan2020dont} compare downstream classification performance with and without additional pretraining on domain text, across four domains. 
 For cross-task composition, AdapterFusion compares fused models with individual task adapters and multitask alternatives. The gains over individual adapters on multiple target tasks support the usefulness of borrowing representations across tasks, although improvements are not uniform \citep{pfeiffer2021adapterfusion}. 

\subsection{Agentic AI for heterogeneous domains and tasks}

Domain and task heterogeneity create different requirements for agent systems. Domains differ in their knowledge resources, tools, and operating rules, while tasks within a domain require different functions and execution procedures. An LLM-based agent combines a language model with instructions, resources, and an execution process to perform a task.

Specialized agents can be configured with domain resources and rules, then assigned roles for particular tasks. For example, the retail and airline environments in Tau-bench provide different tools and policy guidelines to customer-service agents \citep{yao2025taubench}.  MetaGPT gives agents distinct software-development roles, such as product manager, architect, and engineer, together with procedures and structured outputs for their work \citep{hong2024metagpt}. These configurations specify what each agent is responsible for, what information it receives, and what it passes to the next stage.

Selecting among specialists requires matching both domain expertise and task capabilities. 
HuggingGPT decomposes a request into subtasks and assigns specialist models according to their documented functions, allowing an LLM controller to combine capabilities such as language, vision, and speech processing \citep{shen2023hugginggpt}.
Dynamic LLM-Powered Agent Network (DyLAN) selects a team from candidate agents for the target task. Its ablation study shows that the selection can
	improve answer correctness while reducing model calls \citep{liu2024dylan}.

Coordination determines how the selected specialists turn their separate contributions into a task result. MetaGPT follows a prescribed workflow that passes structured outputs between roles \citep{hong2024metagpt}. Learned orchestration instead selects the next agent using the task and accumulated execution state, adapting participation as execution progresses \citep{dang2025evolving}.

Errors can propagate when one specialist's output becomes another's input. MetaGPT uses feedback from executing and testing generated code to guide revision \citep{hong2024metagpt}. In service tasks, Tau-bench examines failed execution sequences to identify incorrect tool use, wrong decisions, and unresolved requests \citep{yao2025taubench}.

Uncertainty-aware escalation allows an agent to seek help before committing to an ambiguous action. KnowNo uses conformal prediction to form sets of candidate actions and requests human clarification when the set does not identify a unique choice \citep{ren2023robots}. Additional system and evaluation details are provided in Online Supplement C: AI methods and evaluation.

\section{Heterogeneity in Reinforcement Learning}
\label{sec:heterogeneous-rl}

Reinforcement learning (RL) extends heterogeneous learning from one-time prediction to sequences of actions. An action changes both the immediate reward and the distribution of future observations, so heterogeneity can enter through response to treatment, state transitions, task objectives, or calendar-time change. Let unit $u$ index an individual, population, or task and write
\[
  \mathcal M_u=(\mathcal S,\mathcal A,P_u,r_u,\mu_u,\gamma),
\]
where $P_u$ is the transition law, $r_u$ is the conditional mean reward, $\mu_u$ is the initial-state distribution, and $\gamma$ is a discount factor. For a policy $\pi$,
\begin{equation}
  V_u^\pi(\mu_u)
  =\mathbb E_{\mathcal M_u,\pi}
    \left\{\sum_{t=0}^{\infty}\gamma^tR_{u,t}\right\}.
  \label{eq:unit-value}
\end{equation}
A population policy maximizes a weighted average of $V_u^\pi$, whereas subgroup- or individual-specific policies optimize value for a more local target.

\begin{figure}[t]
  \centering
  \includegraphics[width=\textwidth]{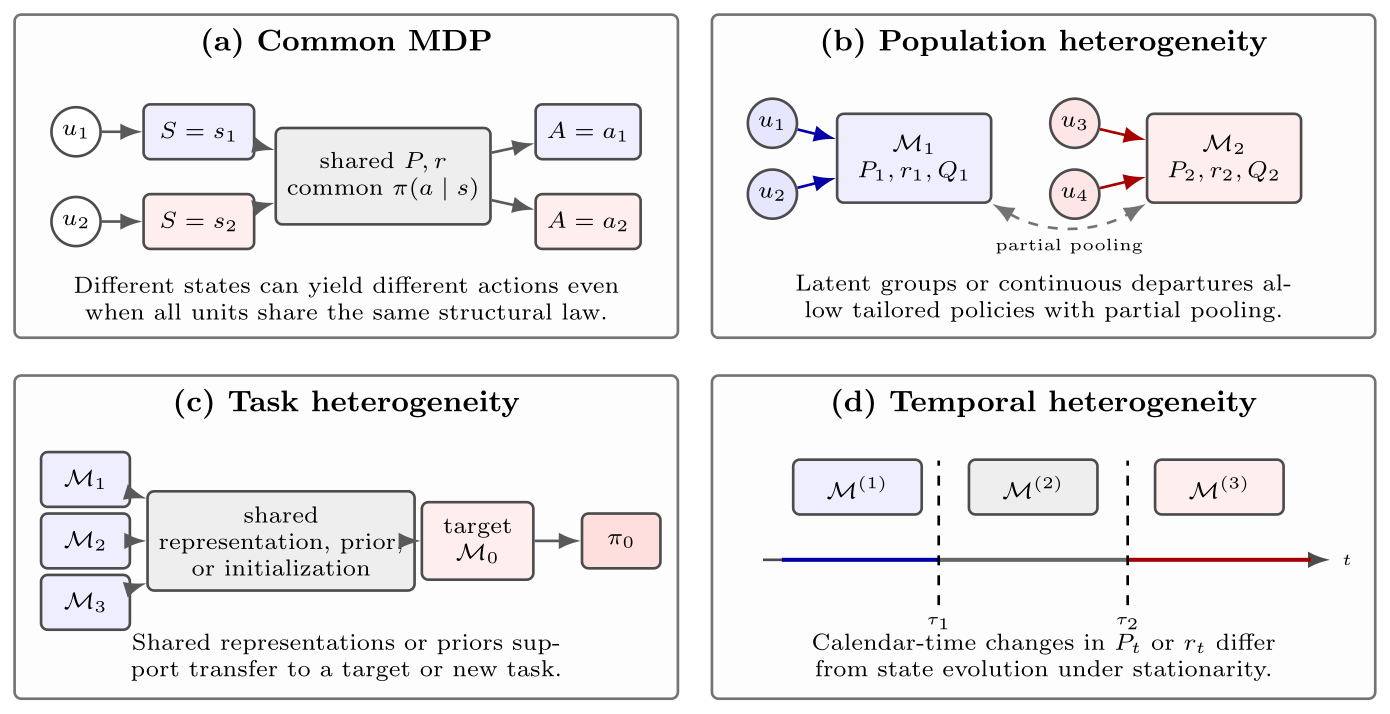}
  \caption{Four forms of heterogeneity in sequential decision-making. Panel (a) distinguishes state-dependent actions under a common Markov decision process from structural population heterogeneity. Panel (b) represents latent or continuous differences across individuals. Panel (c) depicts information sharing across related tasks and adaptation to a target task. Panel (d) separates ordinary state evolution from calendar-time changes in rewards or transition laws.}
  \label{fig:rl-heterogeneity}
\end{figure}

Figure~\ref{fig:rl-heterogeneity} emphasizes a foundational distinction. A common policy $\pi(a\mid s)$ can prescribe different actions because people occupy different observed states; this is state-dependent decision-making within one structural process. Genuine population heterogeneity arises when $P_u$, $r_u$, the value function, or the optimal policy differs across otherwise comparable units. Historical trajectories may also be generated under different behavior policies, which changes state--action coverage and the identification of off-policy values even when the optimal policies are the same.

\subsection{Sequential Decisions within a Common Process}
\label{subsec:dtr-foundations}

Dynamic treatment regimes map an observed history $H_t$ into an action at each decision point. Their causal interpretation requires well-defined interventions, consistency, sequential exchangeability or randomization, and positivity over the histories relevant to the candidate regime \citep{Murphy2003}. This framework captures heterogeneity explained by recorded history, but it usually assumes a common conditional response and transition mechanism after conditioning on that history. Two individuals with the same recorded state therefore receive the same rule even when stable, unmeasured differences imply different long-run responses.

For a fixed policy, a policy-value interval is a confidence interval for $V^\pi$ under a specified target distribution. For a learned policy $\widehat\pi$, regret compares its value with the optimal value, for example $V^{\pi^\star}-V^{\widehat\pi}$. These objects differ from uncertainty in a reward-model coefficient: they depend on sequential identification, behavior-policy coverage, and the population over which value is averaged. General methods for uncertainty and regret are discussed in Section~8; Q-learning, A-learning, V-learning, Bellman equations, and additional policy-evaluation background are summarized in Online Supplement~B.

\subsection{Population, Task, and Temporal Heterogeneity}
\label{subsec:population-task-rl}

A direct subgroup formulation assumes that units belong to latent decision environments,
\begin{equation}
  (P_u,r_u,Q_u^\pi)=(P_{g_u},r_{g_u},Q_{g_u}^\pi),
  \qquad g_u\in\{1,\ldots,G\}.
  \label{eq:latent-mdp-groups}
\end{equation}
Auto-clustered policy learning and fusion-based multistage methods estimate group memberships together with group-specific values or regimes \citep{ChenSongJordan2024,SunZhuSun2026}. The principal identification question is what should be homogeneous: complete transition--reward systems, value functions, advantages, or optimal actions. Distinct environments can share the same optimal policy, while small value differences near an action boundary can imply different actions. Unequal trajectory lengths, behavior-policy differences, serial dependence, and weak overlap further complicate group recovery and policy evaluation.

When variation is gradual, a partial-pooling representation is more natural,
\begin{equation}
  Q_u^\pi(s,a)=Q_0^\pi(s,a)+\Delta_u^\pi(s,a),
  \label{eq:individual-q-departure}
\end{equation}
with departures constrained by latent factors, sparsity, smoothness, or a hierarchical distribution. Individualized offline RL uses this structure to stabilize data-poor individuals while allowing informative histories to support more personalization \citep{MiaoShahbabaQu2025}. The borrowing problem is analogous to random-effects modeling, but weak state--action coverage adds a sequential source of uncertainty. When the number of individuals is large but each trajectory is short, population-level structure may be learnable even when person-specific effects remain weakly identified. Latent effects alone do not resolve unmeasured confounding or partial observability in observational trajectories. Average value or average regret can also conceal poor performance for rare individuals, motivating subgroup, lower-tail, or constrained criteria.

When $u$ indexes a task, multi-task RL shares information symmetrically across observed tasks, transfer RL borrows toward a designated target, and meta-RL learns structure intended to support adaptation to a new task. Successful transfer requires similarity in the components being shared---transitions, rewards, representations, values, or policies---rather than generic similarity of observed states. Recent transfer methods allow target-specific departures or estimate which sources are compatible with the target \citep{ChenLiJordan2025,ChenChenJing2026,ZhangWang2021}. Their central statistical difficulty is negative transfer: source tasks that improve representation learning can still bias values or actions when reward mechanisms differ. Representative deep multi-task and meta-RL constructions are described in Online Supplement~B.

Temporal heterogeneity should be distinguished from ordinary state evolution in a stationary process. If $P_{u,t}$ or $r_{u,t}$ changes with calendar time, pooling across a structural break can bias current policy values, whereas using only recent observations increases variance. Stationarity tests and change-point procedures can be coupled with policy learning \citep{LiEtAl2025Stationarity}. Individual and temporal heterogeneity may also coexist; latent-factor approaches to off-policy evaluation illustrate the additional structure required when neither individual homogeneity nor temporal stationarity holds \citep{BianEtAl2025}.

Across these settings, evaluation must match the deployment target: another decision for an observed individual, a new individual from a known population, a new task, or a changed environment. Behavior-policy heterogeneity and limited overlap can make an apparently favorable subgroup value weakly identified. The section's main statistical message is therefore not merely that policies vary, but that the unit of heterogeneity, the shared mechanism, and the target value must be stated before borrowing across trajectories. Broader questions about adaptive subgroup selection, interval validity, and rare-group guarantees are taken up in Sections~8 and~12.

\section{Uncertainty quantification for heterogeneous models}
\label{sec:uq}

Uncertainty quantification in heterogeneous models addresses three questions: whether individuals or subpopulations differ, how precisely their effects can be estimated, and how reliable predictions and decisions are for the target population. When subgroup structures, borrowing rules, or policies are learned from data, inference must also account for uncertainty in those choices. Table~\ref{tab:uq-guarantees} summarizes the inferential targets and corresponding guarantees.

\begin{table}[!hbp]
\centering
\caption{Inferential objects and guarantees in heterogeneous analysis.}
\label{tab:uq-guarantees}
\begingroup
\small
\renewcommand{\baselinestretch}{1}\selectfont
\renewcommand{\arraystretch}{1.2}
\begin{tabular}{@{}p{0.29\linewidth}@{\hspace{4pt}}p{0.70\linewidth}@{}}
\hline
\textbf{Inferential object} & \textbf{What should be established} \\
\hline
Existence of heterogeneity & Valid tests for differences across individuals or subpopulations \\
Discovered groups & Uncertainty in group memberships, boundaries, and number \\
Selected contrasts & Post-selection or simultaneous error control \\
Predictions & Coverage and calibration within specified subgroups \\[4pt]
Learned decisions & Confidence intervals for policy value and bounds on regret \\
\hline
\end{tabular}
\endgroup
\end{table}

\subsection{Uncertainty in heterogeneity detection and subgroup discovery}

For heterogeneous models, a primary problem is to determine whether individuals or subpopulations differ beyond sampling variation. Three representative testing frameworks can address this problem: fixed-effects tests for prespecified groups, variance-component tests, and mixture-homogeneity tests. For prespecified groups, fixed-effects ANOVA tests whether group means are equal. Interaction tests extend this comparison to assess whether a covariate effect or a treatment-control mean difference varies across groups \citep{scheffe1959analysis}. When random effects represent individual-specific or group-specific differences, variance-component tests assess heterogeneity by testing whether the random-effect variance is zero. 
When mixture components represent latent subpopulations, mixture-homogeneity tests compare a single-component model with a mixture alternative. Testing the number of components is nonregular because some parameters become unidentified when components disappear or coincide. The expectation-maximization (EM) test for normal-mixture homogeneity addresses this problem using a modified likelihood and specially derived limiting distributions \citep{chen2009hypothesis}. 

The inferred number, memberships, and boundaries of subpopulations are themselves uncertain. Model-based inference and stability assessment address different aspects of this uncertainty. Model-based inference describes uncertainty through membership probabilities, probabilities for alternative subgroup models, and credible sets for subgroup structures. Membership probabilities from finite mixtures describe ambiguity in individual assignments \citep{mclachlan2000finite}, while Bayesian model-selection probabilities summarize support for candidate subgroup-effect models \citep{berger2014bayesian}. To describe uncertainty in the overall structure, credible balls collect plausible partitions \citep{wade2018bayesian}, while credible-subgroup pairs give inner and outer bounds for a treatment-benefiting subpopulation \citep{schnell2016bayesian}. For these Bayesian sets, the reported probability describes uncertainty conditional on the model and observed data. 

A complementary question is whether discovered groups persist across repeated analyses. Stability methods address this question by resampling or splitting the data. Bootstrap similarities measure how consistently individual clusters reappear under resampling \citep{hennig2007clusterwise}. Prediction strength evaluates whether a chosen number of clusters produces reproducible group assignments across data splits \citep{tibshirani2005prediction}.

\subsection{Uncertainty in individual- and subgroup-specific estimates}

After a heterogeneity structure is specified, we need to quantify uncertainty in individual effects and subgroup-specific parameters. Two common settings are random-effects models for individual or group differences and finite-mixture models for latent subpopulations. In random-effects models, each random effect describes an individual or group's deviation from a shared population model. Standard-error and mean-squared-error approximations quantify uncertainty in these effects or group-specific means, including the additional error introduced by estimating variance components \citep{kackar1984approximations,datta2000unified}. In finite mixtures, the corresponding targets are component-specific parameters and mixing proportions. Likelihood-based or bootstrap procedures quantify their uncertainty under the fitted mixture model \citep{mclachlan2000finite}.


Data-driven subgroup selection creates an additional source of error in effect estimation. Intervals that treat the selected groups as fixed can understate uncertainty when the same observations support both discovery and estimation. Three approaches address this problem: correcting selection bias, conditioning on the selection procedure, and using separate samples for discovery and estimation. Debiasing bootstrap methods correct the upward bias caused by selecting the subgroup with the largest estimated effect \citep{guo2021inference}. Conditional selective-inference methods account for the fitted tree when testing differences between selected terminal nodes or constructing intervals for their means \citep{neufeld2022treevalues}. Sample splitting separates discovery from estimation, as illustrated by honest causal trees in Section~\ref{sec:causal}. When a search considers many interactions, trees, or partitions, inference must also address multiplicity across the reported claims. Subgroup identification based on differential effect search (SIDES) uses resampling to adjust for multiplicity within its subgroup-search framework \citep{lipkovich2011subgroup}. 

The choice of which groups or sites share information can also be uncertain. This choice involves both the parameters to pool and the weight assigned to each source. Fusion graphs specify the pooling structure, while source weights control how much each population contributes to an estimate. If either is learned from the data, treating it as fixed can understate uncertainty. Robust inference for federated meta-learning (RIFL) illustrates how source-selection uncertainty can be incorporated into inference. It targets a model shared by a majority of sites and uses resampling to account for uncertainty in identifying eligible sites \citep{GuoEtAl2025RIFL}. 

Federated estimation involves several sources of uncertainty beyond source selection. These include errors in estimating site-specific nuisance parameters, noise added for privacy, and variation in client participation. Site-specific nuisance parameters allow local populations to differ, and heterogeneity-aware inference accounts for their estimation when combining information \citep{DuanNingChen2022}. Privacy mechanisms introduce further randomness into the communicated information, as in the differentially private version of stochastic controlled averaging for federated learning (DP-SCAFFOLD) \citep{noble2022differentially}. More details on federated learning are provided in Section~\ref{sec:federated}.

\subsection{Uncertainty in prediction and decision making}

The construction of subgroup-valid prediction sets depends on whether groups are calibrated separately or jointly. Separate calibration uses a conformity-score quantile within each prespecified group, giving group-conditional coverage under exchangeability \citep{romano2020malice}. For potentially overlapping groups, a common prediction rule must satisfy several coverage requirements simultaneously. Multivalid conformal prediction addresses this problem by adjusting score thresholds until coverage errors are controlled across groups and threshold levels \citep{jung2023batch}. Conditional conformal methods express group membership through indicator functions and impose the associated coverage conditions within a shared function class \citep{gibbs2025conditional}.


Prediction for a new patient or user population requires checking whether the calibration data still represent the intended population. The appropriate adjustment depends on what has changed. Under covariate shift, the covariate distribution changes but the conditional outcome law remains the same. Weighted conformal prediction uses the ratio of the target and calibration covariate densities to reweight calibration scores, providing marginal coverage in the target population when this ratio is known \citep{tibshirani2019covariate}. Estimated ratios introduce additional error, and changes in response mechanisms or user preferences fall outside this guarantee. 

A related question is whether predicted probabilities reflect observed frequencies within subpopulations. Multicalibration addresses this question by requiring approximate agreement within each sufficiently large group in a specified class, including among individuals assigned similar predicted probabilities \citep{hebertjohnson2018multicalibration}. For numerical outcomes, calibrating the mean does not by itself describe the variability around that mean. Mean-conditioned moment multicalibration therefore calibrates variances and higher moments together with means \citep{jung2021moment}. 

Personalized AI requires quantifying uncertainty about what an individual user prefers. For example, limited feedback may leave it unclear whether the user prefers concise answers or detailed explanations. This epistemic uncertainty reflects incomplete knowledge of the user and can be reduced by collecting informative feedback. Variational preference learning represents it through a distribution over latent user preferences conditioned on the user's observed feedback and uses that distribution to guide further preference queries \citep{poddar2024personalizing}. 
Even if those preferences were known, the user's feedback on a particular answer could still vary. This variation in feedback is distinct from uncertainty about the underlying preferences. 

Reliable predictions are not sufficient to establish the benefit of an individualized decision rule. A policy maps individual characteristics or histories to actions, so its performance depends on both the decision rule and the population in which it is used. Decision uncertainty has two distinct targets: the expected outcome under a policy and the loss from learning a policy that is worse than the best alternative in a specified class. Value-inference methods address the first target by constructing confidence intervals for the policy's expected outcome, including for fixed and data-dependent policies in infinite-horizon settings under suitable trajectory and estimation conditions \citep{ShiEtAl2022Value}. Regret bounds address the second target by controlling the loss relative to the best policy in the stated class. Doubly robust policy learning provides such bounds under causal identification, overlap, nuisance-estimation, and policy-class conditions \citep{athey2021policy}. Value intervals quantify uncertainty about policy performance, whereas regret bounds assess the quality of the procedure used to learn the policy. 


\section{Data Heterogeneity in Federated Learning}
\label{sec:federated}

Federated learning combines information held at multiple sites without centralizing individual records. We focus on horizontal, cross-silo settings, where institutions observe common variables on different individuals. Because sites often represent distinct populations, an aggregation rule specifies both how information is exchanged and which population the result describes.

\begin{figure}[t]
\centering
\includegraphics[width=0.88\textwidth]{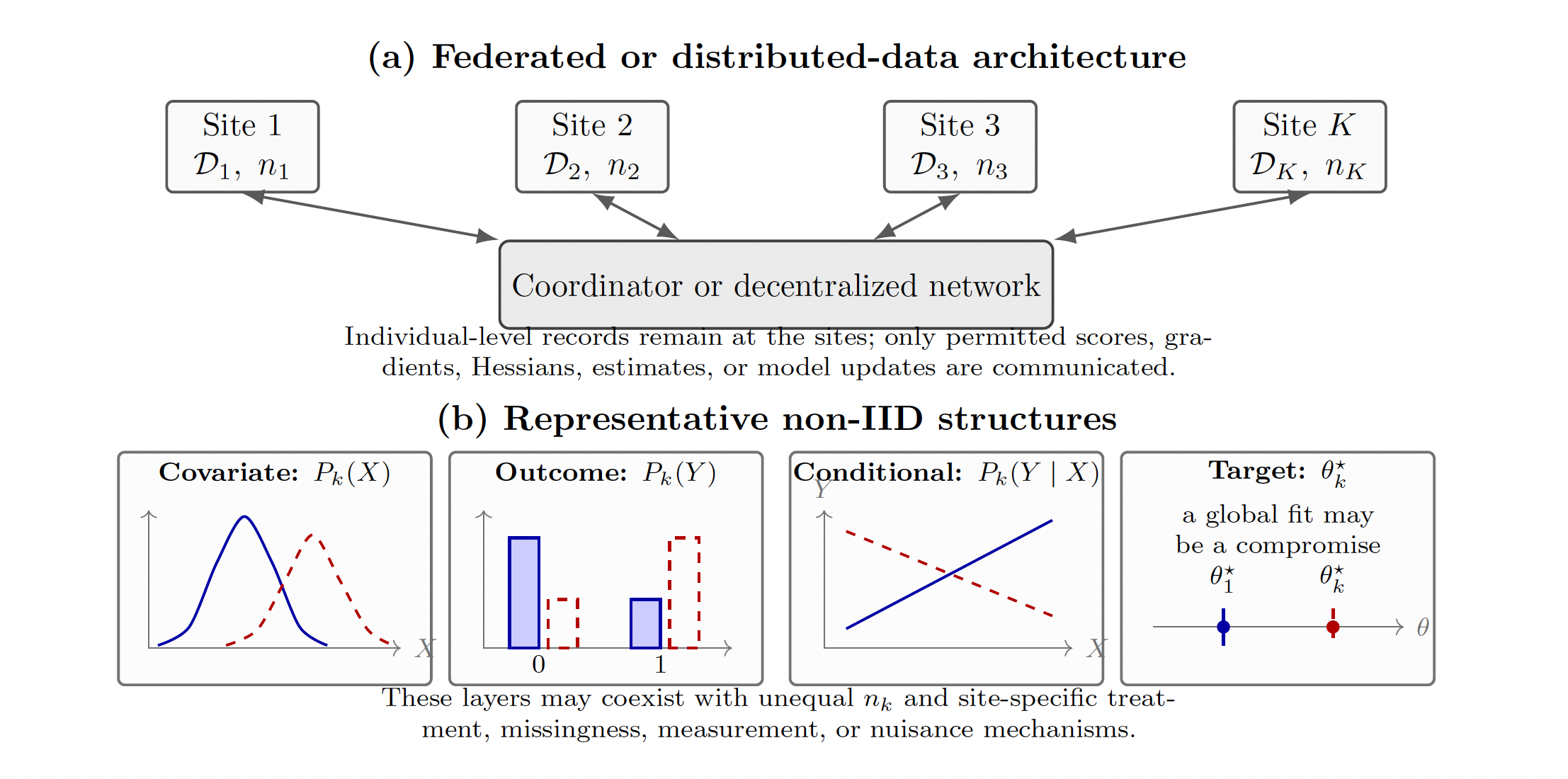}
\caption{Federated architecture and distinct forms of non-IID structure. Aggregation specifies both a communication protocol and a statistical target.}
\label{fig:fl-heterogeneity}
\end{figure}

Let site $k$ observe data from law $P_k$, with risk $R_k(\theta)=\mathbb E_{P_k}\{\ell(\theta;Z)\}$. A conventional global objective is
\begin{equation}
  \theta_w^\star=\argmin_{\theta\in\Theta}\sum_{k=1}^K w_kR_k(\theta),\qquad
  w_k\geq 0,\quad \sum_{k=1}^K w_k=1.
  \label{eq:global-target}
\end{equation}
The weights define the target: sample-size weights approximate a pooled individual-level population, equal site weights target an average institution, and target-specific weights can emphasize a deployment site. The label ``non-IID'' does not identify a particular model: sites may differ in case mix, conditional response, treatment or missingness mechanisms, measurement, or primary parameters. Figure~\ref{fig:fl-heterogeneity} distinguishes these sources.

Consider four hospitals. A and C treat more severe cases than B and D; A and D use a new treatment more often than B and C; outcome mechanisms may instead group A with B and C with D; and every hospital may contain different proportions of frail and robust patients. Hospital D may also be the intended deployment target. These structures cross-cut one another, so ``non-IID'' alone does not determine what should be pooled.

\subsection{Shared and site-specific targets}
Federated optimization methods such as FedAvg, FedProx, and SCAFFOLD address communication and client drift for a chosen objective, but do not establish that the objective is scientifically suitable \citep{McMahanEtAl2017,LiEtAl2020,KarimireddyEtAl2020}. If site-specific optima differ, a well-converged global model can still represent no site. Optimization error should therefore be separated from estimation error and from bias caused by an inappropriate common target.

A common parameter can nevertheless be meaningful when sites share the target mechanism but differ in covariate distributions or nuisance components. Distributed score, likelihood-surrogate, and one-step procedures combine local summaries without exchanging raw records; transport-oriented variants target a specified population under assumptions such as overlap and correct nuisance estimation \citep{JordanLeeYang2019,DuanNingChen2022,GuChen2023}. This setting connects federated learning to fixed-effect evidence synthesis. When primary parameters vary, hierarchical and random-effects models instead allow partial pooling and systematic site differences \citep{DerSimonianLaird1986,ThompsonHiggins2002}.

A general representation is $\theta_k^\star=\theta_0^\star+\delta_k$, where the departures $\delta_k$ are unrestricted under local estimation, set to zero under complete pooling, or shrunk toward shared structure under partial pooling. Borrowing reduces variance when sites are sufficiently similar but can cause negative transfer when their differences are consequential. Minimax results formalize this trade-off: global aggregation can be optimal under small heterogeneity, while local or partially pooled estimation can be preferable as heterogeneity grows \citep{ChenEtAl2023,SuXuYang2023}.

\subsection{Subgroups, mixtures, and target-site transfer}
Two forms of subgroup structure should be distinguished. Client clustering posits groups of sites with shared parameters, $\theta_k^\star=\beta_{g_k}^\star$. Fusion and assignment methods estimate these groups, but separation of local estimates can also reflect sample size, noise, or optimization error; a computational cluster is not automatically a scientific subgroup \citep{ZhangLiuZhu2024,GhoshEtAl2020}. A within-client mixture instead writes $P_k=\sum_g\pi_{kg}Q_g$: each site contains shared latent populations in site-specific proportions. Mixture methods estimate the components and proportions, with the usual challenges of identifiability and weak separation \citep{MarfoqEtAl2021}. These structures differ from target-site transfer, which asks which source institutions can improve estimation at a designated site. Source relevance is estimand-specific and need not be transitive \citep{LiCaiLi2022}.

Site-level groups must also be distinguished from individual-level treatment-effect subgroups. A patient-level target may be $\tau_k(x)=\mathbb E\{Y(1)-Y(0)\mid X=x,\text{site}=k\}$. Its identification requires the usual treatment consistency, exchangeability or randomization, and positivity conditions; borrowing additionally requires support overlap and assumptions linking source and target potential outcomes. Similar hospital-level prediction models alone do not justify pooling causal effects \citep{LiDonohueTang2026,TanEtAl2022}.

\subsection{Adaptive borrowing, inference, and privacy}
The client groups, fusion structure, and source weights are often selected from the same data used for inference. Treating the selected structure as fixed can understate uncertainty when groups are weakly separated or sites are small. Likewise, pooled conformal calibration need not provide a site-specific coverage guarantee when score distributions differ. Federated inference must state whether its target is a random individual from the mixture, a designated site, or each site separately \citep{GuoEtAl2025RIFL,HanEtAl2025FACE,LuEtAl2023FCP}. Evaluation should report site-averaged and worst-site performance, target-site calibration, and sensitivity to alternative borrowing structures.

Federation alone is not a privacy guarantee. Local data retention, secure aggregation, cryptographic computation, and differential privacy address different threat models. The protected unit (for example, a patient or an entire institution), privacy parameters, and subgroup utility costs should be reported explicitly. Formal definitions, distributed algorithm comparisons, fusion formulations, and privacy mechanisms are given in Online Supplement~D.

\section{Data Security in the Presence of Heterogeneity}
\label{sec:security}

Aggregate security guarantees can conceal unequal vulnerability. A poisoning or backdoor defense may mistake benign subgroup differences for malicious behavior, while an attacker may target a naturally occurring subgroup and leave overall performance largely unchanged. Evaluation should therefore report subgroup-specific attack success and defense performance, not only population averages \citep{biggio2012poisoning,gu2017badnets,jagielski2021subpopulation}.

Heterogeneity also supports data-ownership protection. Watermarking embeds a hidden, statistically detectable signal to establish ownership, protect intellectual property, or identify unauthorized use. Fingerprinting can assign source- or recipient-specific marks, making it possible to trace the provenance of a released copy. These methods deliberately introduce differences that a detector can test. For example, \citet{zhao2025tabdrw} propose TAB-DRW, a post-editing framework that embeds signals in the frequency domain.

Attribution can also use source-specific patterns already present in the data or deliberately designed marks. \citet{cui2026stamp} propose observation-level keys for tabular data that support detection of source-specific signals. In either case, attribution depends on sources being distinguishable; overlapping source distributions can make it less reliable. The online supplement reviews attack and defense mechanisms, watermarking, and attribution in greater detail.

\section{Precision Medicine and Mobile Health}
\label{sec:mhealth}

Precision medicine is where the review's three themes, representing heterogeneity, borrowing information selectively, and drawing valid inference, meet clinical and behavioral decisions. We illustrate them with three complementary applications, each specified through the four objects of Section~\ref{sec:foundational-concepts-and-classic-methods}. Molecular subtyping partitions a population of tumors into latent groups. Treatment-effect subgrouping asks whether an average effect conceals variation across patients. Longitudinal mobile health (mHealth) studies, using smartphones, wearables, and ecological momentary assessment (EMA), ask how much each person's pattern should borrow from others and whether that pattern changes over time.

\subsection{Molecular subtyping and multi-omics integration}

Here the unit of heterogeneity is the tumor, and the goal is a partition that will apply to future patients. \citet{ramazzotti2018multiomic} developed CIMLR, which combines gene expression, methylation, mutation, and copy-number data into one similarity measure before clustering. Across 36 cancer types, the resulting subtypes differed significantly in survival for 27. This is a version of the integration problem in Section~\ref{sec:federated}, since no single omic layer is informative enough alone and naive concatenation lets the highest-dimensional layer dominate. A subtype found this way is the output of an unsupervised algorithm, and it does not show that the underlying biology comes in discrete types. Its clinical value depends on external validation against survival or treatment response.

\subsection{Treatment-effect subgrouping}

Here the unit of heterogeneity is the patient, the varying component is the treatment contrast, and the target is the conditional average treatment effect defined in Section~\ref{sec:causal}. \citet{vanvogt2025application} applied causal forests in a secondary analysis of the VANISH trial, a 408-patient comparison of vasopressin and norepinephrine for septic shock. Although the average effect on 28-day survival was not significant, the forest identified a data-derived split on baseline serum potassium along which the estimated risk difference ranged from mildly positive to strongly negative. Classical and data-adaptive methods also flagged potassium, but agreement among methods applied to the same data is not independent validation, and the forest's own heterogeneity test for this split was not significant ($p=0.124$). The authors describe the finding as hypothesis-generating, and it needs confirmation in independent data (Section~\ref{sec:uq}). Sequential treatment decisions, in which the patient's state evolves between choices, are discussed in Section~\ref{sec:heterogeneous-rl}.

\subsection{Longitudinal personalization in mobile health}

In mHealth studies the unit of heterogeneity is a participant observed over many occasions, and the varying component may be an outcome level, an association, or a treatment effect. The central question is which model components should vary by person. \citet{balliu2024personalized} found that personalized models of depressive symptoms, built from smartphone sensing in 183 individuals over 40 weeks, predicted mood up to three weeks ahead and outperformed a population model. \citet{phanord2026digital} showed that the gain depends on what is personalized. In a year-long study of 24 emerging adults, most of the predictive power of a hybrid model of affect and stress came from participant identity, with little added by person-specific behavioral dynamics. A model can predict well by capturing that one participant generally reports more distress than another, without learning how that participant's distress responds to changes in behavior. Which components to share also depends on whether predictions are for existing participants or new ones (Section~\ref{sec:foundational-concepts-and-classic-methods}).

Two small pregnancy studies using wearable and EMA data illustrate individual and time-varying associations in more detail. \citet{rim2025bindividualized} represented each woman's activity-sleep association as a spline coefficient function over gestation and encouraged sharing through subgroups of similar curves, so that women were grouped by the shape of their association over time instead of by average level. Among eight women, the overall association between step count and deep sleep was not statistically significant. The estimated individual functions, however, differed substantially. Some were positive and diminishing, some turned from negative to positive around 30 weeks of gestation, and others increased steadily. Because the subgroups are themselves estimated, inference about any one woman's curve should also account for uncertainty in the learned grouping. \citet{rim2025aheterogeneous} tied variation to measured context instead. Among nine women, each additional 1,000 steps was associated with about 3.5\% higher nighttime HRV on high-distress days (95\% CI: 2.6\% to 4.4\%), with no reliable association on low- or moderate-distress days. Because distress varies both within and between women, this interaction can reflect both kinds of variation.

Heterogeneity can also sit inside a single summary measure. Using wearable ring data from pregnant women, \citet{guo2026daily} found that deep sleep was associated with lower heart rate and higher HRV, while REM sleep showed the reverse. Total sleep duration combines these stages, so an analysis of total sleep alone averages over associations of opposite sign. Unlike the molecular subtypes above, sleep stages are observed physiological components rather than latent groups inferred from the data. Whether to model total sleep or its individual stages should therefore depend on which cardiovascular process the analysis targets.

Randomization gives the most direct way to estimate how responses to treatment vary over time. In the HeartSteps micro-randomized trial, activity suggestions increased step counts over the following 30 minutes on average, but the effect diminished over the course of the study \citep{klasnja2019efficacy}. As with the pregnancy associations, an average across occasions would hide this decline, and classifying a participant as responsive or unresponsive is of limited use when responsiveness changes over time. Causal excursion effects formalize these occasion-specific contrasts \citep{boruvka2018assessing}. This design supports estimating how effects vary across contexts, although that is a different claim from showing that changing the context itself would change the response. Personalized HeartSteps later used online reinforcement learning to decide when to deliver suggestions \citep{liao2020personalized}.

\subsection{Limitations common to these applications}

Apparent heterogeneity in these settings can arise from data collection rather than from the process of interest, through measurement error in wearable-derived outcomes, informative missingness from non-wear periods, device drift, variable adherence to EMA prompts, and behavior change induced by self-monitoring. Dense within-person sampling also does not substitute for independent participants when the target is population-level. The eight-, nine-, and twenty-four-participant studies above are small, and none evaluates whether its fitted curves or sharing structure transport to other devices or populations. The molecular subtypes and the VANISH subgroup require the same external validation, and few of these studies test whether acting on the heterogeneity would improve outcomes.
\section{Open Problems and Future Directions}
\label{sec:open_problems}

The methods reviewed here leave recurring problems: choosing how to represent consequential variation, deciding when a subgroup is reproducible, and accounting for adaptivity in inference and decisions. Methods must distinguish discrete, continuous, overlapping, and time-varying variation; identify compatible sources for borrowing without erasing rare groups; and account for selection in subgroup discovery, source choice, and policy learning. These questions also shape the evaluation of emerging AI systems.

\paragraph{Auditing heterogeneous systems.} Small or poorly characterized subgroups may bear harms that random evaluation rarely detects. Audit design should allocate queries adaptively across populations, environments, and rare states. Evaluators also differ in reliability across domains, languages, and response styles; modeling evaluator variation would show whether conclusions depend on who judged. Methods are needed to quantify uncertainty over both the audited population and its evaluators.

\paragraph{Specialization and coordination in agentic AI.} Task subgroups may require different models, tools, and workflows, and errors can accumulate differently along execution trajectories. A central question is when specialization and collaboration improve outcomes, accounting for task structure, dependence among agent errors, and computational cost. Evaluation should measure reliability across tasks and environments, including failures and uncertainty in workflow selection.

\paragraph{The AI harness as a source of variation.} System prompts, tools and permissions, retrieval and memory, orchestration, guardrails, sampling settings, and model versions can change an agent's behavior even when its underlying model is fixed. Statistical methods for measuring these effects could support reproducibility and help determine when harness adaptation is appropriate. More broadly, AI-enabled analyses should be judged by whether they discover stable structure, preserve consequential differences, and support reliable decisions across intended populations.

\section{Conclusion}
\label{sec:conclusion}

The appropriate degree of personalization depends on the target, the structure shared across units, and the uncertainty a decision can tolerate. The same trade-off appears in clinical subgroup analysis, sequential learning, synthetic data, foundation models, and multi-site work: pooling can stabilize estimates, but it can also conceal differences that matter for a target population.

Subgroups are useful when they provide stable, interpretable contrasts tied to a decision. When differences overlap, evolve, or vary continuously, a fixed partition is inadequate. A sound analysis should state its target, account for adaptive discovery, and show which populations its conclusions do and do not support.



\clearpage
\setcounter{section}{0}
\renewcommand{\thesection}{S\arabic{section}}

\def\spacingset#1{\renewcommand{\baselinestretch}{#1}\small\normalsize} \spacingset{1}
\begin{center}
  {\LARGE\bf Supplementary Material for\\ ``Heterogeneity and Subgrouping in Modern Statistics and AI''}
\end{center}
\medskip
\spacingset{1.45}



\section{Heterogeneity in Single-Cell Data Analysis}
\label{supp:HDClustering}

\label{subsec:singlecell}

Three downstream questions should be separated. Marker discovery compares features across discovered cell populations, for example
\[
  H_{0,j}^{kk'}:\quad F_{kj}^{\mathcal T}=F_{k'j}^{\mathcal T}.
\]
Differential expression or differential state compares a biologically matched population across conditions,
\[
  H_{0,j}^{(g)}:\quad F_{gj}^{(1)}=F_{gj}^{(0)},
\]
whereas differential abundance compares population prevalence,
\[
  H_0:\quad \Pr(Z=g\mid D=1)=\Pr(Z=g\mid D=0).
\]
These targets have different null hypotheses and experimental units. A marker gene is not automatically condition-associated, a condition-associated gene does not establish a change in cell-type abundance, and an estimated cluster may not correspond one-to-one with a stable biological type across datasets.

Double dipping and pseudoreplication are also distinct. Double dipping results from using the same molecular measurements to discover clusters and test the contrasts induced by those clusters. Pseudoreplication arises when treatment or disease is assigned at the donor or specimen level but individual cells are analyzed as independent replicates. Replicate-aware pseudobulk or hierarchical procedures address donor-level sampling variation more appropriately \citep{ZimmermanEtAl2021,SquairEtAl2021,CrowellEtAl2020}; they do not, by themselves, account for adaptive clustering. Conversely, a selective test can address the cluster-selection event while still using the wrong experimental unit.

Some procedures avoid discrete subgroup contrasts. singleCellHaystack searches for genes whose expression is nonrandomly distributed over an estimated cell-state space rather than comparing two named clusters \citep{VandenbonDiez2020}. Such cluster-free methods can be useful for developmental or activation continua, but they answer a different question and inherit uncertainty from the learned representation. Cluster and embedding diagnostics provide another layer of evidence: ROGUE evaluates internal purity of a proposed cell population, while scDEED examines whether local relationships in a two-dimensional embedding are reliable relative to the higher-dimensional data \citep{LiuEtAl2020ROGUE,XiaLeeLi2024}. These diagnostics assess aspects of representation or cluster quality; they do not themselves establish valid marker inference or reproducibility in a new cohort.

The two workflows in this section should therefore be viewed as complementary. Two-stage procedures permit explicit inferential calibration for selected contrasts, whereas joint procedures can improve subgroup recovery and reveal the features or subspaces that define the partition. Neither approach alone resolves all uncertainty created by variable screening, graph construction, resolution tuning, rare populations, and donor-level dependence. A high-dimensional clustering analysis is scientifically complete only when it states the target of the partition, the source of its interpretation, and the uncertainty that remains after subgroup discovery.


\section{Additional Causal Methods}
\label{supp:causal}

This supplement extends Section~4 of the main text. It records estimators, identification results, and fairness formulations that further illustrate the main text's arguments.

\subsection*{Additional CATE estimators}
\label{supp:cate}

Generalized random forests \citep{athey2019generalized} reinterpret the causal forest as an adaptive kernel that weights local moment equations, extending honest forests from CATEs to quantile and instrumental-variable effects.

Neural CATE estimators build on the observation that counterfactual inference is a covariate-shift problem \citep{johansson2016learning}. TARNet and counterfactual regression \citep{shalit2017estimating} supply the canonical architecture, a shared representation feeding separate outcome heads per arm, with a generalization bound in which CATE error is controlled by factual error plus an integral probability metric between the two arms' representation distributions. DragonNet \citep{shi2019adapting} predicts the propensity score from the shared representation and adds a targeted regularization term derived from semiparametric efficiency theory, building the orthogonality corrections of the R- and DR-learners directly into network training. Generative variants model the missing potential outcome itself. CEVAE \citep{louizos2017causal} treats observed covariates as noisy proxies of a latent confounder, and GANITE \citep{yoon2018ganite} generates plausible counterfactual outcomes adversarially. Critical assessments of this literature have proved as valuable as the architectures. \citet{curth2021nonparametric} recast neural CATE estimators in meta-learner terms, clarifying which architectures implement which estimation strategies, and \citet{curth2021inductive} showed on semi-synthetic benchmarks that inductive biases matter more than architectural novelty.

When the full CATE function cannot be estimated at useful precision, \citet{semenova2021debiased} provide debiased inference on its best linear predictor, a valid and interpretable linear summary that complements the generic machine-learning inference of \citet{chernozhukov2025generic}.

\subsection*{Subgroup claims in randomized trials}
\label{supp:rct}

Surveys documented widespread misuse of baseline data in trial subgroup analyses \citep{assmann2000subgroup}, prompting guidance on when subgroup analysis is warranted \citep{rothwell2005treating} and reporting standards for claimed subgroup effects \citep{wang2007statistics}. On the methodological side, \citet{ding2019decomposing} decomposed randomization-based effect variation into a component systematically explained by covariates and an idiosyncratic remainder \citep{DerSimonianLaird1986, higgins2002quantifying}.

\subsection*{Matching, principal strata, and Mendelian randomization}
\label{supp:observational}

\citet{abadie2006large} derived the large-sample theory of matching estimators, including a bias term that requires correction in general. In the instrumental-variable setting, \citet{angrist1996identification} recast the local average treatment effect in explicitly causal terms, with compliers, always-takers, never-takers, and defiers as principal strata and monotonicity as the assumption excluding defiers. Mendelian randomization \citep{sanderson2022mendelian}, which uses genetic variants as instruments for modifiable exposures, is the framework's most active modern application and inherits all of these subtleties, including the reinterpretation of the estimand as complier-specific once effects are heterogeneous.

\subsection*{Mediation and heterogeneity of mechanisms}
\label{supp:mediation}

Mediation analysis asks whether mechanisms, not only effects, differ. The distinction between moderators and mediators drawn by \citet{baron1986moderator} supplied the vocabulary. Moderators index heterogeneity of effects, whereas mediators carry the effects. \citet{imai2010general} put mediation on counterfactual footing, giving nonparametric identification of natural direct and indirect effects under sequential ignorability together with sensitivity analysis for its violation, and \citet{vanderweele2015explanation} unified mediation and interaction in a four-way decomposition of the total effect. Moderated mediation \citep{preacher2007addressing} covers the case most relevant to the main text. When the strength of an indirect pathway varies across subgroups, heterogeneity resides in the mechanism itself, a strictly richer notion than heterogeneity of the total effect.

\subsection*{Causal approaches to algorithmic fairness}
\label{supp:fairness}

Algorithmic fairness concerns whether a model's behavior differs across subgroups defined by protected attributes, and whether such differences are justifiable. Causal formulations make the question precise. Counterfactual fairness \citep{kusner2017counterfactual} requires that a prediction be invariant to a counterfactual change in the protected attribute, holding background variables fixed. \citet{kilbertus2017avoiding} refined the criterion through resolving and proxy variables, distinguishing attribute-influenced paths considered legitimate from those that are not, and \citet{nabi2018fair} formulated path-specific fairness as constrained inference, bounding effects along discriminatory pathways while leaving others free. The causal explanation formula of \citet{zhang2018fairness} decomposes an observed disparity into direct, indirect, and spurious components. The monograph of \citet{plecko2024causal} unifies this literature into a framework connecting fairness criteria to the standard causal hierarchy.

A complementary concern is intersectionality, the question of whether fairness guarantees survive at the intersections of protected attributes. \citet{kearns2018preventing} showed that a classifier can satisfy statistical fairness on every marginal group while concentrating harm on structured intersections, such as a specific combination of race, gender, and age, and gave an auditing-and-learning framework over exponentially many subgroups via a two-player game. Multicalibration \citep{hebertjohnson2018multicalibration} strengthens calibration to hold simultaneously across every subgroup in a rich, efficiently identifiable class, and is developed as a general tool for subgroup-valid prediction in Section~8 of the main text.

\subsection*{Reliability and transport: additional results}
\label{supp:causal-reliability}

Weighted conformal inference for individual treatment effects extends to observational data with guarantees that depend on how well the propensity score or the relevant outcome quantiles are estimated \citep{lei2021conformal}, and \citet{jin2023sensitivity} extend the construction to remain valid under bounded departures from unconfoundedness. \citet{xu2022calibration} define and estimate calibration error for heterogeneous effect estimators, the diagnostic that causal isotonic calibration \citep{vanderlaan2023causal} is designed to drive to zero. Sample splitting or cross-fitting is recommended when the pseudo-outcomes rely on estimated nuisances. \citet{dahabreh2020extending} give the identification conditions and estimators for extending trial-estimated effects to a target population, complementing the review of \citet{degtiar2023review}. Among foundation-model approaches, \citet{dhawan2024end} use large language models to perform end-to-end treatment effect estimation from unstructured text, and prior-data fitted networks trained on synthetic causal worlds \citep{ma2025foundation} deliver zero-shot CATE estimation from tabular data in a single forward pass.

\setcounter{section}{2}
\section{Generative Models and Personalized AI Methods}
\label{supp:generative-ai}
\begingroup
  \let\suppsubsection\subsection
  \let\subsection\subsubsection
  \renewcommand{\section}[1]{\suppsubsection*{#1}}

\section{Generative Models for Heterogeneous Synthetic Data}
\label{supp:synthetic}

This supplement extends Section~5 of the main text with the classical synthesizers and the deep generative architectures summarized there in one table.

\subsection*{Classical and copula-based synthesizers}
\label{supp:synthetic-classical}

Beyond the fully and partially synthetic proposals of \citet{rubin1993statistical} and \citet{reiter2003inference}, \citet{reiter2005using} introduced CART-based sequential synthesis, modeling each variable given its predecessors with flexible trees. The approach proved durable and underlies the widely used \texttt{synthpop} R package \citep{nowok2016synthpop}, with practical large-sample inference developed by \citet{raab2016practical}. Finite mixture models \citep{mclachlan2000finite} and hierarchical Bayesian models generate a heterogeneous population as draws from labeled latent subpopulations or nested levels, so the subgroup structure is a named model component rather than an emergent property. Copula-based synthesis occupies the middle ground between classical and deep methods. The Synthetic Data Vault \citep{patki2016synthetic} pairs estimated marginals with a Gaussian copula for dependence, and copula flows \citep{kamthe2021copula} replace the parametric copula with a normalizing flow.

\subsection*{Variational autoencoders}
\label{supp:vae}

The variational autoencoder \citep{kingma2014auto} learns a latent-variable generative model by maximizing an evidence lower bound, and its conditional variant \citep{sohn2015learning} makes generation controllable. Conditioning on a covariate or subgroup label yields subgroup-specific synthetic samples, providing the basic mechanism for targeted augmentation of underrepresented groups.

For heterogeneous populations, the standard Gaussian prior is the VAE's weakness. A unimodal latent distribution encourages blending of distinct subpopulations. The remedies mirror classical mixture modeling in latent space. Gaussian-mixture VAEs \citep{dilokthanakul2016deep} replace the prior with a mixture, and variational deep embedding \citep{jiang2017variational} learns generation and clustering jointly, so that latent mixture components align with data subpopulations. The component indicator is the latent class, the encoder estimates posterior component probabilities, an operation reminiscent of the E-step but not generally equivalent to it; the same identifiability concerns arise as in finite mixtures.

Tabular data add the challenge of mixed-type heterogeneity. Continuous, count, categorical, and ordinal variables with missingness are the norm in biomedical and social science applications. HI-VAE \citep{nazabal2020handling} equips the decoder with per-feature likelihoods matched to each variable's type and handles incomplete data natively. TVAE \citep{xu2019modeling} pairs a VAE with mode-specific normalization, fitting a mixture to each continuous variable and encoding values relative to their mode. This design tames the multimodal marginals typical of real tables, and TVAE is a widely used tabular VAE baseline.

\subsection*{Conditional GANs}
\label{supp:gan}

Generative adversarial networks \citep{goodfellow2014generative} learn a generator through a minimax objective against a discriminator. Conditional GANs \citep{mirza2014conditional} condition both networks on side information such as a class or subgroup label, and AC-GAN \citep{odena2017conditional} strengthens the conditioning through an auxiliary classification head.

The central GAN pathology for heterogeneous data is mode collapse. Remedies attack the problem from several angles. Unrolled GANs \citep{metz2017unrolled} let the generator anticipate discriminator updates. VEEGAN \citep{srivastava2017veegan} adds a reconstructor network that enforces latent-space invertibility. PacGAN \citep{lin2018pacgan} feeds the discriminator packs of samples, and a hypothesis-testing analysis shows that discriminating on packs detects mode collapse that single-sample discrimination provably cannot.

Subsequent work adapted these ideas to tabular and biomedical data, settings where subgroup structure is often the primary concern. CTGAN \citep{xu2019modeling} addresses categorical imbalance directly through training-by-sampling, which increases exposure to rare categories during training, and it pairs this with mode-specific normalization. CTAB-GAN and its extension \citep{zhao2021ctabgan, zhao2024ctabganplus} handle mixed continuous-categorical columns, long-tailed distributions, and skewed variables. In health care, medGAN \citep{choi2017generating} generates electronic health record code vectors via an autoencoder-coupled GAN. TimeGAN \citep{yoon2019time} combines adversarial and supervised losses in a learned embedding space to preserve temporal dynamics. Privacy-preserving variants integrate formal guarantees into training. PATE-GAN \citep{jordon2019pategan} adapts the PATE teacher-ensemble framework to the discriminator, and DP-GAN \citep{xie2018differentially} injects calibrated noise into discriminator gradients for differential privacy. As Section~5.3 of the main text and Section~9.3 discuss, formal privacy and minority subgroup fidelity are in inherent tension.

\subsection*{Diffusion models}
\label{supp:diffusion}

Diffusion models offer an alternative based on iterative denoising. The family spans denoising diffusion probabilistic models \citep{ho2020denoising}, score-based models in the continuous-time SDE framework of \citet{song2021score}, and models that operate in learned latent spaces \citep{rombach2022high}. Their mode coverage depends on the model, training distribution, and sampling procedure; better coverage of minority modes is an empirical tendency rather than a guarantee. Architecture alone does not establish subgroup fidelity.

Their conditioning mechanisms are directly relevant to subgroup control. Classifier guidance \citep{dhariwal2021diffusion} steers sampling along the gradient of a classifier's log-probability for a target label. Classifier-free guidance \citep{ho2022classifier} obtains the same steering by jointly training conditional and unconditional score networks and extrapolating between them. The guidance scale exposes an explicit fidelity-diversity trade-off. Stronger guidance yields samples more prototypical of the conditioned subgroup at the cost of within-subgroup variance.

Tabular diffusion adapted the framework to mixed types. TabDDPM \citep{kotelnikov2023tabddpm} runs Gaussian diffusion on continuous features and multinomial diffusion on categorical ones. STaSy \citep{kim2023stasy} applies score-based generation with self-paced training. CoDi \citep{lee2023codi} co-evolves two conditioned diffusion processes for the continuous and discrete portions of a table. TabSyn \citep{zhang2024mixed} performs diffusion in a transformer-learned joint latent space and reported leading fidelity and sampling speed on standard tabular benchmarks at publication. Biomedical and temporal applications include EHRDiff \citep{yuan2023ehrdiff} and MedDiff \citep{he2023meddiff} for electronic health records, along with CSDI \citep{tashiro2021csdi} for conditional time-series imputation and generation.

\subsection*{Additional fidelity and coverage methods}
\label{supp:synthetic-fidelity}

Inclusive GAN \citep{yu2020inclusive} combines adversarial training with reconstructive coverage objectives to ensure that minority samples are represented, complementing the reweighting approach of \citet{choi2020fair}. Language models now serve as tabular synthesizers as well. GReaT \citep{borisov2023language} serializes table rows as sentences and fine-tunes an autoregressive language model to generate them, conditioning flexibly on arbitrary feature subsets. It imports both the world knowledge and flexible conditioning of language models and their opaque, prompt-sensitive representation of subgroup structure. For orientation to this fast-moving area, recent reviews cover conditional generation for precision medicine and diverse representation \citep{liu2025conditional}, synthetic data in biomedicine broadly \citep{vanbreugel2024synthetic}, and the conceptual landscape of synthetic data \citep{jordon2022synthetic}.


\section{AI methods and evaluation}
\label{supp:ai-details}

\subsection*{Prompt representations and evaluation designs}

Persona-Plug weights historical records according to the current input and projects their aggregate representation into the LLM's embedding space. It trains the input encoder, projector, and instruction embedding while keeping the history encoder and LLM fixed \citep{liu2025personaplug}. DEP encodes differences between the target user's responses and those of other users facing similar content. Its ablations compare history embeddings alone with their combination with difference-aware embeddings, both with and without retrieved review text, using reference-text generation scores \citep{qiu2025latent}.

Persona-DB's cold-start analysis selects users with the fewest available historical records. Response prediction is evaluated using correlations for response intensity and F1 scores for sentiment polarity \citep{sun2025personadb}. Comparisons with a version omitting collaborative refinement examine the contribution of collaborators' records. The cold-start setting therefore involves limited history rather than no personal information.

LaMP separates users across training, validation, and test sets, with test users supplying their own histories. Its time-based split instead uses older records as profiles for predicting later responses. Comparisons include profile retrieval and a no-retrieval baseline, with accuracy and F1 for classification, mean absolute and root mean squared errors for rating prediction, and ROUGE scores for text generation \citep{salemi2024lamp}. The temporal split evaluates later responses but does not isolate adaptation to preference drift.

For interactive memory, \citet{yuan2025evolving} compare memory mechanisms on synthetic dialogues testing continuation, acquisition of personal knowledge, and use of earlier feedback in later responses. Evaluation uses ROUGE and GPT-4-based scoring or comparisons. These designs assess feedback use within simulated interactions, leaving adaptation to changing preferences in real users unresolved.

\subsection*{Evaluation of parameter specialization}

For domain adaptation, \citet{gururangan2020dont} compare downstream classification performance with and without additional pretraining on domain text across four domains and both high- and low-resource task settings. The varying resource is labeled task data, rather than information about individual users. AdapterFusion compares fused models with individual task adapters and multitask alternatives. Gains over individual adapters on multiple target tasks support cross-task representation borrowing, although improvements are not uniform \citep{pfeiffer2021adapterfusion}.

PLoRA evaluates personalized text understanding through sentiment prediction using accuracy, F1, and mean squared error. Its few-shot cold-start experiments learn only a new user embedding from a few labeled examples, keeping the trained model parameters fixed. Its zero-shot evaluation instead makes predictions without user-specific representations \citep{zhang2024plora}. The two settings assess different amounts of personal information and different adaptation procedures.

\subsection*{Agent systems and trajectory evaluation}

The retail and airline environments in Tau-bench provide different tools and policy guidelines to customer-service agents \citep{yao2025taubench}. MetaGPT assigns software-development roles such as product manager, architect, and engineer, together with procedures and structured intermediate outputs. Generated code is executed and tested, and this feedback guides revision. Comparisons with and without executable feedback report higher code-generation pass rates when it is included \citep{hong2024metagpt}.

HuggingGPT decomposes a request into subtasks and selects specialist models according to their documented functions, combining language, vision, and speech capabilities \citep{shen2023hugginggpt}. These specialists are callable models and need not be autonomous agents. ReConcile instead uses discussion among agents to revise answers to a common problem \citep{chen2024reconcile}. DyLAN selects teams, filters contributors, and stops deliberation early. Its ablations compare optimized and unoptimized teams and examine correctness and model calls. Smaller gains with domain-mismatched roles qualify the evidence for reusing candidate expertise \citep{liu2024dylan}.

Tau-bench compares the database state after a simulated customer conversation with an annotated goal state and measures consistent success across repeated trials. Manual examination of failed retail trajectories distinguishes incorrect tool arguments, incorrect information, wrong decisions, and partially resolved requests \citep{yao2025taubench}. This analysis locates failures within the execution sequence, but does not provide an automatic score for every intermediate step or establish transfer to unseen domains.

KnowNo's robot experiments examine ambiguities involving object attributes, spatial relations, and incompletely specified preferences. Evaluation compares task success with the frequency of human assistance, including intervention rates across multi-step execution \citep{ren2023robots}. Assistance is therefore assessed together with successful execution, rather than treating reduced intervention alone as evidence of a better policy.

\endgroup

\setcounter{section}{3}
\section{Federated Algorithms, Fusion, and Privacy}
\label{supp:federated}

This appendix expands the discussion of federated learning in the main text. Federated protocols constrain what can be transmitted, but statistical validity still depends on the target population, site relationships, and the type of heterogeneity represented.

\subsection{Distributed estimation and objective mismatch}
FedAvg, FedProx, and SCAFFOLD modify local optimization or control client drift, while surrogate likelihood and one-step procedures communicate low-dimensional summaries to approximate pooled estimation \citep{McMahanEtAl2017,LiEtAl2020,KarimireddyEtAl2020,JordanLeeYang2019}. For a shared estimand with site-specific nuisance functions, local scores or Hessians can be aggregated; under covariate shift, density-ratio weighting can target a designated population \citep{DuanNingChen2022,GuChen2023}. These methods address different sources of error and should not be compared solely by convergence speed. A procedure can optimize its stated objective accurately while that objective is misaligned with the scientific or deployment target.

When client parameters differ, summary-statistic integration, hierarchical shrinkage, or target-specific model averaging can replace complete pooling. High-dimensional approaches estimate common and site-specific components from distributed summaries \citep{CaiLiuXia2022}. The bias-variance trade-off depends on the magnitude and structure of between-site variation; theoretical analyses give regimes where global aggregation, local estimation, or partial pooling is minimax preferable \citep{ChenEtAl2023,SuXuYang2023}.

\subsection{Client groups, latent mixtures, and transfer}
Client clustering assumes groups of institutions share parameters. Fusion penalties shrink selected cross-site differences to zero, and distributed spanning-tree or high-dimensional methods provide recovery guarantees under separation and sparsity assumptions \citep{ZhangLiuZhu2024,ChenEtAl2024Clustered}. IFCA alternates client assignment with group-specific model updates \citep{GhoshEtAl2020}. In either case, estimated distances can reflect sample size, curvature, stochastic noise, and optimization error as well as scientific differences.

A different model treats each client as a mixture of shared latent populations with site-specific proportions. Federated EM-type methods estimate common components and local mixture weights \citep{MarfoqEtAl2021}; component identifiability and weak separation remain central challenges. Graph-structured or coefficient-specific sharing provides intermediate choices, allowing selected parameter blocks to be common while others remain local.

Transfer learning designates a target site and selects information from source sites. High-dimensional methods can estimate target-source contrasts and screen sources to protect against negative transfer \citep{LiCaiLi2022}. In causal problems, target-specific transfer additionally requires assumptions connecting conditional potential outcomes and adequate covariate support, beyond similarity of predictive distributions \citep{TanEtAl2022,LiDonohueTang2026}. Robust federated inference and adaptive weighting methods account for uncertainty in selected source sets \citep{GuoEtAl2025RIFL,HanEtAl2025FACE}.

\subsection{Differential privacy and heterogeneous utility}
Federation does not itself guarantee privacy. A randomized mechanism $\mathcal M$ is $(\varepsilon,\delta)$-differentially private if for neighboring datasets $D,D'$ and every measurable output set $A$,
\begin{equation}
 \Pr\{\mathcal M(D)\in A\}\leq e^\varepsilon\Pr\{\mathcal M(D')\in A\}+\delta.
 \label{eq:supp-dp}
\end{equation}
The definition depends on the adjacency relation: patient-level adjacency changes all records for one patient, whereas institution-level adjacency changes an entire site's participation. Institution-level protection concerns a larger unit and typically requires a more demanding privacy-utility trade-off.

Noise may be added locally, centrally after secure aggregation, or through distributed protocols. Secure aggregation and cryptographic computation can conceal individual updates from a coordinator, but do not by themselves bound what the released model reveals. Differential privacy limits output sensitivity but does not automatically protect communication metadata. These mechanisms are complementary and should be matched to a stated adversary and protected unit \citep{dwork2006calibrating}.

Privacy can also affect subgroups unevenly: clipping and noise may disproportionately degrade utility for rare populations or small sites. Reports should pair privacy parameters and adjacency definitions with site- and subgroup-specific performance, rather than relying on a single aggregate utility measure \citep{bagdasaryan2019differential}.

\setcounter{section}{4}


\section{Data Security in the Presence of Heterogeneity}
\label{supp:Security}

\subsection*{Adversarial Attacks and Defenses}
Specifically, we formulate backdoor attacks and defenses, and briefly discuss how they manifest when data heterogeneity or subgroups form naturally.
Let
\[
    \mathcal D_{\mathrm{clean}}
    =
    \{(X_i,Y_i)\}_{i=1}^n,
    \qquad
    (X_i,Y_i)\overset{\mathrm{iid}}{\sim}F_{\mathrm{clean}},
\]
denote a clean training data set drawn from a joint probability distribution $F_{\mathrm{clean}}$. Suppose that the learning objective is to train a prediction model
$f$ that maps each $X_i$ to $Y_i$. We assume that an attack has the capability to modify both $X_i$'s and $Y_i$'s. For each data instance $i$, let $Z_i\sim\operatorname{Bernoulli}(\rho)$ indicate it is selected for poisoning, where $\rho\in(0,1)$ is the poisoning rate. The attacker applies a trigger transformation $T_\eta$ and changes the label to a target class $y^\star$, producing
\[
    \widetilde X_i=
    \begin{cases}
        X_i, & Z_i=0,\\
        T_\eta(X_i), & Z_i=1,
    \end{cases}
    \qquad
    \widetilde Y_i=
    \begin{cases}
        Y_i, & Z_i=0,\\
        y^\star, & Z_i=1.
    \end{cases}
\]
Here $T_\eta (x)$ may represent an image patch, a blended pattern, a spatial warp, or a trigger phrase in text, and is typically small enough to be unnoticeable. In a simple example, $T_\eta(x)$ can take an additive trigger form in an embedding or feature space
\[
    T_\eta(x)=x+\eta,
\]
where $\eta$ is a vector of the same dimension as $x$. In other words, triggered inputs induce an adversarially constructed subpopulation. Then the attacker trains a poisoned model
\[
    \widehat{f}_{\mathrm{poi}}
    =
    \arg\min_{f \in \mathcal{F}}
    \frac{1}{n}
    \sum_{i=1}^n
    \ell\left(
        \widetilde{Y}_i,
        f(\widetilde{X}_i)
    \right),
\]
where $\mathcal F$ is the prespecified class of candidate prediction functions.
A successful backdoor attack seeks to choose the poisoning rate $\rho$, trigger $\eta$, and target label $y^\star$ in order to achieve two simultaneous objectives. First, the
poisoned model should retain its normal predictive performance on clean inputs, that is:
\[
    R^{\mathrm{cl}}(\widehat f_{\mathrm{poi}})
    \approx
    R^{\mathrm{cl}}(\widehat f_{\mathrm{cl}}),
\]
where $R^{\mathrm{cl}}(\widehat f)=
    \mathbb E_{(X,Y)\sim F_{\mathrm{clean}}}
    \left[
        \ell(Y,\widehat f(X))
    \right]$ and $\widehat f_{\mathrm{cl}}$ is the
model trained without poisoning. Second, the triggered inputs should be classified as the target label with high probability:
\[
    P_{(X,Y)\sim F_{\mathrm{clean}}}
    \left(
        \widehat f_{\mathrm{poi}}\bigl(T_\eta(X)\bigr)=y^\star
    \right)
    \approx 1.
\]
It can be seen that conventional validation on
untriggered data may reveal little or no performance loss. The attack remains stealthy unless one can find the trigger and the target class.
\citet{wang2024demystifying} find that the geometry
of clean-data distributions is central to backdoor vulnerability.

Backdoor defense is usually pursued at either the training stage or the
inference stage. Training-stage defenses seek to identify poisoned examples or
malicious model updates before deployment. A general screening rule flags
instances whose representation, loss, or gradient statistic is anomalous. 
Let
$\mathcal D_{\mathrm{train}}
=\{(\widetilde X_i,\widetilde Y_i)\}_{i=1}^n$
denote the observed, potentially poisoned training sample.
For a screening score $s(\cdot,\cdot)$ and threshold $\tau$,
define
\[
    \mathcal I_{\mathrm{sus}}
    =
    \left\{i:s(\widetilde X_i,\widetilde Y_i)>\tau\right\}.
\]
The learner may remove flagged observations and retrain on
\[
    \mathcal D_{\mathrm{cleaned}}
    =
    \left\{(\widetilde X_i,\widetilde Y_i):
    i\in\{1,\ldots,n\}\setminus\mathcal I_{\mathrm{sus}}\right\}
\]
or retain them with reduced weights. Such screening should
be evaluated within benign subgroups, because an elevated
false-positive rate can disproportionately exclude legitimate
observations from an atypical subgroup.
Clustering-based defenses search for small, label-specific clusters in the
latent space, whereas robust methods use anomalous losses, gradients, or
distances from an estimated clean distribution.

Inference-stage defenses assume that a potentially backdoored model has already
been deployed and seek to identify a future query containing a trigger. 
Let $F_{\mathrm{clean}}^X$ denote the marginal distribution of clean inputs, and let $F_{\mathrm{trig},\eta}^X$ denote the distribution of $T_\eta(X)$ when $X\sim F_{\mathrm{clean}}^X$. For a specified trigger transformation, detection can be framed as
\[
\begin{aligned}
H_0 &: X_{\mathrm{test}}\sim F_{\mathrm{clean}}^X,\\
H_1 &: X_{\mathrm{test}}\sim F_{\mathrm{trig},\eta}^X.
\end{aligned}
\]
When the trigger is unknown, the alternative is composite over a family of transformations. Triggered inputs may overlap substantially with clean inputs, so detection cannot generally be reduced to identifying generic out-of-distribution observations.

A detector computes a score $s\{G(X_{\mathrm{test}})\}$, where $G(\cdot)$ may
be the model output, a perturbation-based statistic, or a latent representation,
and rejects the query when
\[
    d(X_{\mathrm{test}})
    =
    \mathbbm{1}
    \left\{
        s\{G(X_{\mathrm{test}})\}\geq\tau
    \right\}.
\]
The threshold $\tau$ is chosen to maintain a small false-positive rate:
\[
    \Pr\left\{
        d(X)=1\mid X\sim F_{\mathrm{clean}}^X
    \right\}
    \leq \alpha.
\]
For example, \citet{xian2023unified} adopt the conformal prediction framework for finite-sample false-positive control and the Mahalanobis distance of a latent representation
$z=\phi(X)$ from the clean validation distribution as the score:
\[
    s(z)
    =
    (z-\widehat\mu)^\top
    \widehat\Sigma^{-1}
    (z-\widehat\mu),
\]
where $\widehat\mu$ and $\widehat\Sigma$ are the empirical mean and covariance matrix, respectively.

\subsection*{Ownership Protection}

\noindent\textbf{Data Watermarking.}
Let $X$ denote an original data object and let $K$ denote a secret watermark key. 
A watermarking mechanism can be represented as
\[
    \widetilde X = \mathcal{W}(X;K),
\]
where $\widetilde X$ is designed to remain close to $X$ in terms of perceptual or statistical utility while carrying information about $K$. 
Given a candidate object $Z$, watermark detection can then be formulated as the hypothesis-testing problem
\[
    H_0: Z \text{ does not contain } K
    \qquad\text{versus}\qquad
    H_1: Z \text{ contains } K.
\]
A successful watermarking procedure therefore needs to balance several competing objectives, including detectability, fidelity, and robustness against transformations or adversarial removal. 
The specific mechanisms for achieving these objectives vary substantially across data modalities.

For text generated by large language models (LLMs), recent watermarking methods often embed the watermark directly into the stochastic token-generation process. 
Representative examples include the Gumbel-max watermark \citep{aaronson2023watermarking} and the inverse-transform watermark \citep{kuditipudi2023robust}. 
Let $P_{t,w}$ denote the probability of token $w \in \mathcal V$ at position $t$, where $\mathcal V$ is the vocabulary. Then let $\bm P_t=(P_{t,w})_{w\in\mathcal V}$ denote the next-token probability distribution at position $t$. 
Under the Gumbel-max watermark, pseudorandom variables
\[
    U_{t,w}\stackrel{\mathrm{iid}}{\sim}\mathrm{Unif}(0,1),
\]
are generated using a secret key, and the next token is sampled according to
\[
    W_t
    =
    \arg\max_{w\in\mathcal V}
    \frac{\log U_{t,w}}{P_{t,w}}.
\]
It is shown that the Gumbel-max construction $W_t$ follows the original next-token distribution $\bm P_t$ while inducing dependence between the generated token and the pseudorandom variables that can subsequently be exploited for watermark detection. 
Similarly, the inverse-transform watermark introduces a random permutation $\pi_t$ of the vocabulary and a pseudorandom variable $U_t\sim\mathrm{Unif}(0,1)$ and generates
\[
    W_t
    =
    \pi_t^{-1}
    \left\{
        F_t^{-1}(U_t;\pi_t)
    \right\},
\]
where $F_t(\cdot;\pi_t)$ is the cumulative distribution function induced by the permuted next-token probabilities. 
This construction also preserves the marginal next-token distribution, while creating a detectable dependence between $W_t$ and $(U_t,\pi_t)$.
From a statistical perspective, the next-token distributions $\bm P_1,\ldots,\bm P_T$ are unknown to the detector and can be highly heterogeneous across positions.  \citet{li2024statistical} develop a unified hypothesis-testing framework for the Gumbel-max and inverse-transform watermarks by constructing pivotal statistics whose null distributions do not depend on these unknown next-token distributions. And \cite{dathathri2024scalable} further develop a watermarking scheme for large-scale-production systems.

For image data, watermarking typically takes advantage of the continuous and high-dimensional nature of the image space to introduce visually imperceptible perturbations. 
For example, \citet{xian2024raw} propose RAW, a plug-and-play framework for watermarking images. 
In contrast to encoder--decoder approaches that encode prespecified binary messages in latent representations, RAW jointly learns a watermark pattern $\bm K$ and a classifier for detecting its presence directly in the image space. 
An image $X$ can therefore be watermarked post hoc through a transformation without modifying or retraining the underlying generative model. 
Related ideas have also been developed for video data. 
\citet{zhang2019rivagan} propose RivaGAN, an attention-based neural watermarking architecture that embeds an arbitrary $D$-bit message
\[
    M\in\{0,1\}^{D}
\]
into a video. 
Unlike a watermark that encodes only a common ownership signal, the arbitrary message $M$ can also be varied across distributed copies, a feature that naturally connects video watermarking to fingerprinting.

Watermarking has more recently been extended to tabular data, where the problem is complicated by heterogeneous variable types, potentially unknown distributions, and the need to preserve statistical properties required by downstream analyses. 
Existing approaches have primarily focused on embedding a detectable signal at the dataset level. For example, \citet{he2024watermarking} propose a simple statistical watermarking scheme based on carefully partitioned data intervals and a secret green list, with detection formulated through statistical hypothesis testing. The method works effectively for continuous data. 
More recently, \citet{zhao2025tabdrw} propose TAB-DRW, a post-editing watermarking framework that embeds signals in the frequency domain. 
Both approaches, however, are principally designed for \emph{dataset-level} watermark detection: evidence for the watermark is accumulated across multiple observations in a released table. \citet{cui2026stamp} take a different perspective and propose STAMP (\emph{Single-observation Tabular Attribution and Marking Procedure}), which introduces watermarking at the level of individual observations. 

\bigskip
\noindent\textbf{Data Fingerprinting.}
A conventional watermark primarily answers the binary question of whether a protected signal is present. 
Fingerprinting goes one step further by assigning distinguishable marks to different data recipients.

Formally, suppose the same underlying data object $X$ is distributed to $m$ recipients. 
Instead of applying a common watermark key $K$, fingerprinting assigns recipient-specific keys
\[
    K_1,\ldots,K_m,
\]
and constructs
\[
    \widetilde X^{(j)}
    =
    \mathcal{W}(X;K_j),
    \qquad
    j=1,\ldots,m.
\]
Given a subsequently observed copy $Z$, the goal is no longer simply to test whether a watermark exists, but to identify its origin:
\[
    \widehat j
    =
    \arg\max_{j\in[m]}
    S(Z,K_j),
\]
where $S(Z,K_j)$ measures the compatibility of $Z$ with the fingerprint assigned to recipient $j$. 
The defining feature of fingerprinting is therefore not simply the presence of a hidden signal, but the heterogeneity of that signal across recipients.

Closely related to fingerprinting is \emph{attribution}, which refers to the inferential task of identifying the specific source from which an observed data object originates. In other words, fingerprinting concerns the construction of recipient-specific signals, whereas attribution concerns the recovery of source identities.
Suppose there are $m$ candidate sources, indexed by $i\in[m]$, and an observed data object $Z$ originates from one of them. 
The goal of attribution is to recover its source identity,
\[
    \widehat{i}
    =
    \arg\max_{i\in[m]} S(Z,i),
\]
where $S(Z,i)$ measures the evidence that $Z$ originates from source $i$. 



\bibliographystyle{agsm}
\bibliography{references}

@article{laird1982random,
  author  = {Laird, Nan M. and Ware, James H.},
  title   = {Random-Effects Models for Longitudinal Data},
  journal = {Biometrics},
  year    = {1982},
  volume  = {38},
  number  = {4},
  pages   = {963--974}
}

@book{mclachlan2000finite,
  author    = {McLachlan, Geoffrey J. and Peel, David},
  title     = {Finite Mixture Models},
  publisher = {John Wiley \& Sons},
  address   = {New York},
  year      = {2000}
}

@article{lipkovich2011subgroup,
  author  = {Lipkovich, Ilya and Dmitrienko, Alex and Denne, Jonathan and Enas, Gregory},
  title   = {Subgroup Identification Based on Differential Effect Search ({SIDES})---{A} Recursive Partitioning Method for Establishing Response to Treatment in Patient Subpopulations},
  journal = {Statistics in Medicine},
  year    = {2011},
  volume  = {30},
  number  = {21},
  pages   = {2601--2621}
}

@article{higgins2002quantifying,
  author  = {Higgins, Julian P. T. and Thompson, Simon G.},
  title   = {Quantifying Heterogeneity in a Meta-Analysis},
  journal = {Statistics in Medicine},
  year    = {2002},
  volume  = {21},
  number  = {11},
  pages   = {1539--1558}
}

@article{rubin1974estimating,
  author  = {Rubin, Donald B.},
  title   = {Estimating Causal Effects of Treatments in Randomized and Nonrandomized Studies},
  journal = {Journal of Educational Psychology},
  year    = {1974},
  volume  = {66},
  number  = {5},
  pages   = {688--701}
}

@article{rosenbaum1983central,
  author  = {Rosenbaum, Paul R. and Rubin, Donald B.},
  title   = {The Central Role of the Propensity Score in Observational Studies for Causal Effects},
  journal = {Biometrika},
  year    = {1983},
  volume  = {70},
  number  = {1},
  pages   = {41--55}
}

@book{imbens2015causal,
  author    = {Imbens, Guido W. and Rubin, Donald B.},
  title     = {Causal Inference for Statistics, Social, and Biomedical Sciences: An Introduction},
  publisher = {Cambridge University Press},
  address   = {New York},
  year      = {2015}
}

@article{hill2011bayesian,
  author  = {Hill, Jennifer L.},
  title   = {Bayesian Nonparametric Modeling for Causal Inference},
  journal = {Journal of Computational and Graphical Statistics},
  year    = {2011},
  volume  = {20},
  number  = {1},
  pages   = {217--240}
}

@article{athey2016recursive,
  author  = {Athey, Susan and Imbens, Guido},
  title   = {Recursive Partitioning for Heterogeneous Causal Effects},
  journal = {Proceedings of the National Academy of Sciences},
  year    = {2016},
  volume  = {113},
  number  = {27},
  pages   = {7353--7360}
}

@article{wager2018estimation,
  author  = {Wager, Stefan and Athey, Susan},
  title   = {Estimation and Inference of Heterogeneous Treatment Effects using Random Forests},
  journal = {Journal of the American Statistical Association},
  year    = {2018},
  volume  = {113},
  number  = {523},
  pages   = {1228--1242}
}

@article{athey2019generalized,
  author  = {Athey, Susan and Tibshirani, Julie and Wager, Stefan},
  title   = {Generalized Random Forests},
  journal = {The Annals of Statistics},
  year    = {2019},
  volume  = {47},
  number  = {2},
  pages   = {1148--1178}
}

@article{kunzel2019metalearners,
  author  = {K{\"u}nzel, S{\"o}ren R. and Sekhon, Jasjeet S. and Bickel, Peter J. and Yu, Bin},
  title   = {Metalearners for Estimating Heterogeneous Treatment Effects using Machine Learning},
  journal = {Proceedings of the National Academy of Sciences},
  year    = {2019},
  volume  = {116},
  number  = {10},
  pages   = {4156--4165}
}

@article{nie2021quasi,
  author  = {Nie, Xinkun and Wager, Stefan},
  title   = {Quasi-Oracle Estimation of Heterogeneous Treatment Effects},
  journal = {Biometrika},
  year    = {2021},
  volume  = {108},
  number  = {2},
  pages   = {299--319}
}

@article{kennedy2023towards,
  author  = {Kennedy, Edward H.},
  title   = {Towards Optimal Doubly Robust Estimation of Heterogeneous Causal Effects},
  journal = {Electronic Journal of Statistics},
  year    = {2023},
  volume  = {17},
  number  = {2},
  pages   = {3008--3049},
  doi     = {10.1214/23-EJS2157}
}

@article{chernozhukov2018double,
  author  = {Chernozhukov, Victor and Chetverikov, Denis and Demirer, Mert and Duflo, Esther and Hansen, Christian and Newey, Whitney and Robins, James},
  title   = {Double/Debiased Machine Learning for Treatment and Structural Parameters},
  journal = {The Econometrics Journal},
  year    = {2018},
  volume  = {21},
  number  = {1},
  pages   = {C1--C68}
}

@article{semenova2021debiased,
  author  = {Semenova, Vira and Chernozhukov, Victor},
  title   = {Debiased Machine Learning of Conditional Average Treatment Effects and Other Causal Functions},
  journal = {The Econometrics Journal},
  year    = {2021},
  volume  = {24},
  number  = {2},
  pages   = {264--289}
}

@article{yusuf1991analysis,
  author  = {Yusuf, Salim and Wittes, Janet and Probstfield, Jeffrey and Tyroler, Herman A.},
  title   = {Analysis and Interpretation of Treatment Effects in Subgroups of Patients in Randomized Clinical Trials},
  journal = {JAMA},
  year    = {1991},
  volume  = {266},
  number  = {1},
  pages   = {93--98}
}

@article{gail1985testing,
  author  = {Gail, Mitchell and Simon, Richard},
  title   = {Testing for Qualitative Interactions between Treatment Effects and Patient Subsets},
  journal = {Biometrics},
  year    = {1985},
  volume  = {41},
  number  = {2},
  pages   = {361--372}
}

@article{assmann2000subgroup,
  author  = {Assmann, Susan F. and Pocock, Stuart J. and Enos, Laura E. and Kasten, Linda E.},
  title   = {Subgroup Analysis and Other (Mis)uses of Baseline Data in Clinical Trials},
  journal = {The Lancet},
  year    = {2000},
  volume  = {355},
  number  = {9209},
  pages   = {1064--1069}
}

@article{rothwell2005treating,
  author  = {Rothwell, Peter M.},
  title   = {Treating Individuals 2. Subgroup Analysis in Randomised Controlled Trials: Importance, Indications, and Interpretation},
  journal = {The Lancet},
  year    = {2005},
  volume  = {365},
  number  = {9454},
  pages   = {176--186}
}

@article{wang2007statistics,
  author  = {Wang, Rui and Lagakos, Stephen W. and Ware, James H. and Hunter, David J. and Drazen, Jeffrey M.},
  title   = {Statistics in Medicine---{R}eporting of Subgroup Analyses in Clinical Trials},
  journal = {New England Journal of Medicine},
  year    = {2007},
  volume  = {357},
  number  = {21},
  pages   = {2189--2194}
}

@techreport{ich2019estimands,
  author      = {{International Council for Harmonisation}},
  title       = {{E9(R1)} Statistical Principles for Clinical Trials: Addendum: Estimands and Sensitivity Analysis in Clinical Trials},
  institution = {ICH Harmonised Guideline},
  year        = {2019}
}

@article{simon2013adaptive,
  author  = {Simon, Noah and Simon, Richard},
  title   = {Adaptive Enrichment Designs for Clinical Trials},
  journal = {Biostatistics},
  year    = {2013},
  volume  = {14},
  number  = {4},
  pages   = {613--625}
}

@article{ding2016randomization,
  author  = {Ding, Peng and Feller, Avi and Miratrix, Luke},
  title   = {Randomization Inference for Treatment Effect Variation},
  journal = {Journal of the Royal Statistical Society: Series B},
  year    = {2016},
  volume  = {78},
  number  = {3},
  pages   = {655--671}
}

@article{ding2019decomposing,
  author  = {Ding, Peng and Feller, Avi and Miratrix, Luke},
  title   = {Decomposing Treatment Effect Variation},
  journal = {Journal of the American Statistical Association},
  year    = {2019},
  volume  = {114},
  number  = {525},
  pages   = {304--317}
}

@article{chernozhukov2025generic,
  author  = {Chernozhukov, Victor and Demirer, Mert and Duflo, Esther and Fern{\'a}ndez-Val, Iv{\'a}n},
  title   = {Fisher--{S}chultz Lecture: Generic Machine Learning Inference on Heterogeneous Treatment Effects in Randomized Experiments, With an Application to Immunization in {India}},
  journal = {Econometrica},
  year    = {2025},
  volume  = {93},
  number  = {4},
  pages   = {1121--1164},
  doi     = {10.3982/ECTA19303}
}

@article{abadie2006large,
  author  = {Abadie, Alberto and Imbens, Guido W.},
  title   = {Large Sample Properties of Matching Estimators for Average Treatment Effects},
  journal = {Econometrica},
  year    = {2006},
  volume  = {74},
  number  = {1},
  pages   = {235--267}
}

@article{imai2014covariate,
  author  = {Imai, Kosuke and Ratkovic, Marc},
  title   = {Covariate Balancing Propensity Score},
  journal = {Journal of the Royal Statistical Society: Series B},
  year    = {2014},
  volume  = {76},
  number  = {1},
  pages   = {243--263}
}

@article{imbens1994identification,
  author  = {Imbens, Guido W. and Angrist, Joshua D.},
  title   = {Identification and Estimation of Local Average Treatment Effects},
  journal = {Econometrica},
  year    = {1994},
  volume  = {62},
  number  = {2},
  pages   = {467--475}
}

@article{angrist1996identification,
  author  = {Angrist, Joshua D. and Imbens, Guido W. and Rubin, Donald B.},
  title   = {Identification of Causal Effects Using Instrumental Variables},
  journal = {Journal of the American Statistical Association},
  year    = {1996},
  volume  = {91},
  number  = {434},
  pages   = {444--455}
}

@article{sanderson2022mendelian,
  author  = {Sanderson, Eleanor and Glymour, M. Maria and Holmes, Michael V. and Kang, Hyunseung and Morrison, Jean and Munaf{\`o}, Marcus R. and Palmer, Tom and Schooling, C. Mary and Wallace, Chris and Zhao, Qingyuan and Davey Smith, George},
  title   = {Mendelian Randomization},
  journal = {Nature Reviews Methods Primers},
  year    = {2022},
  volume  = {2},
  pages   = {6}
}

@article{baron1986moderator,
  author  = {Baron, Reuben M. and Kenny, David A.},
  title   = {The Moderator--Mediator Variable Distinction in Social Psychological Research: Conceptual, Strategic, and Statistical Considerations},
  journal = {Journal of Personality and Social Psychology},
  year    = {1986},
  volume  = {51},
  number  = {6},
  pages   = {1173--1182}
}

@article{imai2010general,
  author  = {Imai, Kosuke and Keele, Luke and Tingley, Dustin},
  title   = {A General Approach to Causal Mediation Analysis},
  journal = {Psychological Methods},
  year    = {2010},
  volume  = {15},
  number  = {4},
  pages   = {309--334}
}

@book{vanderweele2015explanation,
  author    = {VanderWeele, Tyler J.},
  title     = {Explanation in Causal Inference: Methods for Mediation and Interaction},
  publisher = {Oxford University Press},
  address   = {New York},
  year      = {2015}
}

@article{preacher2007addressing,
  author  = {Preacher, Kristopher J. and Rucker, Derek D. and Hayes, Andrew F.},
  title   = {Addressing Moderated Mediation Hypotheses: Theory, Methods, and Prescriptions},
  journal = {Multivariate Behavioral Research},
  year    = {2007},
  volume  = {42},
  number  = {1},
  pages   = {185--227}
}

@inproceedings{johansson2016learning,
  author    = {Johansson, Fredrik D. and Shalit, Uri and Sontag, David},
  title     = {Learning Representations for Counterfactual Inference},
  booktitle = {Proceedings of the 33rd International Conference on Machine Learning},
  year      = {2016},
  series    = {PMLR},
  volume    = {48},
  pages     = {3020--3029}
}

@inproceedings{shalit2017estimating,
  author    = {Shalit, Uri and Johansson, Fredrik D. and Sontag, David},
  title     = {Estimating Individual Treatment Effect: Generalization Bounds and Algorithms},
  booktitle = {Proceedings of the 34th International Conference on Machine Learning},
  year      = {2017},
  series    = {PMLR},
  volume    = {70},
  pages     = {3076--3085}
}

@inproceedings{shi2019adapting,
  author    = {Shi, Claudia and Blei, David M. and Veitch, Victor},
  title     = {Adapting Neural Networks for the Estimation of Treatment Effects},
  booktitle = {Advances in Neural Information Processing Systems 32},
  year      = {2019}
}

@inproceedings{curth2021nonparametric,
  author    = {Curth, Alicia and van der Schaar, Mihaela},
  title     = {Nonparametric Estimation of Heterogeneous Treatment Effects: From Theory to Learning Algorithms},
  booktitle = {Proceedings of the 24th International Conference on Artificial Intelligence and Statistics},
  year      = {2021},
  series    = {PMLR},
  volume    = {130},
  pages     = {1810--1818}
}

@inproceedings{curth2021inductive,
  author    = {Curth, Alicia and van der Schaar, Mihaela},
  title     = {On Inductive Biases for Heterogeneous Treatment Effect Estimation},
  booktitle = {Advances in Neural Information Processing Systems 34},
  year      = {2021}
}

@inproceedings{louizos2017causal,
  author    = {Louizos, Christos and Shalit, Uri and Mooij, Joris and Sontag, David and Zemel, Richard and Welling, Max},
  title     = {Causal Effect Inference with Deep Latent-Variable Models},
  booktitle = {Advances in Neural Information Processing Systems 30},
  year      = {2017}
}

@inproceedings{yoon2018ganite,
  author    = {Yoon, Jinsung and Jordon, James and van der Schaar, Mihaela},
  title     = {{GANITE}: Estimation of Individualized Treatment Effects using Generative Adversarial Nets},
  booktitle = {International Conference on Learning Representations},
  year      = {2018}
}

@inproceedings{kusner2017counterfactual,
  author    = {Kusner, Matt J. and Loftus, Joshua and Russell, Chris and Silva, Ricardo},
  title     = {Counterfactual Fairness},
  booktitle = {Advances in Neural Information Processing Systems 30},
  year      = {2017},
  pages     = {4066--4076}
}

@inproceedings{kilbertus2017avoiding,
  author    = {Kilbertus, Niki and Rojas-Carulla, Mateo and Parascandolo, Giambattista and Hardt, Moritz and Janzing, Dominik and Sch{\"o}lkopf, Bernhard},
  title     = {Avoiding Discrimination through Causal Reasoning},
  booktitle = {Advances in Neural Information Processing Systems 30},
  year      = {2017},
  pages     = {656--666}
}

@inproceedings{nabi2018fair,
  author    = {Nabi, Razieh and Shpitser, Ilya},
  title     = {Fair Inference on Outcomes},
  booktitle = {Proceedings of the 32nd AAAI Conference on Artificial Intelligence},
  year      = {2018},
  pages     = {1931--1940}
}

@inproceedings{zhang2018fairness,
  author    = {Zhang, Junzhe and Bareinboim, Elias},
  title     = {Fairness in Decision-Making---{T}he Causal Explanation Formula},
  booktitle = {Proceedings of the 32nd AAAI Conference on Artificial Intelligence},
  year      = {2018}
}

@article{plecko2024causal,
  author  = {Ple{\v c}ko, Drago and Bareinboim, Elias},
  title   = {Causal Fairness Analysis: A Causal Toolkit for Fair Machine Learning},
  journal = {Foundations and Trends in Machine Learning},
  year    = {2024},
  volume  = {17},
  number  = {3},
  pages   = {304--589},
  doi     = {10.1561/2200000106}
}

@inproceedings{kearns2018preventing,
  author    = {Kearns, Michael and Neel, Seth and Roth, Aaron and Wu, Zhiwei Steven},
  title     = {Preventing Fairness Gerrymandering: Auditing and Learning for Subgroup Fairness},
  booktitle = {Proceedings of the 35th International Conference on Machine Learning},
  year      = {2018},
  series    = {PMLR},
  volume    = {80},
  pages     = {2564--2572}
}

@inproceedings{hebertjohnson2018multicalibration,
  author    = {H{\'e}bert-Johnson, {\'U}rsula and Kim, Michael P. and Reingold, Omer and Rothblum, Guy N.},
  title     = {Multicalibration: Calibration for the (Computationally-Identifiable) Masses},
  booktitle = {Proceedings of the 35th International Conference on Machine Learning},
  year      = {2018},
  series    = {PMLR},
  volume    = {80},
  pages     = {1939--1948}
}

@article{athey2021policy,
  author  = {Athey, Susan and Wager, Stefan},
  title   = {Policy Learning With Observational Data},
  journal = {Econometrica},
  year    = {2021},
  volume  = {89},
  number  = {1},
  pages   = {133--161}
}

@article{lei2021conformal,
  author  = {Lei, Lihua and Cand{\`e}s, Emmanuel J.},
  title   = {Conformal Inference of Counterfactuals and Individual Treatment Effects},
  journal = {Journal of the Royal Statistical Society: Series B},
  year    = {2021},
  volume  = {83},
  number  = {5},
  pages   = {911--938}
}

@article{jin2023sensitivity,
  author  = {Jin, Ying and Ren, Zhimei and Cand{\`e}s, Emmanuel J.},
  title   = {Sensitivity Analysis of Individual Treatment Effects: A Robust Conformal Inference Approach},
  journal = {Proceedings of the National Academy of Sciences},
  year    = {2023},
  volume  = {120},
  number  = {6},
  pages   = {e2214889120},
  doi     = {10.1073/pnas.2214889120}
}

@article{yadlowsky2022bounds,
  author  = {Yadlowsky, Steve and Namkoong, Hongseok and Basu, Sanjay and Duchi, John and Tian, Lu},
  title   = {Bounds on the Conditional and Average Treatment Effect with Unobserved Confounding Factors},
  journal = {The Annals of Statistics},
  year    = {2022},
  volume  = {50},
  number  = {5},
  pages   = {2587--2615}
}

@inproceedings{xu2022calibration,
  author    = {Xu, Yizhe and Yadlowsky, Steve},
  title     = {Calibration Error for Heterogeneous Treatment Effects},
  booktitle = {Proceedings of the 25th International Conference on Artificial Intelligence and Statistics},
  year      = {2022},
  series    = {PMLR},
  volume    = {151}
}

@inproceedings{vanderlaan2023causal,
  author    = {van der Laan, Lars and Ulloa-P{\'e}rez, Ernesto and Carone, Marco and Luedtke, Alex},
  title     = {Causal Isotonic Calibration for Heterogeneous Treatment Effects},
  booktitle = {Proceedings of the 40th International Conference on Machine Learning},
  year      = {2023},
  series    = {PMLR},
  volume    = {202}
}

@article{dahabreh2020extending,
  author  = {Dahabreh, Issa J. and Robertson, Sarah E. and Steingrimsson, Jon A. and Stuart, Elizabeth A. and Hern{\'a}n, Miguel A.},
  title   = {Extending Inferences from a Randomized Trial to a New Target Population},
  journal = {Statistics in Medicine},
  year    = {2020},
  volume  = {39},
  number  = {14},
  pages   = {1999--2014}
}

@article{degtiar2023review,
  author  = {Degtiar, Irina and Rose, Sherri},
  title   = {A Review of Generalizability and Transportability},
  journal = {Annual Review of Statistics and Its Application},
  year    = {2023},
  volume  = {10},
  pages   = {501--524},
  doi     = {10.1146/annurev-statistics-042522-103837}
}

@inproceedings{dhawan2024end,
  author    = {Dhawan, Nikita and Cotta, Leonardo and Ullrich, Karsten and Krishnan, Rahul G. and Maddison, Chris J.},
  title     = {End-To-End Causal Effect Estimation from Unstructured Natural Language Data},
  booktitle = {Advances in Neural Information Processing Systems 37},
  year      = {2024}
}

@misc{ma2025foundation,
  author = {Ma, Yuchen and Frauen, Dennis and Javurek, Emil and Feuerriegel, Stefan},
  title  = {Foundation Models for Causal Inference via Prior-Data Fitted Networks},
  year   = {2025},
  note   = {arXiv preprint arXiv:2506.10914}
}

@article{rubin1993statistical,
  author  = {Rubin, Donald B.},
  title   = {Discussion: Statistical Disclosure Limitation},
  journal = {Journal of Official Statistics},
  year    = {1993},
  volume  = {9},
  number  = {2},
  pages   = {461--468}
}

@article{raghunathan2003multiple,
  author  = {Raghunathan, Trivellore E. and Reiter, Jerome P. and Rubin, Donald B.},
  title   = {Multiple Imputation for Statistical Disclosure Limitation},
  journal = {Journal of Official Statistics},
  year    = {2003},
  volume  = {19},
  number  = {1},
  pages   = {1--16}
}

@article{reiter2003inference,
  author  = {Reiter, Jerome P.},
  title   = {Inference for Partially Synthetic, Public Use Microdata Sets},
  journal = {Survey Methodology},
  year    = {2003},
  volume  = {29},
  number  = {2},
  pages   = {181--188}
}

@article{reiter2005using,
  author  = {Reiter, Jerome P.},
  title   = {Using {CART} to Generate Partially Synthetic Public Use Microdata},
  journal = {Journal of Official Statistics},
  year    = {2005},
  volume  = {21},
  number  = {3},
  pages   = {441--462}
}

@article{nowok2016synthpop,
  author  = {Nowok, Beata and Raab, Gillian M. and Dibben, Chris},
  title   = {synthpop: Bespoke Creation of Synthetic Data in {R}},
  journal = {Journal of Statistical Software},
  year    = {2016},
  volume  = {74},
  number  = {11},
  pages   = {1--26}
}

@article{raab2016practical,
  author  = {Raab, Gillian M. and Nowok, Beata and Dibben, Chris},
  title   = {Practical Data Synthesis for Large Samples},
  journal = {Journal of Privacy and Confidentiality},
  year    = {2016},
  volume  = {7},
  number  = {3},
  pages   = {67--97}
}

@inproceedings{patki2016synthetic,
  author    = {Patki, Neha and Wedge, Roy and Veeramachaneni, Kalyan},
  title     = {The Synthetic Data Vault},
  booktitle = {IEEE International Conference on Data Science and Advanced Analytics (DSAA)},
  year      = {2016},
  pages     = {399--410}
}

@misc{kamthe2021copula,
  author = {Kamthe, Sanket and Assefa, Samuel and Deisenroth, Marc},
  title  = {Copula Flows for Synthetic Data Generation},
  year   = {2021},
  note   = {arXiv preprint arXiv:2101.00598}
}

@inproceedings{kingma2014auto,
  author    = {Kingma, Diederik P. and Welling, Max},
  title     = {Auto-Encoding Variational {Bayes}},
  booktitle = {International Conference on Learning Representations},
  year      = {2014}
}

@inproceedings{sohn2015learning,
  author    = {Sohn, Kihyuk and Lee, Honglak and Yan, Xinchen},
  title     = {Learning Structured Output Representation using Deep Conditional Generative Models},
  booktitle = {Advances in Neural Information Processing Systems 28},
  year      = {2015},
  pages     = {3483--3491}
}

@misc{dilokthanakul2016deep,
  author = {Dilokthanakul, Nat and Mediano, Pedro A. M. and Garnelo, Marta and Lee, Matthew C. H. and Salimbeni, Hugh and Arulkumaran, Kai and Shanahan, Murray},
  title  = {Deep Unsupervised Clustering with {Gaussian} Mixture Variational Autoencoders},
  year   = {2016},
  note   = {arXiv preprint arXiv:1611.02648}
}

@inproceedings{jiang2017variational,
  author    = {Jiang, Zhuxi and Zheng, Yin and Tan, Huachun and Tang, Bangsheng and Zhou, Hanning},
  title     = {Variational Deep Embedding: An Unsupervised and Generative Approach to Clustering},
  booktitle = {Proceedings of the 26th International Joint Conference on Artificial Intelligence},
  year      = {2017},
  pages     = {1965--1972}
}

@article{nazabal2020handling,
  author  = {Naz{\'a}bal, Alfredo and Olmos, Pablo M. and Ghahramani, Zoubin and Valera, Isabel},
  title   = {Handling Incomplete Heterogeneous Data using {VAEs}},
  journal = {Pattern Recognition},
  year    = {2020},
  volume  = {107},
  pages   = {107501}
}

@inproceedings{xu2019modeling,
  author    = {Xu, Lei and Skoularidou, Maria and Cuesta-Infante, Alfredo and Veeramachaneni, Kalyan},
  title     = {Modeling Tabular Data using Conditional {GAN}},
  booktitle = {Advances in Neural Information Processing Systems 32},
  year      = {2019},
  pages     = {7335--7345}
}

@inproceedings{goodfellow2014generative,
  author    = {Goodfellow, Ian and Pouget-Abadie, Jean and Mirza, Mehdi and Xu, Bing and Warde-Farley, David and Ozair, Sherjil and Courville, Aaron and Bengio, Yoshua},
  title     = {Generative Adversarial Nets},
  booktitle = {Advances in Neural Information Processing Systems 27},
  year      = {2014},
  pages     = {2672--2680}
}

@misc{mirza2014conditional,
  author = {Mirza, Mehdi and Osindero, Simon},
  title  = {Conditional Generative Adversarial Nets},
  year   = {2014},
  note   = {arXiv preprint arXiv:1411.1784}
}

@inproceedings{odena2017conditional,
  author    = {Odena, Augustus and Olah, Christopher and Shlens, Jonathon},
  title     = {Conditional Image Synthesis with Auxiliary Classifier {GANs}},
  booktitle = {Proceedings of the 34th International Conference on Machine Learning},
  year      = {2017},
  series    = {PMLR},
  volume    = {70},
  pages     = {2642--2651}
}

@inproceedings{metz2017unrolled,
  author    = {Metz, Luke and Poole, Ben and Pfau, David and Sohl-Dickstein, Jascha},
  title     = {Unrolled Generative Adversarial Networks},
  booktitle = {International Conference on Learning Representations},
  year      = {2017}
}

@inproceedings{srivastava2017veegan,
  author    = {Srivastava, Akash and Valkov, Lazar and Russell, Chris and Gutmann, Michael U. and Sutton, Charles},
  title     = {{VEEGAN}: Reducing Mode Collapse in {GANs} using Implicit Variational Learning},
  booktitle = {Advances in Neural Information Processing Systems 30},
  year      = {2017},
  pages     = {3308--3318}
}

@inproceedings{lin2018pacgan,
  author    = {Lin, Zinan and Khetan, Ashish and Fanti, Giulia and Oh, Sewoong},
  title     = {{PacGAN}: The Power of Two Samples in Generative Adversarial Networks},
  booktitle = {Advances in Neural Information Processing Systems 31},
  year      = {2018},
  pages     = {1505--1514}
}

@inproceedings{choi2017generating,
  author    = {Choi, Edward and Biswal, Siddharth and Malin, Bradley and Duke, Jon and Stewart, Walter F. and Sun, Jimeng},
  title     = {Generating Multi-label Discrete Patient Records using Generative Adversarial Networks},
  booktitle = {Proceedings of the 2nd Machine Learning for Healthcare Conference},
  year      = {2017},
  series    = {PMLR},
  volume    = {68},
  pages     = {286--305}
}

@inproceedings{jordon2019pategan,
  author    = {Jordon, James and Yoon, Jinsung and van der Schaar, Mihaela},
  title     = {{PATE-GAN}: Generating Synthetic Data with Differential Privacy Guarantees},
  booktitle = {International Conference on Learning Representations},
  year      = {2019}
}

@misc{xie2018differentially,
  author = {Xie, Liyang and Lin, Kaixiang and Wang, Shu and Wang, Fei and Zhou, Jiayu},
  title  = {Differentially Private Generative Adversarial Network},
  year   = {2018},
  note   = {arXiv preprint arXiv:1802.06739}
}

@inproceedings{zhao2021ctabgan,
  author    = {Zhao, Zilong and Kunar, Aditya and Birke, Robert and Chen, Lydia Y.},
  title     = {{CTAB-GAN}: Effective Table Data Synthesizing},
  booktitle = {Proceedings of the 13th Asian Conference on Machine Learning},
  year      = {2021},
  series    = {PMLR},
  volume    = {157},
  pages     = {97--112}
}

@article{zhao2024ctabganplus,
  author  = {Zhao, Zilong and Kunar, Aditya and Birke, Robert and Van der Scheer, Hiek and Chen, Lydia Y.},
  title   = {{CTAB-GAN+}: Enhancing Tabular Data Synthesis},
  journal = {Frontiers in Big Data},
  year    = {2024},
  volume  = {6},
  pages   = {1296508},
  doi     = {10.3389/fdata.2023.1296508}
}

@inproceedings{yoon2019time,
  author    = {Yoon, Jinsung and Jarrett, Daniel and van der Schaar, Mihaela},
  title     = {Time-series Generative Adversarial Networks},
  booktitle = {Advances in Neural Information Processing Systems 32},
  year      = {2019},
  pages     = {5508--5518}
}

@inproceedings{ho2020denoising,
  author    = {Ho, Jonathan and Jain, Ajay and Abbeel, Pieter},
  title     = {Denoising Diffusion Probabilistic Models},
  booktitle = {Advances in Neural Information Processing Systems 33},
  year      = {2020},
  pages     = {6840--6851}
}

@inproceedings{song2021score,
  author    = {Song, Yang and Sohl-Dickstein, Jascha and Kingma, Diederik P. and Kumar, Abhishek and Ermon, Stefano and Poole, Ben},
  title     = {Score-Based Generative Modeling through Stochastic Differential Equations},
  booktitle = {International Conference on Learning Representations},
  year      = {2021}
}

@inproceedings{rombach2022high,
  author    = {Rombach, Robin and Blattmann, Andreas and Lorenz, Dominik and Esser, Patrick and Ommer, Bj{\"o}rn},
  title     = {High-Resolution Image Synthesis with Latent Diffusion Models},
  booktitle = {Proceedings of the IEEE/CVF Conference on Computer Vision and Pattern Recognition},
  year      = {2022},
  pages     = {10684--10695}
}

@inproceedings{dhariwal2021diffusion,
  author    = {Dhariwal, Prafulla and Nichol, Alexander},
  title     = {Diffusion Models Beat {GANs} on Image Synthesis},
  booktitle = {Advances in Neural Information Processing Systems 34},
  year      = {2021},
  pages     = {8780--8794}
}

@misc{ho2022classifier,
  author = {Ho, Jonathan and Salimans, Tim},
  title  = {Classifier-Free Diffusion Guidance},
  year   = {2022},
  note   = {arXiv preprint arXiv:2207.12598}
}

@inproceedings{kotelnikov2023tabddpm,
  author    = {Kotelnikov, Akim and Baranchuk, Dmitry and Rubachev, Ivan and Babenko, Artem},
  title     = {{TabDDPM}: Modelling Tabular Data with Diffusion Models},
  booktitle = {Proceedings of the 40th International Conference on Machine Learning},
  year      = {2023},
  series    = {PMLR},
  volume    = {202},
  pages     = {17564--17579}
}

@inproceedings{kim2023stasy,
  author    = {Kim, Jayoung and Lee, Chaejeong and Park, Noseong},
  title     = {{STaSy}: Score-based Tabular Data Synthesis},
  booktitle = {International Conference on Learning Representations},
  year      = {2023}
}

@inproceedings{lee2023codi,
  author    = {Lee, Chaejeong and Kim, Jayoung and Park, Noseong},
  title     = {{CoDi}: Co-evolving Contrastive Diffusion Models for Mixed-type Tabular Synthesis},
  booktitle = {Proceedings of the 40th International Conference on Machine Learning},
  year      = {2023},
  series    = {PMLR},
  volume    = {202},
  pages     = {18940--18956}
}

@inproceedings{zhang2024mixed,
  author    = {Zhang, Hengrui and Zhang, Jiani and Srinivasan, Balasubramaniam and Shen, Zhengyuan and Qin, Xiao and Faloutsos, Christos and Rangwala, Huzefa and Karypis, George},
  title     = {Mixed-Type Tabular Data Synthesis with Score-based Diffusion in Latent Space},
  booktitle = {International Conference on Learning Representations},
  year      = {2024}
}

@misc{yuan2023ehrdiff,
  author = {Yuan, Hongyi and Zhou, Songchi and Yu, Sheng},
  title  = {{EHRDiff}: Exploring Realistic {EHR} Synthesis with Diffusion Models},
  year   = {2023},
  note   = {arXiv preprint arXiv:2303.05656; published in Transactions on Machine Learning Research}
}

@misc{he2023meddiff,
  author = {He, Huan and Zhao, Shifan and Xi, Yuanzhe and Ho, Joyce C.},
  title  = {{MedDiff}: Generating Electronic Health Records using Accelerated Denoising Diffusion Model},
  year   = {2023},
  note   = {arXiv preprint arXiv:2302.04355}
}

@inproceedings{tashiro2021csdi,
  author    = {Tashiro, Yusuke and Song, Jiaming and Song, Yang and Ermon, Stefano},
  title     = {{CSDI}: Conditional Score-based Diffusion Models for Probabilistic Time Series Imputation},
  booktitle = {Advances in Neural Information Processing Systems 34},
  year      = {2021},
  pages     = {24804--24816}
}

@inproceedings{choi2020fair,
  author    = {Choi, Kristy and Grover, Aditya and Singh, Trisha and Shu, Rui and Ermon, Stefano},
  title     = {Fair Generative Modeling via Weak Supervision},
  booktitle = {Proceedings of the 37th International Conference on Machine Learning},
  year      = {2020},
  series    = {PMLR},
  volume    = {119},
  pages     = {1887--1898}
}

@inproceedings{yu2020inclusive,
  author    = {Yu, Ning and Li, Ke and Zhou, Peng and Malik, Jitendra and Davis, Larry and Fritz, Mario},
  title     = {Inclusive {GAN}: Improving Data and Minority Coverage in Generative Models},
  booktitle = {European Conference on Computer Vision},
  year      = {2020},
  pages     = {377--393}
}

@inproceedings{sajjadi2018assessing,
  author    = {Sajjadi, Mehdi S. M. and Bachem, Olivier and Lucic, Mario and Bousquet, Olivier and Gelly, Sylvain},
  title     = {Assessing Generative Models via Precision and Recall},
  booktitle = {Advances in Neural Information Processing Systems 31},
  year      = {2018},
  pages     = {5228--5237}
}

@inproceedings{alaa2022faithful,
  author    = {Alaa, Ahmed and Van Breugel, Boris and Saveliev, Evgeny S. and van der Schaar, Mihaela},
  title     = {How Faithful is your Synthetic Data? {S}ample-level Metrics for Evaluating and Auditing Generative Models},
  booktitle = {Proceedings of the 39th International Conference on Machine Learning},
  year      = {2022},
  series    = {PMLR},
  volume    = {162},
  pages     = {290--306}
}

@inproceedings{vanbreugel2023membership,
  author    = {Van Breugel, Boris and Sun, Hao and Qian, Zhaozhi and van der Schaar, Mihaela},
  title     = {Membership Inference Attacks against Synthetic Data through Overfitting Detection},
  booktitle = {Proceedings of the 26th International Conference on Artificial Intelligence and Statistics},
  year      = {2023},
  series    = {PMLR},
  volume    = {206}
}

@inproceedings{borisov2023language,
  author    = {Borisov, Vadim and Se{\ss}ler, Kathrin and Leemann, Tobias and Pawelczyk, Martin and Kasneci, Gjergji},
  title     = {Language Models are Realistic Tabular Data Generators},
  booktitle = {International Conference on Learning Representations},
  year      = {2023}
}

@inproceedings{decruyenaere2024debiasing,
  author    = {Decruyenaere, Alexander and Dehaene, Heidelinde and Rabaey, Paloma and Polet, Christiaan and Decruyenaere, Johan and Vansteelandt, Stijn and Demeester, Thomas},
  title     = {Debiasing Synthetic Data Generated by Deep Generative Models},
  booktitle = {Advances in Neural Information Processing Systems 37},
  year      = {2024}
}

@inproceedings{decruyenaere2024real,
  author = {Decruyenaere, Alexander and Dehaene, Heidelinde and Rabaey, Paloma and Polet, Christiaan and Decruyenaere, Johan and Vansteelandt, Stijn and Demeester, Thomas},
  title     = {The Real Deal Behind the Artificial Appeal: Inferential Utility of Tabular Synthetic Data},
  booktitle = {Proceedings of the 40th Conference on Uncertainty in Artificial Intelligence},
  year      = {2024},
  series    = {PMLR}
}

@article{liu2025conditional,
  author  = {Liu, Kathleen and Altman, Russ B.},
  title   = {Conditional Generative Models for Synthetic Tabular Data: Applications for Precision Medicine and Diverse Representations},
  journal = {Annual Review of Biomedical Data Science},
  year    = {2025},
  volume  = {8},
  pages   = {21--49},
  doi     = {10.1146/annurev-biodatasci-103123-094844}
}

@article{vanbreugel2024synthetic,
  author  = {Van Breugel, Boris and Liu, Tennison and Oglic, Dino and van der Schaar, Mihaela},
  title   = {Synthetic Data in Biomedicine via Generative Artificial Intelligence},
  journal = {Nature Reviews Bioengineering},
  year    = {2024},
  volume  = {2},
  pages   = {991--1004},
  doi     = {10.1038/s44222-024-00245-7}
}

@misc{jordon2022synthetic,
  author = {Jordon, James and Szpruch, Lukasz and Houssiau, Florimond and Bottarelli, Mirko and Cherubin, Giovanni and Maple, Carsten and Cohen, Samuel N. and Weller, Adrian},
  title  = {Synthetic Data---{W}hat, Why and How?},
  year   = {2022},
  note   = {arXiv preprint arXiv:2205.03257}
}

@inproceedings{salemi2024lamp,
  author    = {Salemi, Alireza and Mysore, Sheshera and Bendersky, Michael and Zamani, Hamed},
  title     = {{LaMP}: When Large Language Models Meet Personalization},
  booktitle = {Proceedings of the 62nd Annual Meeting of the Association for Computational Linguistics},
  year      = {2024},
  pages     = {7370--7392}
}

@inproceedings{salemi2024optimization,
  author    = {Salemi, Alireza and Kallumadi, Surya and Zamani, Hamed},
  title     = {Optimization Methods for Personalizing Large Language Models through Retrieval Augmentation},
  booktitle = {Proceedings of the 47th International ACM SIGIR Conference on Research and Development in Information Retrieval},
  year      = {2024},
  pages     = {752--762}
}

@inproceedings{hu2022lora,
  author    = {Hu, Edward J. and Shen, Yelong and Wallis, Phillip and Allen-Zhu, Zeyuan and Li, Yuanzhi and Wang, Shean and Wang, Lu and Chen, Weizhu},
  title     = {{LoRA}: Low-Rank Adaptation of Large Language Models},
  booktitle = {International Conference on Learning Representations},
  year      = {2022}
}

@inproceedings{tian2024hydralora,
  author    = {Tian, Chunlin and Shi, Zhan and Guo, Zhijiang and Li, Li and Xu, Cheng-Zhong},
  title     = {{HydraLoRA}: An Asymmetric {LoRA} Architecture for Efficient Fine-Tuning},
  booktitle = {Advances in Neural Information Processing Systems 37},
  year      = {2024}
}

@inproceedings{poddar2024personalizing,
  author    = {Poddar, Sriyash and Wan, Yanming and Ivison, Hamish and Gupta, Abhishek and Jaques, Natasha},
  title     = {Personalizing Reinforcement Learning from Human Feedback with Variational Preference Learning},
  booktitle = {Advances in Neural Information Processing Systems 37},
  year      = {2024}
}

@inproceedings{doddapaneni2024userembedding,
  author    = {Doddapaneni, Sumanth and Sayana, Krishna and Jash, Ambarish and Sodhi, Sukhdeep and Kuzmin, Dima},
  title     = {User Embedding Model for Personalized Language Prompting},
  booktitle = {Proceedings of the 1st Workshop on Personalization of Generative AI Systems},
  year      = {2024},
  pages     = {124--131},
  publisher = {Association for Computational Linguistics},
  doi       = {10.18653/v1/2024.personalize-1.12},
}

@misc{hebert2024persoma,
  author = {Hebert, Liam and Sayana, Krishna and Jash, Ambarish and Karatzoglou, Alexandros and Sodhi, Sukhdeep and Doddapaneni, Sumanth and Cai, Yanli and Kuzmin, Dima},
  title  = {{PERSOMA}: {PER}sonalized {SO}ft Pro{M}pt Adapter Architecture for Personalized Language Prompting},
  year   = {2024},
  note   = {arXiv preprint arXiv:2408.00960; presented at GenAIRecP at KDD 2024},
}

@inproceedings{zhang2024guided,
  author    = {Zhang, Jiarui},
  title     = {Guided Profile Generation Improves Personalization with Large Language Models},
  booktitle = {Findings of the Association for Computational Linguistics: EMNLP 2024},
  year      = {2024},
  pages     = {4005--4016},
  publisher = {Association for Computational Linguistics},
  doi       = {10.18653/v1/2024.findings-emnlp.231},
}

@inproceedings{chen2024reconcile,
  author    = {Chen, Justin and Saha, Swarnadeep and Bansal, Mohit},
  title     = {{ReConcile}: Round-Table Conference Improves Reasoning via Consensus among Diverse {LLM}s},
  booktitle = {Proceedings of the 62nd Annual Meeting of the Association for Computational Linguistics},
  year      = {2024},
  pages     = {7066--7085},
  publisher = {Association for Computational Linguistics},
  doi       = {10.18653/v1/2024.acl-long.381},
}

@inproceedings{sun2025personadb,
  author    = {Sun, Chenkai and Yang, Ke and Gangi Reddy, Revanth and Fung, Yi R. and Chan, Hou Pong and Small, Kevin and Zhai, ChengXiang and Ji, Heng},
  title     = {Persona-{DB}: Efficient Large Language Model Personalization for Response Prediction with Collaborative Data Refinement},
  booktitle = {Proceedings of the 31st International Conference on Computational Linguistics},
  year      = {2025},
  pages     = {281--296},
  publisher = {Association for Computational Linguistics},
}

@inproceedings{yuan2025evolving,
  author    = {Yuan, Ruifeng and Sun, Shichao and Li, Yongqi and Wang, Zili and Cao, Ziqiang and Li, Wenjie},
  title     = {Personalized Large Language Model Assistant with Evolving Conditional Memory},
  booktitle = {Proceedings of the 31st International Conference on Computational Linguistics},
  year      = {2025},
  pages     = {3764--3777},
  publisher = {Association for Computational Linguistics},
}

@inproceedings{liu2025personaplug,
  author    = {Liu, Jiongnan and Zhu, Yutao and Wang, Shuting and Wei, Xiaochi and Min, Erxue and Lu, Yu and Wang, Shuaiqiang and Yin, Dawei and Dou, Zhicheng},
  title     = {{LLM}s + Persona-Plug = Personalized {LLM}s},
  booktitle = {Proceedings of the 63rd Annual Meeting of the Association for Computational Linguistics},
  year      = {2025},
  pages     = {9373--9385},
  publisher = {Association for Computational Linguistics},
  doi       = {10.18653/v1/2025.acl-long.461},
}

@inproceedings{qiu2025latent,
  author    = {Qiu, Yilun and Shi, Tianhao and Zhao, Xiaoyan and Zhu, Fengbin and Zhang, Yang and Feng, Fuli},
  title     = {Latent Inter-User Difference Modeling for {LLM} Personalization},
  booktitle = {Proceedings of the 2025 Conference on Empirical Methods in Natural Language Processing},
  year      = {2025},
  pages     = {10599--10617},
  publisher = {Association for Computational Linguistics},
  doi       = {10.18653/v1/2025.emnlp-main.536},
}

@inproceedings{zhang2024plora,
  author    = {Zhang, You and Wang, Jin and Yu, Liang-Chih and Xu, Dan and Zhang, Xuejie},
  title     = {Personalized {LoRA} for Human-Centered Text Understanding},
  booktitle = {Proceedings of the AAAI Conference on Artificial Intelligence},
  year      = {2024},
  volume    = {38},
  pages     = {19588--19596},
  doi       = {10.1609/aaai.v38i17.29931},
}

@inproceedings{gururangan2020dont,
  author    = {Gururangan, Suchin and Marasovi{\'c}, Ana and Swayamdipta, Swabha and Lo, Kyle and Beltagy, Iz and Downey, Doug and Smith, Noah A.},
  title     = {Don{'}t Stop Pretraining: Adapt Language Models to Domains and Tasks},
  booktitle = {Proceedings of the 58th Annual Meeting of the Association for Computational Linguistics},
  year      = {2020},
  pages     = {8342--8360},
  publisher = {Association for Computational Linguistics},
  doi       = {10.18653/v1/2020.acl-main.740},
}

@inproceedings{cheng2024adaptllm,
  author    = {Cheng, Daixuan and Huang, Shaohan and Wei, Furu},
  title     = {Adapting Large Language Models to Domains via Reading Comprehension},
  booktitle = {International Conference on Learning Representations},
  year      = {2024},
}

@article{chung2024scaling,
  author  = {Chung, Hyung Won and Hou, Le and Longpre, Shayne and Zoph, Barret and Tay, Yi and Fedus, William and Li, Yunxuan and Wang, Xuezhi and Dehghani, Mostafa and Brahma, Siddhartha and Webson, Albert and Gu, Shixiang Shane and Dai, Zhuyun and Suzgun, Mirac and Chen, Xinyun and Chowdhery, Aakanksha and Castro-Ros, Alex and Pellat, Marie and Robinson, Kevin and Valter, Dasha and Narang, Sharan and Mishra, Gaurav and Yu, Adams and Zhao, Vincent and Huang, Yanping and Dai, Andrew and Yu, Hongkun and Petrov, Slav and Chi, Ed H. and Dean, Jeff and Devlin, Jacob and Roberts, Adam and Zhou, Denny and Le, Quoc V. and Wei, Jason},
  title   = {Scaling Instruction-Finetuned Language Models},
  journal = {Journal of Machine Learning Research},
  year    = {2024},
  volume  = {25},
  number  = {70},
  pages   = {1--53},
}

@inproceedings{houlsby2019adapters,
  author    = {Houlsby, Neil and Giurgiu, Andrei and Jastrzebski, Stanislaw and Morrone, Bruna and de Laroussilhe, Quentin and Gesmundo, Andrea and Attariyan, Mona and Gelly, Sylvain},
  title     = {Parameter-Efficient Transfer Learning for {NLP}},
  booktitle = {Proceedings of the 36th International Conference on Machine Learning},
  year      = {2019},
  series    = {PMLR},
  volume    = {97},
  pages     = {2790--2799},
}

@inproceedings{pfeiffer2021adapterfusion,
  author    = {Pfeiffer, Jonas and Kamath, Aishwarya and R{\"u}ckl{\'e}, Andreas and Cho, Kyunghyun and Gurevych, Iryna},
  title     = {{AdapterFusion}: Non-Destructive Task Composition for Transfer Learning},
  booktitle = {Proceedings of the 16th Conference of the European Chapter of the Association for Computational Linguistics},
  year      = {2021},
  pages     = {487--503},
  publisher = {Association for Computational Linguistics},
  doi       = {10.18653/v1/2021.eacl-main.39},
}

@inproceedings{shen2023hugginggpt,
  author    = {Shen, Yongliang and Song, Kaitao and Tan, Xu and Li, Dongsheng and Lu, Weiming and Zhuang, Yueting},
  title     = {Hugging{GPT}: Solving {AI} Tasks with {ChatGPT} and Its Friends in Hugging Face},
  booktitle = {Advances in Neural Information Processing Systems 36},
  year      = {2023},
  pages     = {38154--38180},
  doi       = {10.52202/075280-1657},
}

@inproceedings{hong2024metagpt,
  author    = {Hong, Sirui and Zhuge, Mingchen and Chen, Jonathan and Zheng, Xiawu and Cheng, Yuheng and Wang, Jinlin and Zhang, Ceyao and Wang, Zili and Yau, Steven Ka Shing and Lin, Zijuan and Zhou, Liyang and Ran, Chenyu and Xiao, Lingfeng and Wu, Chenglin and Schmidhuber, J{\"u}rgen},
  title     = {Meta{GPT}: Meta Programming for a Multi-Agent Collaborative Framework},
  booktitle = {International Conference on Learning Representations},
  year      = {2024},
  pages     = {23247--23275},
}

@inproceedings{liu2024dylan,
  author    = {Liu, Zijun and Zhang, Yanzhe and Li, Peng and Liu, Yang and Yang, Diyi},
  title     = {A Dynamic {LLM}-Powered Agent Network for Task-Oriented Agent Collaboration},
  booktitle = {First Conference on Language Modeling},
  year      = {2024},
}

@inproceedings{dang2025evolving,
  author    = {Dang, Yufan and Qian, Chen and Luo, Xueheng and Fan, Jingru and Xie, Zihao and Shi, Ruijie and Chen, Weize and Yang, Cheng and Che, Xiaoyin and Tian, Ye and Xiong, Xuantang and Han, Lei and Liu, Zhiyuan and Sun, Maosong},
  title     = {Multi-Agent Collaboration via Evolving Orchestration},
  booktitle = {Advances in Neural Information Processing Systems 38},
  year      = {2025},
  pages     = {165025--165059},
  doi       = {10.52202/085713-5502},
}

@article{McMahanEtAl2017,
  author  = {McMahan, H. Brendan and Moore, Eider and Ramage, Daniel and Hampson, Seth and Ag{\"u}era y Arcas, Blaise},
  title   = {Communication-Efficient Learning of Deep Networks from Decentralized Data},
  journal = {Proceedings of the 20th International Conference on Artificial Intelligence and Statistics},
  series  = {Proceedings of Machine Learning Research},
  volume  = {54},
  pages   = {1273--1282},
  year    = {2017}
}

@article{LiEtAl2020,
  author  = {Li, Tian and Sahu, Anit Kumar and Zaheer, Manzil and Sanjabi, Maziar and Talwalkar, Ameet and Smith, Virginia},
  title   = {Federated Optimization in Heterogeneous Networks},
  journal = {Proceedings of Machine Learning and Systems},
  volume  = {2},
  pages   = {429--450},
  year    = {2020}
}

@article{KarimireddyEtAl2020,
  author  = {Karimireddy, Sai Praneeth and Kale, Satyen and Mohri, Mehryar and Reddi, Sashank J. and Stich, Sebastian U. and Suresh, Ananda Theertha},
  title   = {{SCAFFOLD}: Stochastic Controlled Averaging for Federated Learning},
  journal = {Proceedings of the 37th International Conference on Machine Learning},
  series  = {Proceedings of Machine Learning Research},
  volume  = {119},
  pages   = {5132--5143},
  year    = {2020}
}

@article{JordanLeeYang2019,
  author  = {Jordan, Michael I. and Lee, Jason D. and Yang, Yun},
  title   = {Communication-Efficient Distributed Statistical Inference},
  journal = {Journal of the American Statistical Association},
  volume  = {114},
  number  = {526},
  pages   = {668--681},
  year    = {2019},
  doi     = {10.1080/01621459.2018.1429274}
}

@article{DuanNingChen2022,
  author  = {Duan, Rui and Ning, Yang and Chen, Yong},
  title   = {Heterogeneity-Aware and Communication-Efficient Distributed Statistical Inference},
  journal = {Biometrika},
  volume  = {109},
  number  = {1},
  pages   = {67--83},
  year    = {2022},
  doi     = {10.1093/biomet/asab007}
}

@article{GuChen2023,
  author  = {Gu, Jiaying and Chen, Song Xi},
  title   = {Distributed Statistical Inference under Heterogeneity},
  journal = {Journal of Machine Learning Research},
  volume  = {24},
  number  = {387},
  pages   = {1--57},
  year    = {2023}
}

@article{CaiLiuXia2022,
  author  = {Cai, Tianxi and Liu, Molei and Xia, Yin},
  title   = {Individual Data Protected Integrative Regression Analysis of High-Dimensional Heterogeneous Data},
  journal = {Journal of the American Statistical Association},
  volume  = {117},
  number  = {540},
  pages   = {2105--2119},
  year    = {2022},
  doi     = {10.1080/01621459.2021.1904958}
}

@article{ChenEtAl2023,
  author  = {Chen, Shuxiao and Zheng, Qinqing and Long, Qi and Su, Weijie J.},
  title   = {Minimax Estimation for Personalized Federated Learning: An Alternative between {FedAvg} and Local Training?},
  journal = {Journal of Machine Learning Research},
  volume  = {24},
  number  = {262},
  pages   = {1--59},
  year    = {2023}
}

@article{SuXuYang2023,
  author  = {Su, Lili and Xu, Jiaming and Yang, Pengkun},
  title   = {A Non-Parametric View of {FedAvg} and {FedProx}: Beyond Stationary Points},
  journal = {Journal of Machine Learning Research},
  volume  = {24},
  number  = {203},
  pages   = {1--48},
  year    = {2023}
}

@article{ZhangLiuZhu2024,
  author  = {Zhang, Xin and Liu, Jia and Zhu, Zhengyuan},
  title   = {Learning Coefficient Heterogeneity over Networks: A Distributed Spanning-Tree-Based Fused-Lasso Regression},
  journal = {Journal of the American Statistical Association},
  volume  = {119},
  number  = {545},
  pages   = {485--497},
  year    = {2024},
  doi     = {10.1080/01621459.2022.2126363}
}

@article{ChenEtAl2024Clustered,
  author  = {Chen, Yuanxing and Zhang, Qingzhao and Ma, Shuangge and Fang, Kuangnan},
  title   = {Heterogeneity-Aware Clustered Distributed Learning for Multi-Source Data Analysis},
  journal = {Journal of Machine Learning Research},
  volume  = {25},
  number  = {211},
  pages   = {1--60},
  year    = {2024}
}

@inproceedings{GhoshEtAl2020,
  author    = {Ghosh, Avishek and Chung, Jichan and Yin, Dong and Ramchandran, Kannan},
  title     = {An Efficient Framework for Clustered Federated Learning},
  booktitle = {Advances in Neural Information Processing Systems},
  volume    = {33},
  pages     = {19586--19597},
  year      = {2020}
}

@inproceedings{MarfoqEtAl2021,
  author    = {Marfoq, Othmane and Neglia, Giovanni and Bellet, Aur{\'e}lien and Kameni, Laetitia and Vidal, Richard},
  title     = {Federated Multi-Task Learning under a Mixture of Distributions},
  booktitle = {Advances in Neural Information Processing Systems},
  volume    = {34},
  pages     = {15434--15447},
  year      = {2021}
}

@article{LiCaiLi2022,
  author  = {Li, Sai and Cai, T. Tony and Li, Hongzhe},
  title   = {Transfer Learning for High-Dimensional Linear Regression: Prediction, Estimation and Minimax Optimality},
  journal = {Journal of the Royal Statistical Society: Series B (Statistical Methodology)},
  volume  = {84},
  number  = {1},
  pages   = {149--173},
  year    = {2022},
  doi     = {10.1111/rssb.12479}
}

@article{GuoEtAl2025RIFL,
  author  = {Guo, Zijian and Li, Xiudi and Han, Larry and Cai, Tianxi},
  title   = {Robust Inference for Federated Meta-Learning},
  journal = {Journal of the American Statistical Association},
  volume  = {120},
  number  = {551},
  pages   = {1695--1710},
  year    = {2025},
  doi     = {10.1080/01621459.2024.2443246}
}

@article{HanEtAl2025FACE,
  author  = {Han, Larry and Hou, Jue and Cho, Kelly and Duan, Rui and Cai, Tianxi},
  title   = {Federated Adaptive Causal Estimation ({FACE}) of Target Treatment Effects},
  journal = {Journal of the American Statistical Association},
  volume  = {120},
  number  = {551},
  pages   = {1503--1516},
  year    = {2025},
  doi     = {10.1080/01621459.2025.2453249}
}

@article{LiDonohueTang2026,
  author  = {Li, Jinhong and Donohue, Julie M. and Tang, Lu},
  title   = {Distributed Fusion {$R$}-Learner of Heterogeneous Treatment Effect Using Distributed Medicaid Data},
  journal = {Biometrics},
  volume  = {82},
  number  = {1},
  pages   = {ujag034},
  year    = {2026},
  doi     = {10.1093/biomtc/ujag034}
}

@inproceedings{TanEtAl2022,
  author    = {Tan, Xiaoqing and Chang, Chung-Chou H. and Zhou, Ling and Tang, Lu},
  title     = {A Tree-Based Model Averaging Approach for Personalized Treatment Effect Estimation from Heterogeneous Data Sources},
  booktitle = {Proceedings of the 39th International Conference on Machine Learning},
  series    = {Proceedings of Machine Learning Research},
  volume    = {162},
  pages     = {21013--21036},
  year      = {2022}
}

@article{Murphy2003,
  author  = {Murphy, Susan A.},
  title   = {Optimal Dynamic Treatment Regimes},
  journal = {Journal of the Royal Statistical Society: Series B (Statistical Methodology)},
  year    = {2003},
  volume  = {65},
  number  = {2},
  pages   = {331--355},
  doi     = {10.1111/1467-9868.00389}
}

@article{ShiEtAl2022Value,
  author  = {Shi, Chengchun and Zhang, Sheng and Lu, Wenbin and Song, Rui},
  title   = {Statistical Inference of the Value Function for Reinforcement Learning in Infinite-Horizon Settings},
  journal = {Journal of the Royal Statistical Society: Series B (Statistical Methodology)},
  year    = {2022},
  volume  = {84},
  number  = {3},
  pages   = {765--793},
  doi     = {10.1111/rssb.12465}
}

@article{ChenSongJordan2024,
  author  = {Chen, Elynn Y. and Song, Rui and Jordan, Michael I.},
  title   = {Reinforcement Learning in Latent Heterogeneous Environments},
  journal = {Journal of the American Statistical Association},
  year    = {2024},
  volume  = {119},
  number  = {548},
  pages   = {3113--3126},
  doi     = {10.1080/01621459.2024.2308317}
}

@article{BianEtAl2025,
  author  = {Bian, Zeyu and Shi, Chengchun and Qi, Zhengling and Wang, Lan},
  title   = {Off-Policy Evaluation in Doubly Inhomogeneous Environments},
  journal = {Journal of the American Statistical Association},
  year    = {2025},
  volume  = {120},
  number  = {550},
  pages   = {1102--1114},
  doi     = {10.1080/01621459.2024.2395593}
}

@article{LiEtAl2025Stationarity,
  author  = {Li, Mengbing and Shi, Chengchun and Wu, Zhenke and Fryzlewicz, Piotr},
  title   = {Testing Stationarity and Change Point Detection in Reinforcement Learning},
  journal = {The Annals of Statistics},
  year    = {2025},
  volume  = {53},
  number  = {3},
  pages   = {1230--1256},
  doi     = {10.1214/25-AOS2501}
}

@article{MiaoShahbabaQu2025,
  author  = {Miao, Rui and Shahbaba, Babak and Qu, Annie},
  title   = {Reinforcement Learning for Individual Optimal Policy from Heterogeneous Data},
  journal = {The Annals of Statistics},
  year    = {2025},
  volume  = {53},
  number  = {4},
  pages   = {1513--1534},
  doi     = {10.1214/25-AOS2512}
}

@article{ChenLiJordan2025,
  author  = {Chen, Elynn Y. and Li, Sai and Jordan, Michael I.},
  title   = {Transfer {Q}-Learning for Finite-Horizon Markov Decision Processes},
  journal = {Electronic Journal of Statistics},
  year    = {2025},
  volume  = {19},
  number  = {2},
  pages   = {5289--5312},
  doi     = {10.1214/25-EJS2459}
}

@article{ChenChenJing2026,
  author  = {Chen, Elynn and Chen, Xi and Jing, Wenbo},
  title   = {Data-Driven Knowledge Transfer in Batch {$Q^*$} Learning},
  journal = {Journal of the American Statistical Association},
  year    = {2026},
  volume  = {121},
  number  = {553},
  pages   = {276--288},
  doi     = {10.1080/01621459.2025.2603731}
}

@article{SunZhuSun2026,
  author  = {Sun, Jubo and Zhu, Wensheng and Sun, Guozhe},
  title   = {A Concave Pairwise Fusion Approach to Heterogeneous {Q}-Learning for Dynamic Treatment Regimes},
  journal = {Statistics in Medicine},
  year    = {2026},
  volume  = {45},
  number  = {3--5},
  pages   = {e70415},
  doi     = {10.1002/sim.70415}
}

@inproceedings{ZhangWang2021,
  author    = {Zhang, Chicheng and Wang, Zhi},
  title     = {Provably Efficient Multi-Task Reinforcement Learning with Model Transfer},
  booktitle = {Advances in Neural Information Processing Systems},
  volume    = {34},
  pages     = {19771--19783},
  year      = {2021}
}

@article{FraleyRaftery2002,
  author  = {Fraley, Chris and Raftery, Adrian E.},
  title   = {Model-Based Clustering, Discriminant Analysis, and Density Estimation},
  journal = {Journal of the American Statistical Association},
  year    = {2002},
  volume  = {97},
  number  = {458},
  pages   = {611--631}
}

@article{KiselevEtAl2019,
  author  = {Kiselev, Vladimir Yu. and Andrews, Tallulah S. and Hemberg, Martin},
  title   = {Challenges in Unsupervised Clustering of Single-Cell {RNA}-Seq Data},
  journal = {Nature Reviews Genetics},
  year    = {2019},
  volume  = {20},
  number  = {5},
  pages   = {273--282},
  doi     = {10.1038/s41576-018-0088-9}
}

@article{WittenTibshirani2010,
  author  = {Witten, Daniela M. and Tibshirani, Robert},
  title   = {A Framework for Feature Selection in Clustering},
  journal = {Journal of the American Statistical Association},
  year    = {2010},
  volume  = {105},
  number  = {490},
  pages   = {713--726},
  doi     = {10.1198/jasa.2010.tm09415}
}

@article{SunWangFang2012,
  author  = {Sun, Wei and Wang, Junhui and Fang, Yixin},
  title   = {Regularized {$K$}-Means Clustering of High-Dimensional Data and Its Asymptotic Consistency},
  journal = {Electronic Journal of Statistics},
  year    = {2012},
  volume  = {6},
  pages   = {148--167},
  doi     = {10.1214/12-EJS668}
}

@article{WangEtAl2018SparseConvex,
  author  = {Wang, Binhuan and Zhang, Yilong and Sun, Will Wei and Fang, Yixin},
  title   = {Sparse Convex Clustering},
  journal = {Journal of Computational and Graphical Statistics},
  year    = {2018},
  volume  = {27},
  number  = {2},
  pages   = {393--403},
  doi     = {10.1080/10618600.2017.1377081}
}

@article{GuoEtAl2010,
  author  = {Guo, Jian and Levina, Elizaveta and Michailidis, George and Zhu, Ji},
  title   = {Pairwise Variable Selection for High-Dimensional Model-Based Clustering},
  journal = {Biometrics},
  year    = {2010},
  volume  = {66},
  number  = {3},
  pages   = {793--804},
  doi     = {10.1111/j.1541-0420.2009.01341.x}
}

@article{JinWang2016,
  author  = {Jin, Jiashun and Wang, Wanjie},
  title   = {Influential Features {PCA} for High-Dimensional Clustering},
  journal = {The Annals of Statistics},
  year    = {2016},
  volume  = {44},
  number  = {6},
  pages   = {2323--2359},
  doi     = {10.1214/15-AOS1423}
}

@article{CaiMaZhang2019,
  author  = {Cai, T. Tony and Ma, Jing and Zhang, Linjun},
  title   = {{CHIME}: Clustering of High-Dimensional Gaussian Mixtures with {EM} Algorithm and Its Optimality},
  journal = {The Annals of Statistics},
  year    = {2019},
  volume  = {47},
  number  = {3},
  pages   = {1234--1267},
  doi     = {10.1214/18-AOS1711}
}

@unpublished{ZhangEtAlPARSE2026,
  author = {Zhang, Lyuou and Tang, Xiwei and Zhou, Wen and Wang, Lulu},
  title  = {High-Resolution Feature Identification in High-Dimensional Clustering},
  year   = {2026}, 
   url     = {https://github.com/WatermeloneXT/PARSE-HighResolutionFeatureIdentification}
}

@article{tang2021individualized,
  title={Individualized multidirectional variable selection},
  author={Tang, Xiwei and Xue, Fei and Qu, Annie},
  journal={Journal of the American Statistical Association},
  volume={116},
  number={535},
  pages={1280--1296},
  year={2021},
  publisher={Taylor \& Francis}
}

@article{zhu2021longitudinal,
  title={Longitudinal clustering for heterogeneous binary data},
  author={Zhu, Xiaolu and Tang, Xiwei and Qu, Annie},
  journal={Statistica Sinica},
  volume={31},
  number={2},
  pages={603--624},
  year={2021},
  publisher={JSTOR}
}

@article{ZhangKamathTse2019,
  author  = {Zhang, Jesse M. and Kamath, Govinda M. and Tse, David N.},
  title   = {Valid Post-Clustering Differential Analysis for Single-Cell {RNA}-Seq},
  journal = {Cell Systems},
  year    = {2019},
  volume  = {9},
  number  = {4},
  pages   = {383--392.e6},
  doi     = {10.1016/j.cels.2019.07.012}
}

@article{ChenWitten2023,
  author  = {Chen, Yiqun T. and Witten, Daniela M.},
  title   = {Selective Inference for {$K$}-Means Clustering},
  journal = {Journal of Machine Learning Research},
  year    = {2023},
  volume  = {24},
  number  = {152},
  pages   = {1--41}
}

@article{GaoBienWitten2024,
  author  = {Gao, Lucy L. and Bien, Jacob and Witten, Daniela M.},
  title   = {Selective Inference for Hierarchical Clustering},
  journal = {Journal of the American Statistical Association},
  year    = {2024},
  volume  = {119},
  number  = {545},
  pages   = {332--342},
  doi     = {10.1080/01621459.2022.2116331}
}

@article{ChenGao2025,
  author  = {Chen, Yiqun T. and Gao, Lucy L.},
  title   = {Testing for a Difference in Means of a Single Feature after Clustering},
  journal = {Biostatistics},
  year    = {2025},
  volume  = {26},
  number  = {1},
  pages   = {kxae046},
  doi     = {10.1093/biostatistics/kxae046}
}

@article{NeufeldEtAl2024Biostatistics,
  author  = {Neufeld, Anna and Gao, Lucy L. and Popp, Joshua and Battle, Alexis and Witten, Daniela M.},
  title   = {Inference after Latent Variable Estimation for Single-Cell {RNA} Sequencing Data},
  journal = {Biostatistics},
  year    = {2024},
  volume  = {25},
  number  = {1},
  pages   = {270--287},
  doi     = {10.1093/biostatistics/kxac047}
}

@article{DharamshiEtAl2025,
  author  = {Dharamshi, Ameer and Neufeld, Anna and Motwani, Keshav and Gao, Lucy L. and Witten, Daniela M. and Bien, Jacob},
  title   = {Generalized Data Thinning Using Sufficient Statistics},
  journal = {Journal of the American Statistical Association},
  year    = {2025},
  volume  = {120},
  number  = {549},
  pages   = {511--523},
  doi     = {10.1080/01621459.2024.2353948}
}

@article{VandenbonDiez2020,
  author  = {Vandenbon, Alexis and Diez, Diego},
  title   = {A Clustering-Independent Method for Finding Differentially Expressed Genes in Single-Cell Transcriptome Data},
  journal = {Nature Communications},
  year    = {2020},
  volume  = {11},
  pages   = {4318},
  doi     = {10.1038/s41467-020-17900-3}
}

@unpublished{SongEtAl2023ClusterDE,
  author = {Song, Dongyuan and Li, Kexin and Ge, Xinzhou and Li, Jingyi Jessica},
  title  = {{ClusterDE}: A Post-Clustering Differential Expression Method Robust to False-Positive Inflation Caused by Double Dipping},
  year   = {2023},
  note   = {Research Square preprint},
  doi    = {10.21203/rs.3.rs-3211191/v1}
}

@article{LeeEtAl2026ClusterDE,
  author  = {Lee, Christy and Song, Dongyuan and Chen, Siqi and Li, Jingyi Jessica},
  title   = {{ClusterDE}: A Statistical Software Package for Removing Double-Dipping Bias in Post-Clustering Differential Expression Analysis},
  journal = {Journal of Computational Biology},
  year    = {2026},
  volume  = {33},
  number  = {1},
  pages   = {36--42},
  doi     = {10.1177/15578666251383562}
}

@article{AndrewsEtAl2021,
  author  = {Andrews, Tallulah S. and Kiselev, Vladimir Yu. and McCarthy, Davis and Hemberg, Martin},
  title   = {Tutorial: Guidelines for the Computational Analysis of Single-Cell {RNA} Sequencing Data},
  journal = {Nature Protocols},
  year    = {2021},
  volume  = {16},
  number  = {1},
  pages   = {1--9},
  doi     = {10.1038/s41596-020-00409-w}
}

@article{ButlerEtAl2018,
  author  = {Butler, Andrew and Hoffman, Paul and Smibert, Peter and Papalexi, Efthymia and Satija, Rahul},
  title   = {Integrating Single-Cell Transcriptomic Data across Different Conditions, Technologies, and Species},
  journal = {Nature Biotechnology},
  year    = {2018},
  volume  = {36},
  number  = {5},
  pages   = {411--420},
  doi     = {10.1038/nbt.4096}
}

@article{HaoEtAl2021,
  author  = {Hao, Yuhan and Hao, Stephanie and Andersen-Nissen, Erica and Mauck, William M. and Zheng, Shiwei and Butler, Andrew and Lee, Maddie J. and Wilk, Aaron J. and Darby, Charlotte and Zager, Michael and others},
  title   = {Integrated Analysis of Multimodal Single-Cell Data},
  journal = {Cell},
  year    = {2021},
  volume  = {184},
  number  = {13},
  pages   = {3573--3587.e29},
  doi     = {10.1016/j.cell.2021.04.048}
}

@article{WolfEtAl2018,
  author  = {Wolf, F. Alexander and Angerer, Philipp and Theis, Fabian J.},
  title   = {{SCANPY}: Large-Scale Single-Cell Gene Expression Data Analysis},
  journal = {Genome Biology},
  year    = {2018},
  volume  = {19},
  pages   = {15},
  doi     = {10.1186/s13059-017-1382-0}
}

@article{ZimmermanEtAl2021,
  author  = {Zimmerman, Kip D. and Espeland, Mark A. and Langefeld, Carl D.},
  title   = {A Practical Solution to Pseudoreplication Bias in Single-Cell Studies},
  journal = {Nature Communications},
  year    = {2021},
  volume  = {12},
  pages   = {738},
  doi     = {10.1038/s41467-021-21038-1}
}

@article{SquairEtAl2021,
  author  = {Squair, Jordan W. and Gautier, Matthieu and Kathe, Claudia and Anderson, Mark A. and James, Nicholas D. and Hutson, Thomas H. and Hudelle, Romain and Qaiser, Tahlia and Matson, Kevin J. E. and Barraud, Quentin and others},
  title   = {Confronting False Discoveries in Single-Cell Differential Expression},
  journal = {Nature Communications},
  year    = {2021},
  volume  = {12},
  pages   = {5692},
  doi     = {10.1038/s41467-021-25960-2}
}

@article{CrowellEtAl2020,
  author  = {Crowell, Helena L. and Soneson, Charlotte and Germain, Pierre-Luc and Calini, Daniela and Collin, Ludovic and Raposo, Catarina and Malhotra, Dheeraj and Robinson, Mark D.},
  title   = {{muscat} Detects Subpopulation-Specific State Transitions from Multi-Sample Multi-Condition Single-Cell Transcriptomics Data},
  journal = {Nature Communications},
  year    = {2020},
  volume  = {11},
  pages   = {6077},
  doi     = {10.1038/s41467-020-19894-4}
}

@article{LiuEtAl2020ROGUE,
  author  = {Liu, Baolin and Li, Chenwei and Li, Ziyi and Wang, Dongfang and Ren, Xianwen and Zhang, Zemin},
  title   = {An Entropy-Based Metric for Assessing the Purity of Single Cell Populations},
  journal = {Nature Communications},
  year    = {2020},
  volume  = {11},
  pages   = {3155},
  doi     = {10.1038/s41467-020-16904-3}
}

@article{XiaLeeLi2024,
  author  = {Xia, Lucy and Lee, Christy and Li, Jingyi Jessica},
  title   = {Statistical Method {scDEED} for Detecting Dubious 2D Single-Cell Embeddings and Optimizing {t-SNE} and {UMAP} Hyperparameters},
  journal = {Nature Communications},
  year    = {2024},
  volume  = {15},
  pages   = {1753},
  doi     = {10.1038/s41467-024-45891-y}
}

@article{DerSimonianLaird1986,
  author  = {DerSimonian, Rebecca and Laird, Nan},
  title   = {Meta-Analysis in Clinical Trials},
  journal = {Controlled Clinical Trials},
  year    = {1986},
  volume  = {7},
  number  = {3},
  pages   = {177--188},
  doi     = {10.1016/0197-2456(86)90046-2}
}

@article{ThompsonHiggins2002,
  author  = {Thompson, Simon G. and Higgins, Julian P. T.},
  title   = {How Should Meta-Regression Analyses Be Undertaken and Interpreted?},
  journal = {Statistics in Medicine},
  year    = {2002},
  volume  = {21},
  number  = {11},
  pages   = {1559--1573},
  doi     = {10.1002/sim.1187}
}

@inproceedings{LuEtAl2023FCP,
  author    = {Lu, Charles and Yu, Yaodong and Karimireddy, Sai Praneeth and Jordan, Michael I. and Raskar, Ramesh},
  title     = {Federated Conformal Predictors for Distributed Uncertainty Quantification},
  booktitle = {Proceedings of the 40th International Conference on Machine Learning},
  series    = {Proceedings of Machine Learning Research},
  volume    = {202},
  pages     = {22942--22964},
  year      = {2023},
  publisher = {PMLR},
  url       = {https://proceedings.mlr.press/v202/lu23i.html}
}

@inproceedings{dwork2006calibrating,
  author = {Dwork, Cynthia and McSherry, Frank and Nissim, Kobbi and Smith, Adam},
  title = {Calibrating Noise to Sensitivity in Private Data Analysis},
  booktitle = {Theory of Cryptography Conference},
  pages = {265--284},
  year = {2006},
  publisher = {Springer},
  doi = {10.1007/11681878_14}
}

@inproceedings{bagdasaryan2019differential,
  author = {Bagdasaryan, Eugene and Poursaeed, Omid and Shmatikov, Vitaly},
  title = {Differential Privacy Has Disparate Impact on Model Accuracy},
  booktitle = {Advances in Neural Information Processing Systems},
  volume = {32},
  year = {2019}
}

@inproceedings{noble2022differentially,
  author = {Noble, Maxence and Bellet, Aur{\'e}lien and Dieuleveut, Aymeric},
  title = {Differentially Private Federated Learning on Heterogeneous Data},
  booktitle = {Proceedings of the 25th International Conference on Artificial Intelligence and Statistics},
  series = {Proceedings of Machine Learning Research},
  volume = {151},
  pages = {10110--10145},
  year = {2022},
  url = {https://proceedings.mlr.press/v151/noble22a.html}
}

@inproceedings{biggio2012poisoning,
  author = {Biggio, Battista and Nelson, Blaine and Laskov, Pavel},
  title = {Poisoning Attacks against Support Vector Machines},
  booktitle = {Proceedings of the 29th International Conference on Machine Learning},
  pages = {1807--1814},
  year = {2012}
}

@misc{gu2017badnets,
  author = {Gu, Tianyu and Dolan-Gavitt, Brendan and Garg, Siddharth},
  title = {{BadNets}: Identifying Vulnerabilities in the Machine Learning Model Supply Chain},
  year = {2017},
  eprint = {1708.06733},
  archiveprefix = {arXiv},
  url = {https://arxiv.org/abs/1708.06733}
}

@inproceedings{wang2024demystifying,
  author = {Wang, Ganghua and Xian, Xun and Kundu, Ashish and Srinivasa, Jayanth and Bi, Xuan and Hong, Mingyi and Ding, Jie},
  title = {Demystifying Poisoning Backdoor Attacks from a Statistical Perspective},
  booktitle = {International Conference on Learning Representations},
  year = {2024},
  url = {https://openreview.net/forum?id=BPHcEpGvF8}
}

@inproceedings{xian2023unified,
  author = {Xian, Xun and Wang, Ganghua and Srinivasa, Jayanth and Kundu, Ashish and Bi, Xuan and Hong, Mingyi and Ding, Jie},
  title = {A Unified Detection Framework for Inference-Stage Backdoor Defenses},
  booktitle = {Advances in Neural Information Processing Systems},
  volume = {36},
  pages = {7867--7894},
  year = {2023},
  url = {https://proceedings.neurips.cc/paper_files/paper/2023/hash/1868a3c73d0d2a44c42458575fa8514c-Abstract-Conference.html}
}

@inproceedings{xian2024raw,
  author = {Xian, Xun and Wang, Ganghua and Bi, Xuan and Zhang, Rui and Srinivasa, Jayanth and Kundu, Ashish and Fleming, Charles and Hong, Mingyi and Ding, Jie},
  title = {{RAW}: A Robust and Agile Plug-and-Play Watermark Framework for AI-Generated Images with Provable Guarantees},
  booktitle = {Advances in Neural Information Processing Systems},
  volume = {37},
  year = {2024},
  url = {https://proceedings.neurips.cc/paper_files/paper/2024/hash/ee62ab636066cf45a27246acca9545b7-Abstract-Conference.html}
}

@inproceedings{jagielski2021subpopulation,
  author    = {Jagielski, Matthew and Severi, Giorgio and Pousette Harger, Niklas and Oprea, Alina},
  title     = {Subpopulation Data Poisoning Attacks},
  booktitle = {Proceedings of the ACM SIGSAC Conference on Computer and Communications Security (CCS)},
  year      = {2021}
}

@article{dathathri2024scalable,
  author  = {Dathathri, Sumanth and See, Abigail and Ghaisas, Sumedh and others},
  title   = {Scalable Watermarking for Identifying Large Language Model Outputs},
  journal = {Nature},
  volume  = {634},
  number  = {8035},
  pages   = {818--823},
  year    = {2024},
  doi     = {10.1038/s41586-024-08025-4}
}

@article{he2024watermarking,
  title   = {Watermarking Generative Tabular Data},
  author  = {He, Hengzhi and Yu, Peiyu and Ren, Junpeng and Wu, Ying Nian and Cheng, Guang},
  journal = {arXiv preprint arXiv:2405.14018},
  year    = {2024}
}

@article{zhao2025tabdrw,
  title   = {{TAB-DRW}: A {DFT}-based Robust Watermark for Generative Tabular Data},
  author  = {Zhao, Yizhou and Li, Xiang and Song, Peter and Long, Qi and Su, Weijie J.},
  journal = {arXiv preprint arXiv:2511.21600},
  year    = {2025}
}

@misc{aaronson2023watermarking,
  author       = {Scott Aaronson},
  title        = {Watermarking of Large Language Models},
  year         = {2023},
  note         = {Talk at the Simons Institute for the Theory of Computing}
}

@article{kuditipudi2023robust,
  title        = {Robust Distortion-Free Watermarks for Language Models},
  author       = {Kuditipudi, Rohith and Thickstun, John and Hashimoto, Tatsunori and Liang, Percy},
  journal      = {arXiv preprint arXiv:2307.15593},
  year         = {2023}
}

@article{li2024statistical,
  title        = {A Statistical Framework of Watermarks for Large Language Models:
                  Pivot, Detection Efficiency and Optimal Rules},
  author       = {Li, Xiang and Ruan, Feng and Wang, Huiyuan and Long, Qi and Su, Weijie J.},
  journal      = {arXiv preprint arXiv:2404.01245},
  year         = {2024}
}

@article{zhang2019rivagan,
  title        = {Robust Invisible Video Watermarking with Attention},
  author       = {Zhang, Kevin Alex and Xu, Lei and Cuesta-Infante, Alfredo
                  and Veeramachaneni, Kalyan},
  journal      = {arXiv preprint arXiv:1909.01285},
  year         = {2019}
}

@article{cui2026stamp,
  title        = {Observation-Level Watermarking and Detection for Tabular Data},
  author       = {Cui, Dongyu and Bi, Xuan},
  journal      = {arXiv preprint arXiv:2607.10554},
  year         = {2026}
}

@article{balliu2024personalized,
  author  = {Balliu, Brunilda and Douglas, Chris and Seok, Darsol and Shenhav, Liat and Wu, Yue and Chatzopoulou, Doxa and Kaiser, William and Chen, Victor and Kim, Jennifer and Deverasetty, Sandeep and Arnaudova, Inna and Gibbons, Robert and Congdon, Eliza and Craske, Michelle G. and Freimer, Nelson and Halperin, Eran and Sankararaman, Sriram and Flint, Jonathan},
  title   = {Personalized mood prediction from patterns of behavior collected with smartphones},
  journal = {npj Digital Medicine},
  year    = {2024},
  volume  = {7},
  pages   = {49},
  doi     = {10.1038/s41746-024-01035-6}
}

@article{boruvka2018assessing,
  author  = {Boruvka, Audrey and Almirall, Daniel and Witkiewitz, Katie and Murphy, Susan A.},
  title   = {Assessing Time-Varying Causal Effect Moderation in Mobile Health},
  journal = {Journal of the American Statistical Association},
  year    = {2018},
  volume  = {113},
  number  = {523},
  pages   = {1112--1121},
  doi     = {10.1080/01621459.2017.1305274}
}

@article{guo2026daily,
  author  = {Guo, Yuqing and Xu, Qi and Narvaez, Laura and Dutt, Nikil and Kehoe, Priscilla and Qu, Annie},
  title   = {Daily changes in sleep stages and associated cardiovascular parameters during pregnancy: using a wearable device},
  journal = {Frontiers in Global Women's Health},
  year    = {2026},
  volume  = {6},
  pages   = {1622895},
  doi     = {10.3389/fgwh.2025.1622895}
}

@article{klasnja2019efficacy,
  author  = {Klasnja, Predrag and Smith, Shawna and Seewald, Nicholas J. and Lee, Andy and Hall, Kelly and Luers, Brook and Hekler, Eric B. and Murphy, Susan A.},
  title   = {Efficacy of Contextually Tailored Suggestions for Physical Activity: A Micro-randomized Optimization Trial of {HeartSteps}},
  journal = {Annals of Behavioral Medicine},
  year    = {2019},
  volume  = {53},
  number  = {6},
  pages   = {573--582},
  doi     = {10.1093/abm/kay067}
}

@article{liao2020personalized,
  author  = {Liao, Peng and Greenewald, Kristjan and Klasnja, Predrag and Murphy, Susan},
  title   = {Personalized {HeartSteps}: A Reinforcement Learning Algorithm for Optimizing Physical Activity},
  journal = {Proceedings of the ACM on Interactive, Mobile, Wearable and Ubiquitous Technologies},
  year    = {2020},
  volume  = {4},
  number  = {1},
  pages   = {18},
  doi     = {10.1145/3381007}
}

@article{phanord2026digital,
  author  = {Phanord, Coralie S. and Ruzic, Luka L. and Kalyanasundaram, Sujay and Barnes-Horowitz, Sydney and Friedman, Naomi P. and Chaspari, Theodora and Kaiser, Roselinde H.},
  title   = {Digital phenotyping of affect and stress in emerging adults},
  journal = {Frontiers in Digital Health},
  year    = {2026},
  volume  = {8},
  pages   = {1799541},
  doi     = {10.3389/fdgth.2026.1799541}
}

@article{rim2025aheterogeneous,
  author  = {Rim, Jenifer and Xu, Qi and Tang, Xiwei and Jimah, Tamara and Guo, Yuqing and Qu, Annie},
  title   = {Heterogeneous effects of physical activity on physiological stress during pregnancy},
  journal = {PLOS Digital Health},
  year    = {2025},
  volume  = {4},
  number  = {10},
  pages   = {e0000837},
  doi     = {10.1371/journal.pdig.0000837}
}

@article{rim2025bindividualized,
  author  = {Rim, Jenifer and Xu, Qi and Tang, Xiwei and Guo, Yuqing and Qu, Annie},
  title   = {Individualized Time-Varying Nonparametric Model With an Application in Mobile Health},
  journal = {Statistics in Medicine},
  year    = {2025},
  volume  = {44},
  pages   = {e70005},
  doi     = {10.1002/sim.70005}
}

@article{chen2009hypothesis,
  author  = {Chen, Jiahua and Li, Pengfei},
  title   = {Hypothesis Test for Normal Mixture Models: The {EM} Approach},
  journal = {The Annals of Statistics},
  year    = {2009},
  volume  = {37},
  number  = {5A},
  pages   = {2523--2542},
  doi     = {10.1214/08-AOS651}
}

@article{hennig2007clusterwise,
  author  = {Hennig, Christian},
  title   = {Cluster-wise Assessment of Cluster Stability},
  journal = {Computational Statistics \& Data Analysis},
  year    = {2007},
  volume  = {52},
  number  = {1},
  pages   = {258--271},
  doi     = {10.1016/j.csda.2006.11.025}
}

@article{kackar1984approximations,
  author  = {Kackar, Raghu N. and Harville, David A.},
  title   = {Approximations for Standard Errors of Estimators of Fixed and Random Effects in Mixed Linear Models},
  journal = {Journal of the American Statistical Association},
  year    = {1984},
  volume  = {79},
  number  = {388},
  pages   = {853--862},
  doi     = {10.1080/01621459.1984.10477102}
}

@article{datta2000unified,
  author  = {Datta, Gauri Sankar and Lahiri, Partha},
  title   = {A Unified Measure of Uncertainty of Estimated Best Linear Unbiased Predictors in Small Area Estimation Problems},
  journal = {Statistica Sinica},
  year    = {2000},
  volume  = {10},
  number  = {2},
  pages   = {613--627},
  url     = {https://www3.stat.sinica.edu.tw/statistica/j10n2/j10n214/j10n214.htm}
}

@article{tibshirani2005prediction,
  author  = {Tibshirani, Robert and Walther, Guenther},
  title   = {Cluster Validation by Prediction Strength},
  journal = {Journal of Computational and Graphical Statistics},
  year    = {2005},
  volume  = {14},
  number  = {3},
  pages   = {511--528},
  doi     = {10.1198/106186005X59243}
}

@article{morris1983parametric,
  author  = {Morris, Carl N.},
  title   = {Parametric Empirical {Bayes} Inference: Theory and Applications},
  journal = {Journal of the American Statistical Association},
  year    = {1983},
  volume  = {78},
  number  = {381},
  pages   = {47--55},
  doi     = {10.1080/01621459.1983.10477920}
}

@inproceedings{tibshirani2019covariate,
  author    = {Tibshirani, Ryan J. and Barber, Rina Foygel and Cand{\`e}s, Emmanuel J. and Ramdas, Aaditya},
  title     = {Conformal Prediction Under Covariate Shift},
  booktitle = {Advances in Neural Information Processing Systems},
  year      = {2019},
  volume    = {32},
  publisher = {Curran Associates, Inc.}
}

@article{guo2021inference,
  author  = {Guo, Xinzhou and He, Xuming},
  title   = {Inference on Selected Subgroups in Clinical Trials},
  journal = {Journal of the American Statistical Association},
  year    = {2021},
  volume  = {116},
  number  = {535},
  pages   = {1498--1506},
  doi     = {10.1080/01621459.2020.1740096}
}

@article{neufeld2022treevalues,
  author  = {Neufeld, Anna C. and Gao, Lucy L. and Witten, Daniela M.},
  title   = {{Tree-Values}: Selective Inference for Regression Trees},
  journal = {Journal of Machine Learning Research},
  year    = {2022},
  volume  = {23},
  number  = {305},
  pages   = {1--43},
  url     = {https://www.jmlr.org/papers/v23/21-0722.html}
}

@article{berger2014bayesian,
  author  = {Berger, James O. and Wang, Xiaojing and Shen, Lei},
  title   = {A Bayesian Approach to Subgroup Identification},
  journal = {Journal of Biopharmaceutical Statistics},
  year    = {2014},
  volume  = {24},
  number  = {1},
  pages   = {110--129},
  doi     = {10.1080/10543406.2013.856026}
}

@article{schnell2016bayesian,
  author  = {Schnell, Patrick M. and Tang, Qi and Offen, Walter W. and Carlin, Bradley P.},
  title   = {A Bayesian Credible Subgroups Approach to Identifying Patient Subgroups with Positive Treatment Effects},
  journal = {Biometrics},
  year    = {2016},
  volume  = {72},
  number  = {4},
  pages   = {1026--1036},
  doi     = {10.1111/biom.12522}
}

@article{wade2018bayesian,
  author  = {Wade, Sara and Ghahramani, Zoubin},
  title   = {Bayesian Cluster Analysis: Point Estimation and Credible Balls},
  journal = {Bayesian Analysis},
  year    = {2018},
  volume  = {13},
  number  = {2},
  pages   = {559--626},
  doi     = {10.1214/17-BA1073}
}

@article{romano2020malice,
  author  = {Romano, Yaniv and Barber, Rina Foygel and Sabatti, Chiara and Cand{\`e}s, Emmanuel J.},
  title   = {With Malice Toward None: Assessing Uncertainty via Equalized Coverage},
  journal = {Harvard Data Science Review},
  year    = {2020},
  volume  = {2},
  number  = {2},
  doi     = {10.1162/99608f92.03f00592}
}

@inproceedings{jung2023batch,
  author    = {Jung, Christopher and Noarov, Georgy and Ramalingam, Ramya and Roth, Aaron},
  title     = {Batch Multivalid Conformal Prediction},
  booktitle = {The Eleventh International Conference on Learning Representations},
  year      = {2023},
  url       = {https://openreview.net/forum?id=Dk7QQp8jHEo}
}

@article{gibbs2025conditional,
  author  = {Gibbs, Isaac and Cherian, John J. and Cand{\`e}s, Emmanuel J.},
  title   = {Conformal Prediction with Conditional Guarantees},
  journal = {Journal of the Royal Statistical Society Series B: Statistical Methodology},
  year    = {2025},
  volume  = {87},
  number  = {4},
  pages   = {1100--1126},
  doi     = {10.1093/jrsssb/qkaf008}
}

@inproceedings{jung2021moment,
  author    = {Jung, Christopher and Lee, Changhwa and Pai, Mallesh M. and Roth, Aaron and Vohra, Rakesh},
  title     = {Moment Multicalibration for Uncertainty Estimation},
  booktitle = {Proceedings of the 34th Conference on Learning Theory},
  year      = {2021},
  series    = {Proceedings of Machine Learning Research},
  volume    = {134},
  pages     = {2634--2678},
  publisher = {PMLR},
  url       = {https://proceedings.mlr.press/v134/jung21a.html}
}

@article{ramazzotti2018multiomic,
  author  = {Ramazzotti, Daniele and Lal, Avantika and Wang, Bo and Batzoglou, Serafim and Sidow, Arend},
  title   = {Multi-omic tumor data reveal diversity of molecular mechanisms that correlate with survival},
  journal = {Nature Communications},
  year    = {2018},
  volume  = {9},
  pages   = {4453},
  doi     = {10.1038/s41467-018-06921-8}
}

@article{vanvogt2025application,
  author  = {Van Vogt, Eleanor and Gordon, Anthony C. and Diaz-Ordaz, Karla and Cro, Suzie},
  title   = {Application of causal forests to randomised controlled trial data to identify heterogeneous treatment effects: a case study},
  journal = {BMC Medical Research Methodology},
  year    = {2025},
  volume  = {25},
  number  = {1},
  pages   = {50},
  doi     = {10.1186/s12874-025-02489-2}
}

@book{scheffe1959analysis,
  author = {Scheff{\'e}, Henry},
  title = {The Analysis of Variance},
  publisher = {John Wiley \& Sons},
  year = {1959}
}

@inproceedings{staab2024beyond,
  author    = {Staab, Robin and Vero, Mark and Balunovi{\'c}, Mislav and Vechev, Martin},
  title     = {Beyond Memorization: Violating Privacy via Inference with Large Language Models},
  booktitle = {International Conference on Learning Representations},
  year      = {2024}
}

@inproceedings{ren2023robots,
  author    = {Ren, Allen Z. and Dixit, Anushri and Bodrova, Alexandra and Singh, Sumeet and Tu, Stephen and Brown, Noah and Xu, Peng and Takayama, Leila and Xia, Fei and Varley, Jake and Xu, Zhenjia and Sadigh, Dorsa and Zeng, Andy and Majumdar, Anirudha},
  title     = {Robots That Ask For Help: Uncertainty Alignment for Large Language Model Planners},
  booktitle = {Proceedings of the 7th Conference on Robot Learning},
  year      = {2023}
}

@inproceedings{yao2025taubench,
  author    = {Yao, Shunyu and Shinn, Noah and Razavi, Pedram and Narasimhan, Karthik},
  title     = {{$\tau$}-bench: A Benchmark for Tool-Agent-User Interaction in Real-World Domains},
  booktitle = {International Conference on Learning Representations},
  year      = {2025}
}

\end{document}